\documentstyle[11pt,epsfig]{article}

\def\appendix{\par
 \setcounter{section}{0}
 \setcounter{subsection}{0}
 \def\thesection{Appendix \Alph{section}}
 \def\theequation{\Alph{section}.\arabic{equation}}
 \setcounter{equation}{0}}

\begin{document}

\begin{flushright}
hep-ph/9605433\\
revised version (November 1996)
\end{flushright}
\centerline{\Large \bf Regularization}
\centerline{\Large \bf at the next-to-leading order in the}
\centerline{\Large \bf top-mode standard model without gauge bosons}

\vspace{1.2cm}

\begin{center}
{{\bf G.~Cveti\v c}\\Inst.~f\"ur Physik, Universit\"at Dortmund, 
44221 Dortmund, Germany\\}
\end{center}

\vspace{1.2cm} 

\centerline{\bf Abstract}

We study Higgs condensation ${\cal{H}} \sim {\bar t} t $ in the 
top-mode standard model at the next-to-leading (NTL) order in 
$1/N_{\mbox{\footnotesize c}}$, by calculating the effective 
potential as a function of a hard mass term ${\sigma}_0$ of 
the top quark. We include the effects of the third generation 
quarks, the Higgs and the 
Goldstone fields, and the leading QCD effects, but not the effects 
of the transverse components of the electroweak gauge bosons. 
The resulting effective theory contains finite energy cutoff 
parameters (${\Lambda}_{\mbox{\footnotesize f}}$,
${\Lambda}_{\mbox{\footnotesize b}}$) for the fermionic and the bosonic
degrees of freedom. Condensation is supposed to take place at
energies ${\Lambda} \sim {\Lambda}_{\mbox{\footnotesize f}} \sim 
{\Lambda}_{\mbox{\footnotesize b}}$. The paper describes how
to regularize the integrals over the fermionic momenta
in a way free of momentum branching ambiguities and
how to treat the terms of $1/N_{\mbox{\footnotesize c}}$ expansion
mutually consistently. This is achieved by the proper time
approach, employing specifically the proper time
cutoff (PTC) or a Pauli-Villars (PV) regulator. For comparison, 
we use also the covariant spherical (S) cutoff. 
On the other hand, for the bosonic momenta we have to use the 
covariant spherical cutoff. We discuss how 
to ensure the validity of Goldstone theorem. Demanding that 
the NTL contributions not ``wash out'' the leading 
ones, we obtain rather low bounds for the cutoffs: $\Lambda =
{\cal{O}}(1 \mbox{TeV})$. The results for the corresponding cases with
PTC, PV and S regularization differ only marginally. Thus, in the
described framework,
$1/N_{\mbox{\footnotesize c}}$ expansion approach has 
a predictive power only if $\Lambda={\cal{O}}(1 \mbox{TeV})$,
a result largely independent of the regularization procedure.

\newpage

\section{Introduction}

Higgs meson could be a bound state of heavy quark 
pairs~\cite{nambu}-~\cite{yamawaki} (and references 
therein). This idea originates from an earlier work 
of Nambu and Jona-Lasinio (NJL)~\cite{njl} where it was 
applied to low energy QCD, and independently from
Vaks and Larkin~\cite{vl}. The bound states (condensates) are 
treated in these works either in the 
leading-$N_{\mbox{\footnotesize c}}$ approximation, or in a form 
that takes into account part of the effects beyond the 
leading-$N_{\mbox{\footnotesize c}}$ 
-- by using improved Schwinger-Dyson equations, or  
renormalization group equations (RGEs). A particularly 
straightforward NJL-type framework, containing the essential 
features of the idea of ${\bar t} t$ condensation, is the top-mode
standard model (TSM) Lagrangian, in its simplified form
known also as the BHL  
(Bardeen-Hill-Lindner) Lagrangian~\cite{bhl}.

In recent works, we studied the next-to-leading order
(NTL) contributions in $1/N_{\mbox{\footnotesize c}}$ expansion
in TSM by including quadratic fluctuations of the composite 
Higgs ${\cal{H}} \sim {\bar t} t $~\cite{cpv},
as well as those of the composite Goldstones~\cite{cv} in the
effective potential $V_{\mbox{\footnotesize eff}}$. We were
using for the fermionic (quark) and the bosonic momenta the simple  
covariant spherical cutoffs ${\Lambda}_{\mbox{\footnotesize f}}$,
${\Lambda}_{\mbox{\footnotesize b}}$.
These two cutoffs are indicative of the energy at which 
the condensation is supposed to occur: $\Lambda \sim
\Lambda_{\mbox{\footnotesize f}} \sim \Lambda_{\mbox{\footnotesize b}}$.
QCD effects were included, but their impact was 
found to be small. 
We considered the effective potential as a function
of a hard bare mass term $\sigma_0$ of the top quark,
i.e., the expectation value $\sigma_0$ of a composite
(initially auxiliary) Higgs field $\tilde{\sigma} \propto {\cal{H}}$.
We demanded that the NTL contributions not ``wash out''
the leading-$N_{\mbox{\footnotesize c}}$ ones, or equivalently,
that the $1/N_{\mbox{\footnotesize c}}$ expansion
approach in this framework have predictive power. As a
direct consequence of this demand, we obtained that the
energy at which the ${\bar t} t$ pair is supposed to condense
into a Higgs is rather low: $\Lambda = {\cal{O}}(1 \mbox{TeV})$.

In the present work, we continue the work of Refs.~\cite{cpv}-\cite{cv}.
We concentrate on various regularization procedures and the dependence
of the mentioned results on these procedures. 
The reasons for this are at least two:

Firstly, objections may be raised against the results of~\cite{cpv},
\cite{cv}, on the grounds that they may be largely the
consequence of our choosing only one specific regularization
-- the simple covariant spherical cutoff, and that other
regularizations may give substantially different results.

Secondly, theoretical objections had been raised in the past
against using simple covariant spherical cutoffs for
the quark momenta~\cite{willey}, particularly because of the
momentum branching ambiguity which has its origin in the
translational noninvariance of the procedure. 

One possibility to remedy this problem is to use dispersion 
relations (DR) with finite cutoff
for the bubble-chain-corrected scalar propagators -- this 
method was applied by the authors of Refs.~\cite{njl}
and~\cite{gherghetta} at the  leading-$N_{\mbox{\footnotesize c}}$ order.
The leading-$N_{\mbox{\footnotesize c}}$ gap equation,
determining the relationship between the four-fermion coupling,
the dynamical quark mass and the cutoff is then obtained
by requiring that the pole of the Goldstone propagator be
located at the zero momentum. However, when we want to include 
the NTL corrections to the gap equation with the method of the
effective potential, it is unclear how to relate the
DR cutoff appearing in the NTL part $V^{(1)}$ of the effective
potential (containing as integrands the logarithms of the
bubble-chain-corrected scalar propagators) and
the cutoff in the leading-$N_{\mbox{\footnotesize c}}$ part $V^{(0)}$.
For the latter part, there seems to be no dispersion relation
representation available, because the integrand there contains
only one (quark) momentum. 

Therefore, we apply in this paper the so-called proper time
regularization approach for the fermionic momenta, using the
Schwinger representation for the ``bosonized'' effective action.
This regularization, in contrast to the spherical cutoff, does not
suffer from the momentum branching ambiguity problems, since
it is translationally invariant in the momentum space. Furthermore,
in contrast to the DR cutoff regularization, the present procedure
treats mutually consistently the 
leading-$N_{\mbox{\footnotesize c}}$ and the NTL contributions,
due to the introduction of a fermion momentum regulator
function that is common to both contributions. 

In Sec.~II, we calculate the effective potential, including
the NTL contributions. We take into account the effects of the third
generation quarks, and at the NTL level in addition the effects of 
the Higgs and the three Goldstones. The latter degrees of freedom
represent, in the Landau gauge, the longitudinal degrees of
freedom of the $Z$ and $W^{\pm}$.
The effects of the transverse components of the
electroweak gauge bosons are not included.
Several technical details of the derivation, in the proper time
regularization framework, are given in Appendices A and B.
We employ two different fermionic regulators
in the proper time regularization framework: the proper time 
cutoff (PTC) regulator, and the Pauli-Villars (PV)
two-subtractions regulator; for additional comparison, we also employ the
simple covariant spherical cutoff (S) as used in Ref.~\cite{cv}.
For the bosonic momenta, which appear only at the NTL level
and possess no momentum branching ambiguity in our problem,
the proper time approach turns out not to lead to a
regularization, and therefore
we always use the covariant spherical cutoff there.

In Sec.~III, we derive the NTL gap equation, i.e., the requirement
of the minimum of the effective potential in the vacuum.
The solutions of the NTL gap equation give us the ratio
$(m_t(\Lambda)/\Lambda_{\mbox{\footnotesize f}})$
of the bare mass of the top quark and the fermionic cutoff parameter.
Furthermore, we discuss in detail how to eliminate the singularities
in the NTL integrals over the bosonic momenta in all (PTC, PV, S) 
cases, thus ensuring that Goldstone theorem is respected at the NTL 
level.

In Sec.~IV, we include the mass renormalization
effects. We also include the leading QCD effects -- in the gap
equation and in the mass renormalization.

In Sec.~V, we investigate numerically the gap equation solutions
($\mapsto m_t(\Lambda)/\Lambda_{\mbox{\footnotesize f}})$) and their
mass renormalization 
($\mapsto m_t^{\mbox{\scriptsize ren.}}/
\Lambda_{\mbox{\footnotesize f}}$).
The obtained values of the latter ratios lead to the values of the cutoff
parameters ${\Lambda}_{\mbox{\footnotesize f}}$ and
${\Lambda}_{\mbox{\footnotesize b}}$ (using: 
$m_t^{\mbox{\scriptsize ren.}} \approx 180 \mbox{ GeV}$).
We also compare in Sec.~V how the choice of regularization of
the fermionic momenta (PTC, PV, S) influences the numerical
results, in particular the NTL corrections to the solutions
of the gap equation. It turns out that these corrections are in
the S regularization cases somewhat weaker and in the PV cases
somewhat stronger than in the corresponding PTC cases.
We demand that the leading-$N_{\mbox{\footnotesize c}}$ and
the NTL solutions of the gap equation not differ drastically,
i.e., that the NTL corrections not ``wash out'' the
leading-$N_{\mbox{\footnotesize c}}$ solutions. 
As a consequence, it turns out that the cutoff parameters are
rather low:
$\Lambda_{\mbox{\footnotesize f}}, \Lambda_{\mbox{\footnotesize b}}
 \leq {\cal{O}}(1 \mbox{TeV})$.

In Sec.~VI, we recapitulate the basic conclusions of the paper
and compare it with related works of other authors and, in particular,
with the works~\cite{mty}-\cite{king}.

\section{The model, the effective potential, proper time regularization}

The top-mode standard model (TSM) Lagrangian~\cite{bhl}, known also
as the BHL (Bardeen-Hill-Lindner) Lagrangian, contains a 
truncated $SU(2)_L \times U(1)_Y$ invariant four-fermion interaction
at a high energy scale 
$E \sim \Lambda$. This term is assumed to be responsible for the 
creation of 
a composite Higgs field ${\cal{H}} \sim {\bar t} t $
\begin{equation}
{\cal {L}}^{(\Lambda)}={\cal {L}}_{\mbox{\footnotesize kin}}^0
+G \left( \bar \Psi _{\mbox{\scriptsize L}}^{ia}
 t_{\mbox{\scriptsize R}a}\right) 
 \left( \bar t_{\mbox{\scriptsize R}}^b
 \Psi _{\mbox{\scriptsize L}b}^i\right) \qquad 
\mbox{for}\ E\sim \Lambda \ .  
\label{TSM}
\end{equation}
In this expression, $a$ and $b$ are the color and $i$ the isospin indices, 
$\Psi_{\mbox{\scriptsize L}}^T=(t_{\mbox{\scriptsize L}},
b_{\mbox{\scriptsize L}})$. 
${\cal {L}}_{\mbox {\footnotesize kin}}^0$ contains the usual
gauge invariant kinetic terms for fermions and gauge bosons.
The model (\ref{TSM}) is a specific Nambu--Jona-Lasinio
(NJL) type model with the Standard Model symmetry
$SU(2)_L \times U(1)_Y$.
It leads to an effective framework for the 
minimal Standard Model. It can be rewritten in terms of an 
additional, as yet auxiliary, scalar $SU(2)_L$
isodoublet $\Phi $, by adding 
to it the following quadratic term
\[
{\cal {L}}^{(\Lambda)}_{\mbox {\footnotesize new}}=
{\cal {L}}^{(\Lambda)}_{\mbox{\footnotesize old}}-\left[ 
 M_0 \tilde \Phi ^{i\dagger }+\sqrt{G}
 \bar \Psi _{\mbox{\scriptsize L}}^{ia} 
 t_{\mbox{\scriptsize R}a}\right] \left[ 
 M_0{\tilde \Phi }^i+\sqrt{G}\bar 
 t_{\mbox{\scriptsize R}}^b{\Psi }_{\mbox{\scriptsize L} b}^i
 \right] \ ,
\]
\begin{equation}
\label{defPhi}\mbox{where:}\quad 
\tilde \Phi =i\tau _2\Phi ^{*}\ ,\qquad
\Phi =\frac 1{\sqrt{2}}\left( 
\begin{array}{c}
\sqrt{2}{\cal {G}}^{(+)} \\ {\cal H}+i{\cal {G}}^{(0)}
\end{array}
\right) \ ,\qquad 
{\cal {G}}^{(\pm)}= \frac{1}{\sqrt{2}}
 ({\cal {G}}^{(1)} \pm i {\cal {G}}^{(2)}) \ .
\end{equation}

The addition of such a term changes the generating functional
only by a source independent factor~\cite{kugo}.
Here, ${\cal H}$, ${\cal G}^{(0)}$, ${\cal G}^{(1)}$ and 
${\cal G}^{(2)}$ are the Higgs and the three real Goldstone 
components of the auxiliary complex isodoublet field $\Phi $, 
and $M_0$ is an unspecified bare mass term for $\Phi $ at 
$E\sim \Lambda $. 
The physical results will turn out to be independent of the 
specific value of $M_0$.
These scalar fields eventually become the physical Higgs and the
``scalar'' longitudinal components of the massive electroweak bosons 
through quantum effects. We ignore the transverse
components of $W^{\pm }$ and $Z$, and all the lighter quarks which 
we assume to be and remain massless. It can be shown that the 
massless Goldstones discussed here correspond to the Goldstone 
degrees of freedom of $W^{\pm }$ and $Z$ in the Landau gauge 
($\xi \to \infty $). In this gauge, 
the ghosts do not couple to the scalar degrees of freedom and 
therefore the ghosts do not contribute to the effective 
potential~\cite{weinberg}, at least not at the 
leading-$N_{\mbox{\footnotesize c}}$ and the NTL level.

The resulting effective Lagrangian now reads
\begin{equation}
{\cal {L}}^{(\Lambda)} = -{\bar \Psi}^a {\hat {\cal{D}}} \Psi_a -
M_0^2 \Phi^{\dagger} \Phi  \ ,
\label{efflagr}
\end{equation}
where
\begin{equation}
{\hat{\cal{D}}} = - {{\hat{P}} \llap /} +
\left[ 
\begin{array}{cc}
\left( {\tilde \sigma}_0 - i {\gamma}_5 {\tilde \sigma}_1 \right)
& \left[ - \frac{1}{2} \left( 1- \gamma_5 \right)
 \left( {\tilde \sigma}_2 +i {\tilde \sigma}_3 \right) \right] \\
\left[ - \frac{1}{2} \left( 1+ \gamma_5 \right)
 \left( {\tilde \sigma}_2 -i {\tilde \sigma}_3 \right) \right]
& 0 
\end{array}
\right] \ .
\label{Ddef}
\end{equation}
We used for the auxiliary scalar fields the following notation:
\begin{equation}
{\tilde{\sigma}}_0  =  \sqrt{ \frac{g}{2} } {\cal{H}} \ ,
\qquad {\tilde{\sigma}}_j = \sqrt{ \frac{g}{2} } {\cal{G}}^{(j-1)}
\quad (j=1,2,3)  ,
\label{notat1}
\end{equation}
where we introduced a dimensionless coupling constant $g = G M_0^2$.
Equations of motion give
\begin{equation}
{\tilde{\sigma}}_0 = - \frac{G}{2} \left( \bar t^a t_a \right) \ , \qquad
{\tilde \sigma}_1 =  i \frac{G}{2} \left( \bar t^a \gamma_5 t_a \right) \ ,
\qquad
{\tilde{\sigma}}^{(-)} \left( = \frac{1}{\sqrt{2}} ({\tilde \sigma}_2 - i
{\tilde \sigma}_3) \right) =
 \frac{G}{\sqrt{2}} \left[ \bar t^a \left( \frac{1-\gamma_5}{2} \right)
b_a \right] \ ,
\label{eqmot}
\end{equation}
thus signaling that the introduced auxiliary fields, once they become
physical (dynamical) through quantum effects, will represent the
composite Higgs and the composite Goldstone 
(i.e., the composite longitudinal $Z$ and $W$) degrees of freedom.

The effective potential $V_{\mbox{\footnotesize eff}}(\Phi_0)$,
as a function of the expectation values $\Phi_0$ of the scalar
fields, can then be calculated in the Wick rotated
Euclidean space by means of the following general formula 
\begin{eqnarray}
\lefteqn{
\exp \left[ -\Omega V_{\mbox{\footnotesize eff}}({\cal{H}}_0,
{\cal{G}}_0^{(j)} )  \right]
=\mbox{const.} \times
\int \prod_{j=0}^2 \left[ {\cal{D}}{\cal{G}}^{(j)}
\delta \left( \int d^4 {\bar y} {\cal{G}}^{(j)}({\bar y})
-\Omega {\cal{G}}_0^{(j)}  \right)
\right] \times }
\nonumber \\
&&
 \int {\cal{D}}{\cal{H}}
\delta \left( \int d^4 {\bar y} {\cal{H}} ({\bar y})-\Omega {\cal{H}}_0 \right)
\int {\cal{D}} {\bar \Psi} {\cal{D}} \Psi
\exp \left[ +\int d^4 {\bar x} {\cal{L}}({\bar x}) \right] \ ,
\label{pathint1}
\end{eqnarray}
where we set $\hbar = 1$.
The bars over space-time components, derivatives and momenta from 
now on denote Euclidean quantities. $\Omega $ is the four-dimensional 
volume (formally infinite). The effective potential is 
the energy density of the physical ground state when the order parameters 
${\cal{H}}_0= \langle {\cal{H}} \rangle $ and
${\cal{G}}^{(j)}_0 = \langle {\cal{G}}^{(j)} \rangle = 0$ ($j=0,1,2$) 
are kept fixed.
In the expression (\ref{pathint1}) we explicitly kept
also the expectation values ${\cal{G}}^{(j)}_0$, in order to
show later the explicit $SU(2)_L \times U(1)_Y$ invariance of 
the expression. This formula is equivalent to a conventional
expression for the effective potential obtained by using 
Legendre transformation
of the ground state energy density, as shown by 
Higashijima~\cite{higa}~\footnote{
Higashijima has shown the equivalence of the two approaches for the
case of one scalar component (Higgs); however, the extension of his
argument to several scalar components with nonzero expectation
values is straightforward.}.

As the next step, we could simply integrate out the quark degrees of 
freedom,
as was done in Ref.~\cite{cv} (for the case of ${\cal{G}}_0^{(j)}=0$).
This would then naturally lead us to the imposition of the covariant
spherical cutoff on the fermionic momenta. At this stage, however,
we decide to follow the proper time method which will give us
the possibility of regularizing the fermionic momenta in the
leading-$N_{\mbox{\footnotesize c}}$ 
and in the NTL terms in a mutually consistent way, as mentioned earlier.
First we note that the fermionic operator ${\hat{\cal{D}}}$ is not
positive definite and not hermitian, while
${\hat{\cal{D}}}^{\dagger}{\hat{\cal{D}}}$ is.
Furthermore, the integration over the fermionic degrees of freedom
would result in no imaginary part for the effective action, i.e.,
the Lagrangian (without the electroweak gauge bosons) has no
anomalous terms. Therefore, the following identities can be used
\begin{equation}
\int {\cal{D}}{\bar \Psi} {\cal{D}} \Psi
\exp \left[ -\int d^4 {\bar x}
{\bar \Psi} {\hat{\cal{D}}} \Psi \right] =
\exp \left[ \mbox{Tr} \ln {\hat{\cal{D}}} \right]  =
\mbox{Re} \exp \left[ \mbox{Tr} \ln {\hat{\cal{D}}} \right]
= \exp \left[ \frac{1}{2} \mbox{Tr}
\ln {\hat{\cal{D}}}^{\dagger} {\hat{\cal{D}}} \right] \ .
\label{Dsquare}
\end{equation}
Furthermore, this ``bosonized'' expression can be written 
in the Schwinger representation,
as an integral over a ``proper time'' $\tau$
\begin{equation}
\mbox{Tr} \ln {\hat{\cal{D}}}^{\dagger}{\hat{\cal{D}}} \left( \Phi \right)
-\mbox{Tr} \ln {\hat{\cal{D}}}^{\dagger}{\hat{\cal{D}}} \left( 0 \right) =
 - \int_0^{\infty} \frac{d \tau}{\tau} \rho_{\mbox{\scriptsize f}}(\tau)
{\Big \{ } \mbox{Tr} \exp \left[
-\tau {\hat{\cal{D}}}^{\dagger}{\hat{\cal{D}}} \left( \Phi \right) \right]
 -  \mbox{Tr} \exp \left[
- \tau {\hat{\cal{D}}}^{\dagger}{\hat{\cal{D}}} \left( 0 \right) \right]
{\Big \} } \ .
\label{schwing}
\end{equation}
The fermionic proper time regulator
$\rho_{\mbox{\scriptsize f}}(\tau)$ has been introduced,
satisfying the conventional boundary conditions~\footnote{
The proper time techniques are described, for example, in a review 
article by R.D.~Ball~\cite{ball}; the Schwinger representation
(\ref{schwing}) has its origin in the mathematical identity
(\ref{li}).}
\begin{equation}
\rho_{\mbox{\scriptsize f}}(\tau)  =  1 \quad
\mbox{for: } \tau \gg 1/\Lambda_{\mbox{\footnotesize f}}^2 \ ;
\quad \rho_{\mbox{\scriptsize f}}(\tau)  =  0 \quad
\mbox{for: } \tau \ll 1/\Lambda_{\mbox{\footnotesize f}}^2 \ ,
\label{rhofbc}
\end{equation}
where ${\Lambda}_{\mbox{\footnotesize f}}$ represents an effective
cutoff parameter for the fermionic momenta; it is of the same order
of magnitude as the cutoff of the discussed effective theory
(\ref{TSM})-(\ref{Ddef}): ${\Lambda}_{\mbox{\footnotesize f}}
\sim {\Lambda}$.
After integrating out the fermionic (quark) degrees of freedom in
this way, we obtain
\begin{eqnarray}
\lefteqn{
\exp \left[ - \Omega V_{\mbox{\footnotesize eff}}
\left( \lbrace {\sigma_j} \rbrace \right) \right]  =
\mbox{const.} \times \prod_{j=0}^3 \int_{-\infty}^{+\infty} dJ_j
\int {\cal{D}}s_0 {\cal{D}}s_1 {\cal{D}}s_2{\cal{D}}s_3 \times 
\exp {\Big \{ } - \frac{1}{2} \int_0^{\infty} \frac{d \tau}{\tau}
\rho_{\mbox{\scriptsize f}}( \tau ) \times
}
\nonumber\\
& & \left[ \mbox{Tr} e^{- \tau {\hat{\cal{D}}}^{\dagger}{\hat{\cal{D}}}
\left( \lbrace \sigma_i +s_i \rbrace \right) }
- \mbox{Tr} e^{- \tau {\hat{\cal{D}}}^{\dagger}{\hat{\cal{D}}}
\left( 0 \right) } \right]
- \frac{1}{G}
\int d^4 {\bar x} \sum_{k=0}^3 \left( \sigma_k + s_k \right)^2
- i \sum_{k=0}^3 J_k \int d^4 {\bar x} s_k \left( {\bar x} \right)
{\Big \} } \ .
\label{veffsc}
\end{eqnarray}
We denoted by $s_k({\bar x})$ the rescaled quantum fluctuations
of the Higgs ($k=0$) and the Goldstones ($k=1,2,3$)
\begin{equation}
{\sigma}_0 + s_0 ({\bar x}) = 
{\tilde{\sigma}}_0 ({\bar x}) = 
\sqrt{ \frac{g}{2} } {\cal{H}} \left( {\bar x} \right) \ ,
\qquad
{\sigma}_j + s_j ({\bar x}) = 
{\tilde{\sigma}}_j ({\bar x}) = 
\sqrt{ \frac{g}{2} } {\cal{G}}^{(j-1)} \left( {\bar x} \right) \ ,
\quad (j=1,2,3) \ .
\label{fluct}
\end{equation}
Furthermore, we rewrote in (\ref{veffsc}) 
the $\delta$ functions of (\ref{pathint1})
as integrals over the ``sources'' $J_k$
\begin{equation}
\delta \left( \int d^4 {\bar x} s_k({\bar x}) \right) =
\frac{1}{ 2 \pi}
\int_{-\infty}^{+\infty} dJ_k \exp
\left[- i J_k \int d^4 {\bar x} s_k({\bar x}) \right] \ .
\end{equation}
We note that in the physical vacuum we have the following expectation
values:
\begin{equation}
{\sigma}_0 
\left( = \sqrt{ \frac{g}{2} } \langle {\cal{H}} \rangle \right)
= m_t(\Lambda) \ , \qquad
\sigma_k = 0 \ (\mbox{for: } k=1,2,3) \ ,
\label{phyvac}
\end{equation}
where $m_t({\Lambda})$ is the bare mass of the top quark in
the effective theory (\ref{TSM})-(\ref{Ddef}) after the
dynamical symmetry breaking.

Now we perform the expansion of the action
in the exponent of (\ref{veffsc}) in
powers of the scalar quantum fluctuations $s_k({\bar x})$
($k=0,1,2,3$) up to and including the quadratic terms. It turns out
that the terms linear in $s_k$, except for those proportional to
$J_k s_k$ (i.e., the $\delta$ functions), do not contribute to
the effective action in the exponent of (\ref{veffsc}), precisely
because of the $\delta$ functions. The terms with no fluctuations
then yield the leading-$N_{\mbox{\footnotesize c}}$ contribution
$N_{\mbox{\footnotesize c}} V^{(0)}$, and
the terms quadratic in fluctuations yield the next-to-leading
(NTL) contribution $V^{(1)}$ of the formal 
$(1/N_{\mbox{\footnotesize c}})$ expansion
\begin{equation}
V_{\mbox{\footnotesize eff}} \left( \lbrace \sigma_j \rbrace \right)
= N_{\mbox{\footnotesize c}} V^{(0)}
+ V^{(1)} + {\cal{O}} \left( \frac{1}{N_{\mbox{\footnotesize c}}} \right) \ .
\label{veffexp}
\end{equation}
The leading-$N_{\mbox{\footnotesize c}}$
contribution can therefore be read off immediately
from (\ref{veffsc})
\begin{equation}
N_{\mbox{\footnotesize c}} V^{(0)} \left( \lbrace \sigma_j \rbrace \right)
= \frac{1}{G} \sum_{j=0}^3 \sigma^2_j +
\frac{1}{2 \Omega} \int_0^{\infty} \frac{d \tau}{\tau} 
\rho_{\mbox{\scriptsize f}} \left( \tau \right) \left[
\mbox{Tr} e^{- \tau {\hat{\cal{D}}}^{\dagger}{\hat{\cal{D}}}
\left(\lbrace \sigma_i \rbrace \right) }
- \mbox{Tr} e^{- \tau {\hat{\cal{D}}}^{\dagger}{\hat{\cal{D}}}
\left( 0 \right) }
\right] \ .
\label{veff0}
\end{equation}
In Appendix A we wrote down the entire expression for the hermitean
operator matrix ${\hat{\cal{D}}}^{\dagger} {\hat{\cal{D}}}
( \lbrace \sigma_j \rbrace )$ in the Euclidean metric. The matrix
has dimension $8 \times 8$ in the combined spinor and isospin space
[cf.(\ref{Ddef})], and we diagonalized it there by means
of a unitary matrix $U$
\begin{equation}
U^{\dagger} {\hat{\cal{D}}}^{\dagger} {\hat{\cal{D}}}
\left( \lbrace \sigma_j \rbrace \right) U =
{\hat{\bar{P}}} \cdot {\hat{\bar{P}}} + \sum_{j=0}^3 \sigma^2_j
\left[ 
\begin{array}{cc}
1 & 0 \\
0 & 0
\end{array}
\right] \ .
\label{Utransf}
\end{equation}
The matrix above is a block matrix made up of blocks of
dimension $4 \times 4$.
From here we see that only the top quark isospin component contributes
to $N_{\mbox{\footnotesize c}} V^{(0)}$ in the physical vacuum,
since the off-diagonal elements in $U$ matrix of (\ref{U})
are zero then: $g_0^{(\pm)}=0$.
Inserting (\ref{Utransf}) into (\ref{veff0}),
performing the tracing over the coordinates in the momentum basis,
and tracing also in the color and isospin space,
we end up with the general leading-$N_{\mbox{\footnotesize c}}$
contribution in the proper time approach
\begin{eqnarray}
N_{\mbox{\footnotesize c}} V^{(0)}
\left( \Phi_0^{\dagger} \Phi_0 \right) &=&
M^2_0 \Phi_0^{\dagger} \Phi_0 +
\frac{1}{2 \Omega} \int_0^{\infty} \frac{d \tau}{\tau}
\rho_{\mbox{\scriptsize f}} \left( \tau \right)
N_c 4 \Omega \int \frac{d^4 {\bar q}}{ \left( 2 \pi \right)^4 }
{\Big \{}
\exp \left[- \tau \left( {\bar q}^2 + g \Phi_0^{\dagger} \Phi_0 \right)
\right] - \exp \left[ - \tau {\bar q}^2 \right]
{\Big \}}
\nonumber\\
&=&
M^2_0 \Phi_0^{\dagger} \Phi_0 +
\frac{N_{\mbox{\footnotesize c}}}{8 \pi^2}
\int_0^{\infty} \frac{d \tau}{\tau^3}
\rho_{\mbox{\scriptsize f}} \left( \tau \right)
\left[
\exp \left(- \tau g \Phi_0^{\dagger} \Phi_0 \right) - 1
\right] \ ,
\label{veff0res}
\end{eqnarray}
where the factors $N_{\mbox{\footnotesize c}}$ and $4$ come from tracing
the identity matrix in the color space and in the spinor space of the
top quark, respectively.
We denoted by $\Phi_0$ simply the expectation value of the
scalar isodoublet, i.e.,
\begin{equation}
\Phi_0^{\dagger} \Phi_0 = \frac{1}{2} \left[
{\cal{H}}_0^2 + \sum_{k=0}^2 {\cal{G}}^{(k)2}_0 \right]
= \frac{1}{g} \sum_{j=0}^3 {\sigma}_j^2 = \frac{1}{g} {\sigma}^2 \ ,
\label{expval1}
\end{equation}
and we introduced the notation ${\sigma}^2 = {\sum}_0^3 {\sigma}_j^2$.
From (\ref{veff0res}) we see explicitly that the
leading-$N_{\mbox{\footnotesize c}}$
contribution is a function of the $SU(2)_L \times U(1)_Y$ invariant
$\Phi_0^{\dagger} \Phi_0$, and is therefore itself an 
$SU(2)_L \times U(1)_Y$ invariant expression, as it should be. 
Certainly, in our calculations we will take at the end
$\Phi_0^{\dagger} \Phi_0 = \sigma_0^2/g = m_t^2(\Lambda)/g$,
since we will be searching for the physical vacuum
that dynamically breaks the $SU(2)_L \times U(1)_Y$ to 
$U(1)_{\mbox{\footnotesize em}}$.

The leading-$N_{\mbox{\footnotesize c}}$ expression (\ref{veff0res})
can be obtained also diagrammatically, in its nonregularized form,
by calculating the 1-PI one-loop Green functions corresponding
to the diagrams of Fig.~1 with outer legs of zero momentum.
This is described in detail in~\cite{cp}. 

At this point, we introduce two specific choices for the
fermionic (quark) proper time regulator
$\rho_{\mbox{\scriptsize f}}(\tau)$ (i.e.,
the regularization for the fermionic momenta of the model):
proper time cutoff (PTC), and the Pauli-Villars (PV) regularization
with two subtractions:
\begin{eqnarray}
\rho_{\mbox{\scriptsize f}}^{\mbox{\scriptsize (PTC)}}(\tau) =
\Theta( \tau - 1/{\Lambda}_{\mbox{\footnotesize f}}^2) & = &
\left\lbrace
\begin{array}{ll}
1 & \mbox{for: } \tau > 1/\Lambda_{\mbox{\footnotesize f}}^2 \\
0 & \mbox{for: } \tau < 1/\Lambda_{\mbox{\footnotesize f}}^2 
\end{array}
\right\rbrace \ ,
\label{PTC}
\end{eqnarray}
\begin{equation}
\rho_{\mbox{\scriptsize f}}^{\mbox{\scriptsize (PV)}}(\tau) =
\left[ 1 - \exp \left(- \tau \Lambda_{\mbox{\footnotesize f}}^2 \right)
\right]^2 =
1 - 2 e^{- \tau \Lambda_{\mbox{\scriptsize f}}^2} +
e^{- \tau 2 \Lambda_{\mbox{\scriptsize f}}^2} \ .
\label{PV}
\end{equation}
The leading-$N_{\mbox{\footnotesize c}}$ part
$N_{\mbox{\footnotesize c}} V^{(0)}( {\sigma}^2 )$ of the effective
potential was calculated explicitly for these two cases (PTC, PV)
in Appendix B. We note that the two 
${\Lambda}_{\mbox{\footnotesize f}}$'s in (\ref{PTC}) and (\ref{PV})
are not equal in the corresponding cases with equal four-fermion
coupling $G = 8 {\pi}^2 a/ (N_{\mbox{\footnotesize c}}
{\Lambda}_{\mbox{\footnotesize f}}^2)$ of the Lagrangian (\ref{TSM}).
In fact, if we require
that ${\Lambda}_{\mbox{\footnotesize f}}^2$ terms in
$ N_{\mbox{\footnotesize c}} {\partial} V^{(0)} / 
{\partial} {\sigma}^2$ be equal in the two cases, we obtain
[cf.(\ref{dXi0PTC})-(\ref{dXi0PV})]:
${\Lambda}_{\mbox{\footnotesize f}}^2 (\mbox{\scriptsize PTC}) =
(2 \ln 2) {\Lambda}_{\mbox{\footnotesize f}}^2 (\mbox{\scriptsize PV})$.
The $\ln {\Lambda}_{\mbox{\footnotesize f}}^2$ terms then agree
automatically with each other.

Next we turn to the derivation of the next-to-leading (NTL) contribution
$V^{(2)}$. First we expand ${\hat{\cal{D}}}^{\dagger}{\hat{\cal{D}}}$
up to quadratic terms in the scalar fluctuations 
$\lbrace s_j( {\bar x} ) \rbrace$ (\ref{fluct})
\begin{equation}
{\hat{\cal{D}}}^{\dagger}{\hat{\cal{D}}}
\left( \lbrace \sigma_j + s_j({\bar x}) \rbrace \right) =
{\hat \triangle}_0 \left( \lbrace \sigma_j  \rbrace \right) +
{\hat \triangle}_1 \left( \lbrace \sigma_j ; s_j({\bar x}) \rbrace \right)
+{\hat \triangle}_2 \left( \lbrace \sigma_j ; s_j({\bar x}) \rbrace \right)
+ \cdots \ .
\label{expanDD}
\end{equation}
Operators ${\hat \triangle}_1$ and ${\hat \triangle}_2$ are linear
and quadratic in the scalar fluctuations $\lbrace s_j({\bar x}) \rbrace$,
and the dots denote terms which are at least cubic in 
$\lbrace s_j({\bar x}) \rbrace$.
Explicit expressions for ${\hat \triangle}_1$ and
${\hat \triangle}_2$ are given in 
(\ref{triang1})-(\ref{notattri})~\footnote{
Expressions in (\ref{expanDD}) are local functions in the
${\bar x}$ space. Formally, this should be understood in the
$| {\bar x} \rangle$ basis as: $ \langle {\bar x} |
{\hat{\cal{D}}}^{\dagger} {\hat{\cal{D}}} \left( 
\langle {\sigma}_j + s_j
\rangle \right) | {\bar x}^{\prime} \rangle = 
{\delta}^{(4)} ( {\bar x} - {\bar x}^{\prime} )
{\hat{\cal{D}}}^{\dagger} {\hat{\cal{D}}} \left( \langle {\sigma}_j 
+ s( {\bar x} )_j \rangle \right)$. }.
Furthermore, we denoted:
${\hat \triangle}_0 =
{\hat{\cal{D}}}^{\dagger} {\hat{\cal{D}}}( \lbrace \sigma_j \rbrace )$.
Inserting expansion (\ref{expanDD}) into
(\ref{veffsc}), we then obtain the NTL contribution $V^{(1)}$ to
the effective potential by keeping only the contributions quadratic
in the fluctuations (and by keeping the $\delta$ functions, as mentioned
earlier)
\begin{eqnarray}
\lefteqn{
\exp \left[ - \Omega V^{(1)}
\left( \lbrace \sigma_j \rbrace \right) \right] =
\mbox{const.} \times \prod_{j=0}^3 \int_{-\infty}^{+\infty} dJ_j
\int {\cal{D}}s_0 {\cal{D}}s_1 {\cal{D}}s_2{\cal{D}}s_3 \times
}
\nonumber\\
&&
\exp {\Bigg \{ } + \frac{1}{2} \int_0^{\infty} d {\tau}
\rho_{\mbox{\scriptsize f}} ( \tau )
 \mbox{Tr} \left[
 e^{- \tau {\hat \triangle}_0 \left( \lbrace \sigma_k \rbrace \right) }
{\hat \triangle}_2 \left( \lbrace \sigma_k ; s_k \rbrace \right) \right]
\nonumber\\
&& \quad
- \frac{1}{4} \int_0^{\infty} d {\tau}_1 \int_0^{\infty} d {\tau}_2
\rho_{\mbox{\scriptsize f}} ( \tau_1+\tau_2 ) \mbox{Tr} \left[
e^{- \tau_1 {\hat \triangle}_0 \left( \lbrace \sigma_k \rbrace \right) }
{\hat \triangle}_1 \left( \lbrace \sigma_k; s_k \rbrace \right)
e^{- \tau_2 {\hat \triangle}_0 \left( \lbrace \sigma_k \rbrace \right) }
{\hat \triangle}_1 \left( \lbrace \sigma_k; s_k \rbrace \right) \right]
\nonumber\\
&& \quad
- \frac{1}{G} \int d^4 {\bar x} \sum_{k=0}^3 s_k({\bar x})^2
- i \sum_{k=0}^3 J_k \int d^4 {\bar x} s_k ({\bar x})
{\Bigg \} } \ .
\label{veff1}
\end{eqnarray}
We calculate the above traces in the momentum basis and in the
unitarily rotated spinor-isospin basis in which the $8 \times 8$
matrix ${\hat{\triangle}}_0 ( \lbrace \sigma_j \rbrace )
= {\hat{\cal{D}}}^{\dagger}{\hat{\cal{D}}} ( \lbrace \sigma_j \rbrace )$
is diagonal [cf.(\ref{Utransf})]. The steps and some details are 
explained in Appendix B.
It turns out that the proper time action in the curly brackets of
the exponent in (\ref{veff1}) can be written in the following form:
\begin{equation}
- \frac{1}{2} \int \int d^4 {\bar x} d^4 {\bar y}
v_j({\bar y}) {\hat{\cal{A}}}_j \left( {\bar y}, {\bar x};
{\sigma}^2 \right) v_j({\bar x})
- i I_j \int d^4 {\bar x} v_j({\bar x}) \ ,
\label{action}
\end{equation}
where the summation over all $j= 0, \ldots, 3$ is implied, the
fluctuations $v_j({\bar x})$ are obtained from the original scalar
fluctuations $s_j({\bar x})$ by an orthonormal transformation
${\cal{O}}$ given in (\ref{vs})-(\ref{O}), 
and the ``sources'' $I_j$'s are
obtained from $J_j$'s by the same transformation. Matrix
${\cal{O}}$ is constructed in such a way that the action terms
quadratic in fluctuations are now diagonal. Furthermore, the
diagonal kernel elements ${\hat{\cal{A}}}_j ( {\bar y}, {\bar x} ;
{\sigma}^2 )$ depend on the $SU(2)_L \times U(1)_Y$ invariant
expectation value
${\sigma}^2 = {\sum}_0^3 {\sigma}_j^2 = g \Phi_0^{\dagger} \Phi_0$ and
are translationally invariant in the (Euclidean) configuration
space, i.e., they are functions of the difference ${\bar x}-{\bar y}$.
They are obtained explicitly in Appendix B.
Since the Jacobian of any orthonormal transformation is equal to
one, we can replace in the path integral (\ref{veff1}) the integrations
over the fluctuations $s_j$ and sources $J_j$ by integrations
over $v_j$ and $I_j$, respectively. Therefore, we end up with
the diagonal integrals of the Gaussian type which can be solved
\begin{eqnarray}
\lefteqn{
\int {\cal{D}} v_j \int_{-\infty}^{+\infty} dI_j
\exp {\Big \{ }
- \frac{1}{2} \int \int d^4 {\bar x} d^4 {\bar y}
v_j({\bar y}) {\hat{\cal{A}}}_j \left( {\bar y}, {\bar x};
{\sigma}^2 \right) v_j({\bar x})
- i I_j \int d^4 {\bar x} v_j({\bar x})  {\Big \} }  }
\nonumber\\
&=& \exp \left[ - \frac{1}{2} \mbox{Tr} \ln {\hat{\cal{A}}}_j
\left( {\sigma}^2 \right) \right]
\int_{-\infty}^{+\infty} dI_j \exp \left[ - \frac{1}{2} I_j^2
\int \int d^4 {\bar x} d^4 {\bar y} {\hat{\cal{A}}}_j^{-1}
\left( {\bar y}, {\bar x} ; {\sigma}^2 \right) \right]
\nonumber\\
&=& \exp \left[ - \frac{1}{2} \mbox{Tr} \ln {\hat{\cal{A}}}_j
\left( {\sigma}^2 \right) \right]
\sqrt{\frac{2 \pi}{\Omega}}
\sqrt{ {\tilde A}_j \left( {\bar p}^2=0; {\sigma}^2 \right) } \ ,
\label{veff12}
\end{eqnarray}
where we denoted by ${\tilde A}$ the Fourier transform of
${\hat{\cal{A}}}_j ({\bar y}, {\bar x}; {\sigma}^2)$
\begin{equation}
{\tilde A}_j \left( {\bar p}^2; {\sigma}^2 \right)
= \int d^4 {\bar x} \exp \left[ -i {\bar p} \cdot {\bar x} \right]
  {\hat{\cal{A}}}_j \left(0, {\bar x};  {\sigma}^2 \right) \ .
 \label{ft}
\end{equation}
Taking the logarithm, we end up with the following expression for the
NTL part $V^{(1)}$ of the effective potential
\begin{equation}
V^{(1)} ( {\sigma}^2 ) =
\frac{1}{2 \Omega} \sum_{j=0}^3 \mbox{{\large Tr}}
\ln {\hat{\cal{A}}}_j \left( {\sigma}^2 \right)
- \frac{1}{2 \Omega} \sum_{j=0}^3 \ln {\tilde A}
\left( {\bar p}^2 = 0; {\sigma}^2 \right) \ .
\label{veff13}
\end{equation}
The second sum on the RHS of (\ref{veff13}), which is the
remnant of the $\delta$ function conditions, is evidently zero
in the infinite volume limit ($\Omega = \int d^4 {\bar x} \to \infty$).
On the other hand, the first sum is finite. Performing the tracing
in the (bosonic) momentum basis, we obtain finally
\begin{equation}
V^{(1)} ( {\sigma}^2 ) =
\frac{1}{ 32 \pi^2} \sum_{j=0}^3 \int_0^{\Lambda_{\mbox{\scriptsize b}}^2}
d {\bar p}^2 {\bar p}^2
\ln {\tilde A}_j \left( {\bar p}^2; {\sigma}^2 \right) \ .
\label{veff1fin}
\end{equation}
Here we introduced a finite spherical cutoff 
${\Lambda}_{\mbox{\footnotesize b}}$ 
(${\Lambda}_{\mbox{\footnotesize b}} \sim
{\Lambda}_{\mbox{\footnotesize f}} \sim {\Lambda}$)
for the bosonic momenta ${\bar p}$.
If we hadn't cut off the integration this way,
and if we had tried to regularize the bosonic momenta again
with the proper time method, we would have
ended up with severely divergent integrals over the
bosonic proper times, since the effective
actions ${\tilde A}_j$ in momentum space are not
proportional to ${\bar p}^2$ as ${\bar p}^2 \to \infty$
(i.e., no kinetic terms for bosons at ${\bar p}^2 \gg
{\Lambda}_{\mbox{\footnotesize f}}^2$),
but rather converge to a constant. However, the spherical
cutoff for the bosonic momenta is not as problematic
as it is for the fermionic momenta. Namely, unlike
the fermionic spherical cutoff case, the integrals over the
bosonic momenta evidently
don't suffer in our expressions from the momentum branching
ambiguities -- the integrands in (\ref{veff1fin}) don't depend
on scalar products ${\bar p} \cdot {\bar k}$
(${\bar k}$ being a fermionic momentum), but only on
${\bar p}^2$.

The kernels ${\hat{\cal{A}}}_j$ have the following structure
in the ${\bar x}$ basis:
\begin{eqnarray}
\left[
 \begin{array}{c}
 {\hat{\cal{A}}}_0 ( 0, {\bar x}; {\sigma}^2 ) \\
 {\hat{\cal{A}}}_1 ( 0, {\bar x}; {\sigma}^2 )
 \end{array}
\right]
& = & {\alpha}^{(1)} ( {\sigma}^2 )
\delta ( {\bar x} ) + \left( {\beta}_1^{(1)} \pm
{\beta}_2^{(1)} \right) 
( {\bar x}; {\sigma}^2 ) \ ,
\nonumber\\
 {\hat{\cal{A}}}_2 ( 0, {\bar x}; {\sigma}^2 )
& = & \left[ {\alpha}^{(1)} ( {\sigma}^2 )
+ {\alpha}^{(2)} ( {\sigma}^2 ) \right]
\delta ( {\bar x} )
+  \beta^{(2)} ( {\bar x}; {\sigma}^2 )
=  {\hat{\cal{A}}}_3 ( 0, {\bar x}; {\sigma}^2 )
\ ,
\label{veff1argx}
\end{eqnarray}
where the explicit expressions for the functions $\alpha^{(k)}$
and $\beta^{(k)}$ in terms of the integrals over the proper time
are given in (\ref{alphas})-(\ref{betasx}). 
In the momentum basis, applying (\ref{veff1fin}),
we finally get for the NTL part of the effective potential
\begin{eqnarray}
V^{(1)} ( {\sigma}^2 ) =
\frac{1}{32 \pi^2}
\int_0^{\Lambda_{\mbox{\scriptsize b}}} d {\bar p}^2 {\bar p}^2
{\Big \{ }&&
 + \ln \left[ {\alpha}^{(1)} ( {\sigma}^2 )
 + \left( {\tilde{\beta}}_1^{(1)} + {\tilde{\beta}}_2^{(1)} \right)
 ( {\bar p}^2 ; {\sigma}^2 ) \right]
\nonumber\\
&& + \ln \left[ {\alpha}^{(1)} ( {\sigma}^2 )
 + \left( {\tilde{\beta}}_1^{(1)} - {\tilde{\beta}}_2^{(1)} \right)
 ( {\bar p}^2 ; {\sigma}^2 ) \right]
\nonumber\\
&&+ 2 \ln \left[ {\alpha}^{(1)} ( {\sigma}^2 )
 + {\alpha}^{(2)} ( {\sigma}^2 )
 +  {\tilde{\beta}}^{(2)}
  ( {\bar p}^2 ; {\sigma}^2 ) \right]
{\Big \} } \ ,
\label{veff1mom}
\end{eqnarray}
where ${\tilde{\beta}}^{(k)}_j$'s are the Fourier transforms of
${\beta}^{(k)}_j$'s. Note that the three terms on the RHS of
(\ref{veff1mom}) correspond to the NTL contributions of the Higgs,
the neutral Goldstone and the two charged Goldstones, respectively.
From now on, we will take the $SU(2)_L \times U(1)_Y$ 
invariant square of the field
expectation value $g \Phi_0^{\dagger} \Phi_0 = {\sigma}^2
= {\sum}_0^3 {\sigma}_j^2$ in
these formulas to be equal to the square of the (rescaled) Higgs
expectation value ${\sigma}_0^2 = g {\cal{H}}_0^2/2$, i.e.,
we will search for the physical vacuum.

It turns out, as shown in Appendix B, that
$\alpha^{(1)}(\sigma_0^2)$ is in general exactly
twice the derivative of the leading-$N_{\mbox{\footnotesize c}}$
part of the effective potential
\begin{equation}
\alpha^{(1)}(\sigma_0^2) = 2 \frac{\partial}{\partial \sigma_0^2}
\left[ N_{\mbox{\footnotesize c}} V^{(0)} ( \sigma_0^2 )
\right] \ .
\label{alpha1}
\end{equation}
This quantity is zero by definition if $\sigma_0$ is the solution
${\sigma}_0=m_t^{(0)}$ of
the leading-$N_{\mbox{\footnotesize c}}$ gap equation. However,
if $\sigma_0$ is equal (or in the vicinity) of the NTL gap equation
solution, i.e., if ${\sigma}_0^2 = m_t^{(0)2}
( 1 + {\cal{O}}(1/N_{\mbox{\footnotesize c}}) )$, then
$\alpha^{(1)}(\sigma_0^2)$ is formally of order
${\cal{O}}(N_{\mbox{\footnotesize c}} \ast (1/N_{\mbox{\footnotesize c}}))
= {\cal{O}}(N_{\mbox{\footnotesize c}}^0)$
in the $(1/N_{\mbox{\footnotesize c}})$
expansion, and it changes sign from minus to plus as $\sigma_0$
increases across $m_t^{(0)}$. On the other hand, the rest of the
arguments in each of the logarithms in (\ref{veff1mom}) is always positive
and formally of order ${\cal{O}}(N_{\mbox{\footnotesize c}})$, as
can be immediately seen from (\ref{FT2B1})-(\ref{FT2B2}). 
For any specific regulator 
${\rho}_{\mbox{\scriptsize f}} ( {\tau} )$, the explicit expressions 
for all these arguments can be calculated from the corresponding
general formulas (\ref{alphas}) and (\ref{FT2B1})-(\ref{FT2B2}).
Here we write down the results for the PV case (\ref{PV}), as an
explicit series in inverse powers of the cutoff parameter
${\Lambda}_{\mbox{\footnotesize f}}$:
\begin{eqnarray}
{\alpha}^{(1)} ( {\sigma}_0^2; G)^{\mbox{\scriptsize (PV)}} &=& 
\frac{2}{G}
- \frac{ N_{\mbox{\footnotesize c}} }{ 4 {\pi}^2 }
{\Bigg \{ } ( 2 \ln 2 ) 
{\Lambda}_{\mbox{\footnotesize f}}^2
- {\sigma}_0^2 \ln \left( \frac { {\Lambda}_{\mbox{\footnotesize f}}^2 }
{ {\sigma}_0^2 } \right)
+ {\sigma}_0^2 ( \ln 2 -1 ) - \frac{3}{4} \frac{ {\sigma}_0^4 }
{ {\Lambda}_{\mbox{\footnotesize f}}^2 } 
+ \cdots
\nonumber\\
& & + \cdots +
\frac{ \left( -1 \right)^{n+1} }{ n \left( n - 1 \right) }
 \frac{ \left( 2^n -1 \right) }
{ 2^{n-1} } \frac{ {\sigma}_0^{2n} }
{ {\Lambda}_{\mbox{\footnotesize f}}^{2n-2} }
+ \cdots
{\Bigg \} } \ ,
\label{al1PV}
\end{eqnarray}
\begin{eqnarray}
\lefteqn{
\left( {\tilde \beta}_1^{(1)} \pm {\tilde \beta}_2^{(1)} \right)
\left( {\bar p}^2; \sigma_0^2 \right)^{\mbox{\footnotesize (PV)}}
 = \frac{N_{\mbox{\footnotesize c}}}{8 \pi^2}
\left[ {\bar p}^2 + 2 \sigma_0^2 \left( 1 \pm 1 \right) \right]
{\Bigg \{ }
\ln \left( \frac{ \Lambda_{\mbox{\footnotesize f}}^2 }{ 2 \sigma_0^2 } \right)
- \frac{2}{3} z F(z)
{\Bigg |}_{ z=
{\bar p}^2/\left( {\bar p}^2 + 4 \sigma_0^2 \right) }
 }  \nonumber\\
 & &
 + \left[
 \frac{3}{2} \frac{\sigma_0^2}{ \Lambda_{\mbox{\footnotesize f}}^2 }
 + \frac{1}{4} \frac{ {\bar p}^2 }{ \Lambda_{\mbox{\footnotesize f}}^2 }
   \right]
- \frac{7}{8} \left[
 \left( \frac{\sigma_0^2}{ \Lambda_{\mbox{\footnotesize f}}^2 } \right)^2
 + \frac{1}{3}
 \frac{ {\sigma}_0^2 {\bar p}^2 }{ \Lambda_{\mbox{\footnotesize f}}^4 }
 + \frac{1}{30}
 \left( \frac{{\bar p}^2}{ \Lambda_{\mbox{\footnotesize f}}^2 } \right)^2
 \right]
\nonumber\\
&&
+ \sum_{n=3}^{\infty}
\frac{ \left( -1 \right)^{n+1} }{ n }
\frac{ \left( 2^{n+1} - 1 \right) }{ 2^n }
\left( \frac{ {\bar p}^2 }{ {\Lambda}_{\mbox{\footnotesize f}}^2 }
\right)^n
\int_0^1 du \left[ u (1-u) + \frac{ {\sigma}_0^2 }{ {\bar p}^2 }
\right]^n
 {\Bigg \} } \ ,
\label{bePVn}
\end{eqnarray}
\begin{eqnarray}
\lefteqn{
\left[
 \alpha^{(2)} \left( \sigma_0^2 \right) +
 {\tilde \beta}^{(2)}
\left( {\bar p}^2; \sigma_0^2 \right)
\right]^{\mbox{\footnotesize (PV)}}
 = \frac{N_{\mbox{\footnotesize c}}}{8 \pi^2}
{\bar p}^2 {\Bigg \{ }
\ln \left( \frac{\Lambda_{\mbox{\footnotesize f}}^2}{2 \sigma_0^2} \right)
 + {\Bigg [ } - \left( 1 + \frac{\sigma_0^2}{{\bar p}^2} \right)^2
 \ln \left( 1 + \frac{ {\bar p}^2 }{ {\sigma}_0^2 } \right) +
 }  \nonumber\\
& &
 + \frac{ \sigma_0^2 }{ {\bar p}^2 } + 2 {\Bigg ] }
 + \left[ 
\frac{\sigma_0^2}{ \Lambda_{\mbox{\footnotesize f}}^2 }
 + \frac{1}{4} \frac{{\bar p}^2}{ \Lambda_{\mbox{\footnotesize f}}^2 } 
\right]
 - \frac{7}{16} \left[ 
 \left( \frac{\sigma_0^2}{ \Lambda_{\mbox{\footnotesize f}}^2 } \right)^2
 + \frac{2}{5}
 \frac{ {\sigma}_0^2 {\bar p}^2 }{ \Lambda_{\mbox{\footnotesize f}}^4 }
 + \frac{1}{15}
 \left( \frac{{\bar p}^2}{\Lambda_{\mbox{\footnotesize f}}^2} \right)^2
 \right]
\nonumber\\
& &
+ \sum_{n=3}^{\infty} \frac{ \left( -1 \right)^{n+1} }{ n }
\frac{ \left( 2^{n+1} - 1 \right) }{ 2^{n-1} }
\left( \frac{ {\bar p}^2 }{ {\Lambda}_{\mbox{\footnotesize f}}^2 }
\right)^n
\int_0^1 du u^{n+1} \left( 1+ \frac{ {\sigma}_0^2 }{ {\bar p}^2 }
- u \right)^n
 {\Bigg \} } \ .
\label{bePVch}
\end{eqnarray}
Function $F(z)$, appearing in
${\Lambda}_{\mbox{\footnotesize f}}^2$-independent part of
${\tilde{\beta}}_1^{(1)} \pm {\tilde{\beta}}_2^{(1)}$, is
given in Appendix B in (\ref{F}).

The corresponding expressions for the PTC case are similar and were
calculated explicitly in (\ref{al1PTC}) and 
(\ref{be1PTC})-(\ref{be2PTC}).
As mentioned toward the end of Appendix B, the PV results
(\ref{bePVn})-(\ref{bePVch}) can be 
calculated from formulas (\ref{FT2B1})-(\ref{FT2B2})
in a way very similar as in the PTC case. Moreover, we can even
obtain closed analytical expressions for these quantities in
the PV case [cf.~(\ref{PVann})-(\ref{PVannot})].

The NTL part of the effective potential, in its nonregularized form,
can be calculated also diagrammatically, by evaluating the 1-PI
multi-loop Green functions of the ``beads'' diagrams of Fig.~2.
The calculation is described in detail 
in Appendix C of Ref.~\cite{cpv},
and its extension to the case when Goldstone bosons are included
is mentioned in~\cite{cv}.

\section{Gap equation at the next-to-leading order}

In order to write down the NTL gap equation that is suitable for
numerical evaluations, it is useful to define the dimensionless 
analogues
of: the momenta, the scalar expectation values, the proper time,
the four-fermion coupling and the effective potential.
We rescale the bosonic momenta
$ {\bar p}^2 \to
\Lambda_{\mbox{\footnotesize f}}^2 {\bar p}^2$,
and introduce the following dimensionless analogues: 
\begin{equation}
{\varepsilon}^2=
\frac{\sigma_0^2}{\Lambda_{\mbox{\footnotesize f}}^2}=
\frac{G M_0^2}{2 \Lambda_{\mbox{\footnotesize f}}^2}
\langle {\cal{H}} \rangle^2\ , \qquad
a=\frac{(GN_{\mbox{\footnotesize c}}
\Lambda_{\mbox{\footnotesize f}}^2)}{(8\pi^2)} \ , \qquad
z= \tau \Lambda_{\mbox{\footnotesize f}}^2 \ ,
\label{defka1}
\end{equation}
\begin{equation}
\Xi_{\mbox{\footnotesize eff}}=
8\pi ^2 V_{\mbox{\footnotesize eff}}/
(N_{\mbox{\footnotesize c}}\Lambda_{\mbox{\footnotesize f}}^4)=
\Xi ^{(0)}+\frac 1{N_{\mbox{\footnotesize c}}}\Xi^{(1)}
+ {\cal {O}}(\frac{1}{N^2_{\mbox{\footnotesize c}}}) \ .  
\label{defka2}
\end{equation}
The dimensionless coupling parameter $a$ is of order 1 by the
leading-$N_{\mbox{\footnotesize c}}$ gap equation, i.e.,
$a$ is formally of order ${\cal{O}}(N_{\mbox{\footnotesize c}}^0)$
in $1/N_{\mbox{\footnotesize c}}$ expansion, as will be shown
explicitly below.
The resulting expressions for the 
leading-$N_{\mbox{\footnotesize c}}$ term ${\Xi}^{(0)}$ 
and the NTL term ${\Xi}^{(1)}$ are 
\begin{eqnarray}
\Xi ^{(0)}( {\varepsilon}^2; a ) =
\frac{{\varepsilon}^2}{a}
+ \int_0^{\infty} \frac{dz}{z^3} \mbox{reg}_{\mbox{\scriptsize f}}(z)
\left( e^{-z {\varepsilon}^2} - 1 \right)
\label{Xi0}
\end{eqnarray}
\begin{eqnarray}
\lefteqn{
\Xi^{(1)} ( {\varepsilon}^2;
{\Lambda }_{\mbox{\footnotesize b}}^2/
{\Lambda }_{\mbox{\footnotesize f}}^2;a ) =
\frac{1}{4} \int_0^{{\Lambda }_{\mbox{\scriptsize b}}^2/
{\Lambda }_{\mbox{\scriptsize f}}^2} 
 d {\bar p}^2 {\bar p}^2 
 {\Bigg \{ }
 \ln \left[ A^{(1)} ( {\varepsilon}^2; a ) +
 \left( {\tilde B}_1^{(1)} + {\tilde B}_2^{(1)} \right)
 ( {\bar p}^2; {\varepsilon}^2 ) \right] 
}
  \nonumber\\
  & &
+ \ln \left[ A^{(1)} ( {\varepsilon}^2; a ) +
 \left( {\tilde B}_1^{(1)} - {\tilde B}_2^{(1)} \right)
  ( {\bar p}^2; {\varepsilon}^2 ) \right]
 + 2 \ln \left[ A^{(1)} ( {\varepsilon}^2; a ) +
  A^{(2)} ( {\varepsilon}^2 ) +
  {\tilde B}^{(2)} 
 ( {\bar p}^2; {\varepsilon}^2 ) \right]
 {\Bigg \} } \ ,
\label{Xi1}
\end{eqnarray}
where $\mbox{reg}_{\mbox{\scriptsize f}}(z) = \rho_{\mbox{\scriptsize f}}
(\tau = z/\Lambda_{\mbox{\footnotesize f}}^2)$, i.e.,
$\mbox{reg}_{\mbox{\scriptsize f}}(z) = \theta(z-1)$ in the PTC case,
and $\mbox{reg}_{\mbox{\scriptsize f}}(z) = (1-e^{-z})^2$ in the PV case.
The dimensionless functions $A^{(j)}$ and ${\tilde B}^{(j)}_k$
are
\begin{eqnarray}
A^{(j)} ( {\varepsilon}^2 ) & = &
\frac{8 \pi^2}{ N_{\mbox{\footnotesize c}} 
{\Lambda}_{\mbox{\footnotesize f}}^2 } 
{\alpha}^{(j)} ( {\sigma}_0^2 ) \ ,
\quad A^{(1)} = 2 \frac{ {\partial} {\Xi}^{(0)} }{ 
{\partial} {\varepsilon}^2 } \ ,
\nonumber\\
{\tilde{B}}^{(j)}_k ( {\bar p}^2; {\varepsilon}^2 ) & = &
\frac{8 \pi^2}{ N_{\mbox{\footnotesize c}} 
{\Lambda}_{\mbox{\footnotesize f}}^2 } 
{\tilde{\beta}}^{(j)}_k 
( {\bar q}^2= {\Lambda}_{\mbox{\footnotesize f}}^2 {\bar p}^2;
 {\sigma}_0^2 = {\Lambda}_{\mbox{\footnotesize f}}^2 {\varepsilon}^2
 ) \ .
\label{AB}
\end{eqnarray}
The explicit expressions for $A^{(1)}( {\varepsilon}^2; a )$,
$( {\tilde B}_1^{(1)} \pm {\tilde B}_2^{(1)} ) ( {\bar p}^2;
{\varepsilon}^2 )$ and $(A^{(2)} + {\tilde B}^{(2)})
( {\bar p}^2; {\varepsilon}^2 )$ for the PTC and PV cases
can therefore be read off
directly from (\ref{dXi0PTC})-(\ref{dXi0PV})  
[or equivalently from: (\ref{al1PTC}) and (\ref{al1PV})]
and from (\ref{be1PTC})-(\ref{be2PTC}), 
(\ref{bePVn})-(\ref{bePVch}). With the exception of $A^{(1)}$,
they are formally ${\cal{O}}( N_{\mbox{\footnotesize c}}^0 )$;
$A^{(1)}$ is ${\cal{O}}(1/N_{\mbox{\footnotesize c}})$.  

The general leading-$N_{\mbox{\footnotesize c}}$ gap equation is
obtained by minimizing the potential ${\Xi}^{(0)}$ with
respect to ${\varepsilon}^2$
\begin{equation}
\frac{ {\partial} {\Xi}^{(0)} \left( {\varepsilon}^2; a \right) }{
 {\partial} {\varepsilon}^2 } 
{\Bigg | }_{ {\varepsilon}^2 = {\varepsilon}_0^2 } =
\frac{1}{a}  - \int_0^{\infty} \frac{dz}{z^2}
\mbox{reg}_{\mbox{\scriptsize f}} \left( z \right)
e^{- z {\varepsilon}^2 } 
{\Bigg | }_{ {\varepsilon}^2 = {\varepsilon}_0^2 } = 0 \ ,
\label{gaplead}
\end{equation}
where ${\varepsilon}_0 = m_t^{(0)}/{\Lambda}_{\mbox{\footnotesize f}}$,
and $m_t^{(0)}$ is the top quark mass approximation as obtained 
from this leading-$N_{\mbox{\footnotesize c}}$ gap equation. From
(\ref{gaplead}) we also see explicitly that the dimensionless
coupling constant $a$, as defined by (\ref{defka1}), is really
of order 1.

At this point, we note that the NTL contribution in (\ref{Xi1}) is
ill-defined for all ${\varepsilon}^2$'s that are smaller than
the leading-$N_{\mbox{\footnotesize c}}$ gap equation solution
${\varepsilon}_0^2 = (m_t^{(0)}/\Lambda_{\mbox{\footnotesize f}})^2$.
The reason for this is that, as already mentioned in the previous
Section, ${\alpha}_1$ and hence $A^{(1)}$ become 
negative for ${\varepsilon}^2 <
{\varepsilon}_0^2$, while ${\tilde{B}}_1^{(1)} - {\tilde{B}}_2^{(1)}$
and $A^{(2)}+ {\tilde{B}}^{(2)}$, being nonnegative always,
go to zero when ${\bar p}^2 \to 0$, as seen from 
(\ref{FT2B1})-(\ref{FT2B2}). Therefore, the arguments
of the logarithms for the Goldstone contributions in (\ref{Xi1})
become negative in such a case. This problem is
manifest already in formula (\ref{veff12}) where the argument
in the exponent in the integral over the sources $I_j$ becomes
positive for ${\varepsilon}^2 < {\varepsilon}_0^2$ (for $j=1,2,3$)
and the integral becomes divergent. Furthermore, for the same
reasons, the argument of the logarithm for the Higgs contribution
in (\ref{Xi1}) becomes negative for small ${\varepsilon}^2
< {\varepsilon}_{\ast}^2$
(substantially smaller than ${\varepsilon}_0^2$) when
${\bar p}^2 \to 0$.

As mentioned in the previous Section, however, the problematic
part $A^{(1)}({\varepsilon}^2; a)$ of the arguments of the
logarithms in the NTL part (\ref{Xi1}) of the effective
potential is formally suppressed by 
$(1/N_{\mbox{\footnotesize c}})$ in comparison to the
other parts. Therefore, we may be tempted to simply
ignore that term there, on the grounds that it gives
formally the next-to-next-to-leading order (NNTL) corrections
which have not been taken into account in its entirety
here anyway. However, this argument is dangerous and misleading,
because the physically relevant relations are obtained from
the gap equation involving the derivative of the
effective potential, and not the effective potential itself.
It is straightforward to see from (\ref{Xi0}) and (\ref{gaplead})
that the derivative 
${\partial} A^{(1)}/ {\partial} {\varepsilon}^2
= 2 {\partial}^2 {\Xi}^{(0)}/ {\partial} ({\varepsilon}^2)^2$  
at ${\varepsilon}^2 = {\varepsilon}_0^2 ( 1 +
{\cal{O}}(1/N_{\mbox{\footnotesize c}}) )$
is {\it not}
suppressed by $1/N_{\mbox{\footnotesize c}}$, unlike
$A^{(1)}$. In other words, ${\partial} A^{(1)} / {\partial}
{\varepsilon}^2$ is ${\cal{O}}(N_{\mbox{\footnotesize c}}^0 )$
just like ${\partial} ( {\tilde B}_1^{(1)} \pm {\tilde B}_2^{(1)} )
/ {\partial} {\varepsilon}^2$ and $ {\partial} ( A^{(2)} +
{\tilde B}^{(2)} ) / {\partial} {\varepsilon}^2$ are.
Therefore, we are forced to keep this derivative
in the NTL gap equation~\footnote{ 
On the other hand, we are allowed to ignore $A^{(1)} = 
{\cal{O}}(1/N_{\mbox{\scriptsize c}})$ in the NTL gap equation;
in fact, we must do this because otherwise the three integrands
in the NTL part of (\ref{NTLgap}) would have singularities
[cf.~discussion following Eqs.~(\ref{alpha1}) and (\ref{gaplead})]
and Goldstone theorem would be violated.}. 
If we ignored $A^{(1)}$ in ${\Xi}^{(1)}$, we would lose
its derivative in the gap equation, and would thus lose
an important part of the NTL effects there. On these grounds,
we get the NTL gap equation in the following form which
is now free of any singularities and is thus numerically
well defined:
\begin{eqnarray}
\lefteqn{
\frac{\partial \Xi_{\mbox{\footnotesize eff}} \left(
{\varepsilon}^2; a \right) }{\partial {\varepsilon}^2 }
= \frac{1}{a} - \int_0^{\infty} \frac{dz}{z^2}
\mbox{reg}_{\mbox{\scriptsize f}} \left( z \right) e^{- z {\varepsilon}^2 }
}
\nonumber\\
& & + \frac{1}{4 N_{\mbox{\footnotesize c}}}
 \int_0^{{\Lambda }_{\mbox{\scriptsize b}}^2/
{\Lambda }_{\mbox{\scriptsize f}}^2} 
 d {\bar p}^2 {\bar p}^2 
 {\Bigg \{ }
  \left[ \frac{\partial A^{(1)} \left( {\varepsilon}^2; a \right) }{
 \partial {\varepsilon}^2 } +
\frac{ \partial  \left( {\tilde B}_1^{(1)} + {\tilde B}_2^{(1)} \right)
 \left( {\bar p}^2; {\varepsilon}^2 \right) }{ \partial {\varepsilon}^2 }
 \right] 
 \left[  \left( {\tilde B}_1^{(1)} + {\tilde B}_2^{(1)} \right) 
 \left( {\bar p}^2; {\varepsilon}^2 \right) \right]^{-1}
 \nonumber\\
  & &
 + \left[ \frac{\partial A^{(1)} \left( {\varepsilon}^2; a \right) }{
 \partial {\varepsilon}^2 } +
\frac{ \partial  \left( {\tilde B}_1^{(1)} - {\tilde B}_2^{(1)} \right)
 \left( {\bar p}^2; {\varepsilon}^2 \right) }{ \partial {\varepsilon}^2 }
 \right] 
 \left[  \left( {\tilde B}_1^{(1)} - {\tilde B}_2^{(1)} \right)
 \left( {\bar p}^2; {\varepsilon}^2 \right) \right]^{-1}
   \nonumber\\
  & &
 + 2 \left[ \frac{\partial A^{(1)} \left( {\varepsilon}^2; a \right) }{
 \partial {\varepsilon}^2 } +
\frac{ \partial  \left( A^{(2)} + {\tilde B}^{(2)} \right)
 \left( {\bar p}^2; {\varepsilon}^2 \right) }{ \partial {\varepsilon}^2 }
 \right] 
 \left[  \left( A^{(2)} + {\tilde B}^{(2)} \right)
 \left( {\bar p}^2; {\varepsilon}^2 \right) \right]^{-1}
 {\Bigg \} }  = 0 \ .
\label{NTLgap}
\end{eqnarray}
The fact that ${\tilde{B}}_1^{(1)} - {\tilde{B}}_2^{(1)}$ and
$A^{(2)}+ {\tilde{B}}^{(2)}$ are proportional to ${\bar p}^2$ when
${\bar p}^2 \to 0$ is in (\ref{NTLgap}) a manifestation
of Goldstone theorem (masslessness of Goldstones).
We will denote the solution to the NTL gap equation simply as:
${\varepsilon}^2 = {\varepsilon}_{\mbox{\scriptsize gap}}^2$, or 
$({\varepsilon}_{\mbox{\scriptsize gap}}^{\mbox{\tiny (NTL)}})^2$.
This NTL gap equation 
is ``exact'' in the sense that it includes
all the NTL effects of the composite scalars (Higgs and
Goldstones). This means that it includes also the contributions 
of the longitudinal components of the
electroweak gauge bosons (in the Landau gauge); it does not 
include the effects of the other -- transverse -- components
of these gauge bosons. It does not yet include the effects
of the gluons - this will be done in the next Section.
The NTL gap equation (\ref{NTLgap}) 
can be solved numerically, for any given value of
${\Lambda}_{\mbox{\footnotesize b}}^2/
{\Lambda}_{\mbox{\footnotesize f}}^2$ and any allowed given value of the
input parameter $a$  [ $a > 1$ for the PTC, and $a > 1/(2 \ln 2)$
for the PV case]. 

There exists an alternative 
way of solving this equation -- namely in the
``$(1/N_{\mbox{\footnotesize c}})$ perturbative'' approach. We can
make for the solution ${\varepsilon}_{\mbox{\scriptsize gap}}^2$
the familiar $1/N_{\mbox{\footnotesize c}}$ expansion ansatz
\begin{equation}
\frac{\partial \Xi _{\mbox{\footnotesize eff}}
\left( {\varepsilon}^2;a \right) }
{\partial {\varepsilon}^2}
{\Bigg |}_{{\varepsilon}^2=
{\varepsilon}^2_{\mbox{\scriptsize gap}}} = 0 \ ,
\label{gapeq}
\end{equation}
\begin{equation}
\qquad \mbox{with:} \qquad 
{\varepsilon}^2_{\mbox{\scriptsize gap}}
=\frac{m_t^2({\Lambda})}
{{\Lambda}_{\mbox{\footnotesize f}}^2}=
{\varepsilon}_0^2+\frac{1}{N_{\mbox{\footnotesize c}}}
\kappa _{1\mbox{\footnotesize g}}+
{\cal {O}}(\frac{1}{N_{\mbox{\footnotesize c}}^2}) \ ,  
\label{epsexpan}
\end{equation}
where ${\varepsilon}_0^2$ is the solution of the 
leading-$N_{\mbox{\footnotesize c}}$ gap equation. Inserting
expansion (\ref{epsexpan}) into (\ref{gapeq}), using expansion
(\ref{defka2}) for ${\Xi}_{\mbox{\footnotesize eff}}$, and demanding
that coefficients at each power of $1/N_{\mbox{\footnotesize c}}$
be zero, we obtain the following relations:
\begin{equation}
\frac{\partial \Xi ^{(0)}}{\partial {\varepsilon}^2}
{\Bigg |}_{{\varepsilon}^2={\varepsilon}_0^2}=0 \ ,
\qquad
\kappa_{1\mbox{\footnotesize g}} \frac{\partial^2\Xi^{(0)}}
{\partial ({\varepsilon}^2)^2}
{\Bigg |}_{{\varepsilon}^2={\varepsilon}_0^2}+
\frac{\partial \Xi ^{(1)}}{\partial {\varepsilon}^2}
{\Bigg |}_{{\varepsilon}^2={\varepsilon}_0^2} = 0 \ .
\label{gap1}
\end{equation}
The ``$(1/N_{\mbox{\footnotesize c}})$ perturbative''
NTL gap equation (\ref{gap1}) determines the approximate change of 
the ratio 
${\varepsilon}^2_{\mbox{\scriptsize gap}}=
m_t^2({\Lambda})/{\Lambda}^2_{\mbox{\footnotesize f}}$ 
due to NTL effects 
\begin{equation}
\delta ({\varepsilon}^2)^{\mbox{\scriptsize (NTL)}}
_{\mbox{\scriptsize gap}} =
\frac{\kappa _{1\mbox{\footnotesize g}}}
{N_{\mbox{\footnotesize c}}}=
-\left[ 
\frac{\partial \Xi ^{(1)}}{\partial {\varepsilon}^2}
{\Big |}_{{\varepsilon}^2={\varepsilon}_0^2}
\right] {\Bigg /}
\left[ N_{\mbox{\footnotesize c}}
  \frac{\partial ^2\Xi ^{(0)}}{\partial ({\varepsilon}^2)^2}
{\Big |}_{{\varepsilon}^2={\varepsilon}_0^2}\right] \ .  
\label{deps2gap}
\end{equation}
Incidentally, in Eq.~(\ref{deps2gap})
we do not have to worry about setting the problematic
terms $A^{(1)}({\varepsilon}^2;a)$ to zero, since they are
zero automatically for ${\varepsilon}^2={\varepsilon}_0^2$. 
This latter approach was also
applied in Ref.~\cite{cv}, where we employed
the spherical covariant cutoff. We also discussed there briefly
the problem of singularities that would occur otherwise.
These problems of singularities in the NTL derivatives
${\partial} {\Xi}^{(1)} / {\partial} {\varepsilon}^2$ at 
${\varepsilon}^2 < {\varepsilon}_0^2$ have also been
discussed by the authors of Ref.~\cite{nikolov}, in the
context of an $SU(2)$ invariant NJL model which they
regarded as a model for low energy QCD.

Expression (\ref{deps2gap}) is a reasonably good approximation
to the actual NTL change as determined by the ``exact'' NTL
gap equation (\ref{NTLgap}) only as long as the values of
$|{\delta} ( {\varepsilon}^2 )_{\mbox{\scriptsize gap}}^
{\mbox{\scriptsize (NTL)}}/{\varepsilon}_0^2|$, ${\varepsilon}_0^2$, and
${\Lambda}_{\mbox{\footnotesize b}}^2/{\Lambda}_{\mbox{\footnotesize f}}^2$
are all sufficiently small ($\stackrel{<}{\sim} 0.5$). Even in such
cases, expression (\ref{deps2gap}) can frequently overestimate
$|{\delta} ( {\varepsilon}^2 )_{\mbox{\scriptsize gap}}^
{\mbox{\scriptsize (NTL)}}|$ by more than ten percent. 
This expression is less precise in the
PV cases than in the corresponding PTC and S cases. 
For all these reasons,
we use the ``exact'' NTL gap equation (\ref{NTLgap}) in our
calculations.

In this context, we mention that it is possible also in the
case of the covariant spherical cutoff (S) of fermionic
momenta to write down the ``exact'' NTL gap equation free
of any singularities, in close analogy with
(\ref{NTLgap}). The formal expression for ${\partial} {\Xi}^{(1)}
/ {\partial} {\varepsilon}^2$ in the S case was obtained
in Ref.~\cite{cv}. That expression contains as integrands
the S-regulated bubble-chain-corrected scalar propagators
$a [ 1 - a {\cal{J}}_X ({\bar p}^2; {\varepsilon}^2) ]^{-1}/2$ 
($X=H, Gn, Gch$)~\footnote{
Subscripts $H$, $Gn$ and $Gch$ correspond to NTL contributions from 
the Higgs, neutral Goldstone and  charged Goldstone degrees of 
freedom, respectively.}, 
which are represented in Fig.~3 as dashed
lines with a blob. Their analogues in the proper time approach
are the inverses of the arguments of logarithms in (\ref{Xi1}):
$ [ ( A^{(1)} + {\tilde B}_1^{(1)} \pm {\tilde B}_2^{(1)} )
({\bar p}^2; {\varepsilon}^2) ]^{-1}$ and $[ ( A^{(1)} + A^{(2)} +
{\tilde B}^{(2)} ) ({\bar p}^2; {\varepsilon}^2) ]^{-1}$.
The S-regulated scalar propagators as integrands for
${\partial} {\Xi}^{(1)}/ {\partial} {\varepsilon}^2$
have the same kind of singularities as their analogues
in the proper time framework: for $X=Gn$, $Gch$ singularities
occur for ${\varepsilon}^2 < {\varepsilon}_0^2$; for $X=H$
singularities occur for ${\varepsilon}^2 < {\varepsilon}_{\ast}^2
\ ( < {\varepsilon}_0^2 )$.   
The trick now is to replace them in
${\partial} {\Xi}^{(1)}/ {\partial} {\varepsilon}^2$ by
$ [1 - {\varepsilon}^2 \ln ( {\varepsilon}^{-2} + 1 )
 - {\cal{J}}_X ({\bar p}^2; {\varepsilon}^2) ]^{-1}/2$, and the
difference between the inverse of the old and of the latter 
expression is equal to
$2 {\partial} {\Xi}^{(0)}/ {\partial} {\varepsilon}^2$,
as was also the case in the proper time approach.
This represents
formally a next-to-next-to-leading (NNTL) modification, which
is again completely legitimate at the level of the NTL gap
equation. The resulting modified inverse propagators of the
scalars can be shown to be positive everywhere, and therefore
the NTL gap equation written below is free of any singularities:
\begin{eqnarray}
\lefteqn{
\frac{ {\partial } {\Xi}_{\mbox{\footnotesize eff}}
\left( {\varepsilon}^2 ; a \right) }{
{\partial} {\varepsilon}^2 }^{\mbox{\scriptsize (S)}} =
\frac{1}{a} - \left[ 1 - {\varepsilon}^2
\ln \left( {\varepsilon}^{-2} + 1 \right) \right]
} \nonumber\\
& &
 - \frac{1}{ 4 N_{\mbox{\footnotesize c}} }
\int_0^{ {\Lambda}_{\mbox{\scriptsize b}}^2/
 {\Lambda}_{\mbox{\scriptsize f}}^2 }
d {\bar p}^2 {\bar p}^2 
{\Bigg \{ } \sum_X A_X
\frac{ {\partial} {\cal{J}}_X \left( {\bar p}^2; 
{\varepsilon}^2 \right) }{ {\partial} {\varepsilon}^2 }
\left[ 1 - {\varepsilon}^2 \ln \left( {\varepsilon}^{-2} + 1 \right)
- {\cal{J}}_X \left( {\bar p}^2; {\varepsilon}^2 \right) 
\right]^{-1}
{\Bigg \} } = 0 \ .
\label{NTLgapS}
\end{eqnarray}
The subscripts in the sum above are: $X = H, Gn, Gch$; and the
respective multiplicity factors are: $A_X= 1,1,2$.
The corresponding dimensionless two-point 
Green functions with the fermionic S-cutoff are
\begin{eqnarray}
{\cal J}_H\left( {\bar p}^2;{\varepsilon}^2\right)  &=&
\frac 1{\pi ^2}\int_{{\bar k}^2\leq 1}d^4{\bar k}
\frac{\left[ {\bar k}\cdot ({\bar p}+{\bar k})-{\varepsilon}^2
 \right] }
{\left( {\bar k}^2+{\varepsilon}^2\right) 
\left[ ({\bar p}+{\bar k})^2+{\varepsilon}^2\right] }\ ,  
\nonumber \\
{\cal J}_{Gn}\left( {\bar p}^2;{\varepsilon}^2\right)  &=&
\frac 1{\pi ^2}\int_{{\bar k}^2\leq 1}d^4{\bar k}
\frac{\left[ {\bar k}\cdot ({\bar p}+{\bar k})+{\varepsilon}^2
 \right] }
{\left( {\bar k}^2+{\varepsilon}^2\right) 
\left[ ({\bar p}+{\bar k})^2+{\varepsilon}^2\right] }\ ,  
\nonumber \\
{\cal J}_{Gch}\left( {\bar p}^2;{\varepsilon}^2\right)  &=&
\frac 1{\pi^2}\int_{{\bar k}^2\leq 1}d^4{\bar k}
\frac{{\bar k}\cdot ({\bar p}+{\bar k})}
{\left( {\bar k}^2+{\varepsilon}^2\right) ({\bar p}+{\bar k})^2} \ ,  
\label{Js}
\end{eqnarray}
and their explicit expressions were given in Ref.~\cite{cv}.
Also in the S case, we don't use the approximate NTL gap 
equation (\ref{deps2gap}), but the ``exact'' S-cutoff
NTL gap equation (\ref{NTLgapS}).

\section{Mass renormalization, QCD corrections}

We calculate and include at this point also the top quark
mass renormalization effects
$m_t({\Lambda})$ 
$\mapsto m_t^{\mbox{\scriptsize  ren.}}$. In this
context, we stress that the
hard mass of the NTL gap equation (\ref{NTLgap}) is
the fixed (nonrunning) mass in the dynamically broken
effective theory (\ref{TSM}) with a cutoff $\Lambda
\sim \Lambda_{\mbox{\footnotesize f}} \sim
\Lambda_{\mbox{\footnotesize b}}$, i.e., it is
the {\it bare} mass $m_t(\Lambda)$ 
(${\varepsilon}_{\mbox{\scriptsize gap}}=
m_t(\Lambda)/\Lambda_{\mbox{\footnotesize f}}$).
In order to obtain the values of the cutoff parameters
${\Lambda}_{\mbox{\footnotesize f}}$ and
${\Lambda}_{\mbox{\footnotesize b}}$, we therefore have
to perform, after having solved the NTL gap equation
($\mapsto m_t({\Lambda})/{\Lambda}_{\mbox{\footnotesize f}}$),
also the mass renormalization 
($m_t({\Lambda})/{\Lambda}_{\mbox{\footnotesize f}} \mapsto
m_t^{\mbox{\scriptsize ren.}}/{\Lambda}_{\mbox{\footnotesize f}}$),
since the mass of the top quark is more or less known
($m_t^{\mbox{\scriptsize ren.}} \approx 180 
\mbox{ GeV}$)~\cite{CDFD0}.
It is straightforward to check that there are no
leading-$N_{\mbox{\footnotesize c}}$ contributions to these
renormalization corrections, and that only the 1-PI 
diagrams shown in Fig.~3
account for the NTL renormalization effects 
(cf.~also~\cite{cpv},~\cite{cv} ). Therefore
\begin{equation}
\delta ({\varepsilon}^2)_{\mbox{\scriptsize ren.}}=
\frac{(m_t^2)_{\mbox{\scriptsize ren.}}}
 {{\Lambda}_{\mbox{\footnotesize f}}^2}-
 \frac{m_t^2({\Lambda}_{\mbox{\footnotesize f}})}
 {{\Lambda}_{\mbox{\footnotesize f}}^2}
\ ( = {\varepsilon}_{\mbox{\scriptsize ren.}}^2 -
    {\varepsilon}_{\mbox{\scriptsize gap}}^2 ) \
= \frac{1}{N_{\mbox{\footnotesize c}}}
 \kappa _{1\mbox{\footnotesize r}}
 +{\cal {O}}(\frac{1}{N_{\mbox{\footnotesize c}}^2}) \ .
\label{deps1ren}
\end{equation}
There are three separate contributions, coming 
from the Higgs, neutral Goldstone and the charged Goldstone
``coated'' (i.e., bubble-chain-corrected) propagators of Fig.~3
\begin{equation}
\delta ( {\varepsilon}^2 )_{\mbox{\scriptsize ren.}}
^{\mbox{\scriptsize (NTL)}} =
\frac{1}{N_{\mbox{\footnotesize c}}} 
\kappa_{1\mbox{\footnotesize r}} =
\frac{1}{N_{\mbox{\footnotesize c}}} \left(
\kappa _{1\mbox{\footnotesize r}}^{(H)}
+\kappa _{1\mbox{\footnotesize r}}^{(Gn)}
+\kappa _{1\mbox{\footnotesize r}}^{(Gch)}
\right) \ .
\label{kappar}
\end{equation}
In order to obtain the corresponding ``coated'' propagators
of the scalars within the framework of the discussed
proper time regularized results, we note that the
two-point 1PI Green function of the scalar ${\tilde {\sigma}}_j$
in the Euclidean ${\bar x}$ space is
\begin{equation}
\Gamma_j^{(2)} \left( {\bar x}_2 -  {\bar x}_1 \right)
= \frac{ g }{2} \left[ 
\frac{\delta}{\delta s_j \left( {\bar x}_2 \right) }
\frac{\delta}{\delta s_j \left( {\bar x}_1 \right) }
\triangle \Gamma \left( \lbrace \sigma_k; s_k \rbrace \right) 
\right] \ , \quad  ( g = G M_0^2 ) \ ,
\label{Greenx}
\end{equation}
where 
$\triangle \Gamma \left( \lbrace \sigma_k; s_k \rbrace \right)$ 
is the part of the scalar action that is quadratic in the scalar
fluctuations $s_k({\bar y})$. Incidentally, the scalar action
is the expression in the curly brackets of the exponent in
formula (\ref{veff1}). These two-point Green functions are
therefore proportional to the action kernels (\ref{veff1argx})
(multiplied by $g/2$).
In the momentum space, the propagators 
are inversely proportional to the Fourier transforms
of the above expressions
\begin{eqnarray}
\frac{(-i)}{\tilde \Gamma_0^{(2)} ( {\bar p}^2; {\sigma}_0^2 ) } & = &
\frac{ - 2 i}{g} 
 \left[ \alpha^{(1)} ( \sigma_0^2 )
 + \left( {\tilde \beta}_1^{(1)} + {\tilde \beta}_2^{(1)} \right)
 ( {\bar p}^2 ; \sigma_0^2 ) \right]^{-1} \ ,
\nonumber\\
\frac{(-i)}{\tilde \Gamma_1^{(2)} ( {\bar p}^2; {\sigma}_0^2 ) } & = &
\frac{ - 2 i}{g} 
 \left[ \alpha^{(1)} ( \sigma_0^2 )
 + \left( {\tilde \beta}_1^{(1)} - {\tilde \beta}_2^{(1)} \right)
 ( {\bar p}^2 ; \sigma_0^2 ) \right]^{-1} \ ,
\nonumber\\
\frac{(-i)}{\tilde \Gamma_2^{(2)} ( {\bar p}^2; {\sigma}_0^2 ) } & = &
\frac{ - 2 i}{g} 
 \left[ \alpha^{(1)} ( \sigma_0^2 )
 + \alpha^{(2)} ( \sigma_0^2 )
 +  {\tilde \beta}^{(2)}
  ( {\bar p}^2 ; \sigma_0^2 ) \right]^{-1} =
\frac{(-i)}{\tilde \Gamma_3^{(2)} ( {\bar p}^2; {\sigma}_0^2 ) } \ .
\label{Greenp}
\end{eqnarray}
Using these ``coated'' scalar propagators in the diagrams of Fig.~3 in
Euclidean space, and again setting the term $\alpha^{(1)}(\sigma_0^2)$
[$\propto A^{(1)}({\varepsilon}^2)$] equal to zero since it
represents formally NNTL effects and would otherwise give us
singular integrals, we get the following expressions
for the mass renormalization terms $\kappa_{1\mbox{\footnotesize r}}$:
\begin{equation}
\kappa _{1\mbox{\footnotesize r}}^{(H)}=-\frac{1}{2}
\int_0^{{\Lambda }_{\mbox{\scriptsize b}}^2/
{\Lambda}_{\mbox{\scriptsize f}}^2}
\frac{d{\bar p}^2}
{\left( {\tilde B}_1^{(1)}+ {\tilde B}_2^{(1)} \right) 
\left( {\bar p}^2; {\varepsilon}^2 \right) }
\left[ \left( \sqrt{ {\bar p}^2 ( {\bar p}^2+4{\varepsilon}^2 ) }
-{\bar p}^2 \right) 
\left( 2+\frac{{\bar p}^2}{2{\varepsilon}^2}\right) 
- {\bar p}^2 \right] \ ,  
\label{kapHr}
\end{equation}
\begin{equation}
\kappa _{1\mbox{\footnotesize r}}^{(Gn)}=
+\frac{1}{2} \int_0^{{\Lambda }_{\mbox{\scriptsize b}}^2/
{\Lambda}_{\mbox{\scriptsize f}}^2}
\frac{d{\bar p}^2}
{\left( {\tilde B}_1^{(1)} - {\tilde B}_2^{(1)} \right) 
\left( {\bar p}^2; {\varepsilon}^2 \right) }
\frac{1}{2{\varepsilon}^2}
{\bar p}^2
\left[ {\bar p}^2+2{\varepsilon}^2
-\sqrt{ {\bar p}^2 ( {\bar p}^2+4{\varepsilon}^2 ) }
\right] \ ,  
\label{kapGnr}
\end{equation}
\begin{eqnarray}
\kappa _{1\mbox{\footnotesize r}}^{(Gch)} &=&
{\Big \{}
\frac{1}{2} \int_0^{-{\varepsilon}^2} 
\frac{ d{\bar p}^2{\bar p}^2 
 \left[ 2+{\bar p}^2/{\varepsilon}^2 \right] }
{\left( A^{(2)} + {\tilde B}^{(2)} \right)
\left( {\bar p}^2; {\varepsilon}^2 \right) }
\nonumber\\
&&
-\frac{ {\varepsilon}^2}{2} 
 \int_{-{\varepsilon}^2}^{
{\Lambda }_{\mbox{\scriptsize b}}^2/{\Lambda}_{\mbox{\scriptsize f}}^2}
d{\bar p}^2 \left[ 
\frac{1}{ \left( A^{(2)} + {\tilde B}^{(2)} \right)
\left( {\bar p}^2; {\varepsilon}^2 \right) }
-\frac{1}{{\bar p}^2} 
 \lim_{ {\bar{q^{\prime}}}^2 \to 0 } 
\left( \frac{ {\bar{q^{\prime}}}^2 }{
\left( A^{(2)} + {\tilde B}^{(2)} \right)
\left( {\bar{q^{\prime}}}^2; {\varepsilon}^2 \right) }
\right) \right]
\nonumber \\
&& 
-\frac{ {\varepsilon}^2 }{2}
\lim_{ {\bar{q^{\prime}}}^2 \to 0 } 
\left( \frac{ {\bar{q^{\prime}}}^2 }{
\left( A^{(2)} + {\tilde B}^{(2)} \right)
\left( {\bar{q^{\prime}}}^2; {\varepsilon}^2 \right) }
\right) 
\left[ -\ln {\varepsilon}^2+
 \ln \left( {\Lambda }_{\mbox{\footnotesize b}}^2/
{\Lambda}_{\mbox{\footnotesize f}}^2\right) \right] 
{\Big \}} \ .  
\label{kapGchr}
\end{eqnarray}
Here, we denoted by ${\varepsilon}^2$ the ``bare'' mass 
value ${\varepsilon}_{\mbox{\scriptsize gap}}^2 =
m_t^2({\Lambda})/{\Lambda}_{\mbox{\footnotesize f}}^2$,
i.e., the solution of the NTL gap equation. 
The expressions above were obtained 
by assuming first a (normalized) Euclidean
momentum ${\bar q}^2>0$ for the external top quark line in the
diagrams of Fig.~3. Then the 
analytic continuation to the (approximate) on-shell values 
${\bar q}^2=-q^2=-{\varepsilon}^2$ 
[$ = -m_t^2({\Lambda})/{\Lambda}_{\mbox{\footnotesize f}}^2$] 
had to be performed. 
In the cases of $\kappa^{(H)}_{1\mbox{\footnotesize r}}$ and 
$\kappa^{(Gn)}_{1\mbox{\footnotesize r}}$,
this continuation is equivalent to the simple
substitution in the Euclidean integrands: 
${\bar q}^2 \mapsto -{\varepsilon}^2$. The contribution
$\kappa^{(Gch)}_{1\mbox{\footnotesize r}}$ 
of the charged Goldstones leading to 
(\ref{kapGchr}) is somewhat more complicated due to the fact that
the pole at ${\bar p}^2 = 0$ of the massless charged Goldstone 
generates a logarithmic branch cut in 
$\kappa^{(Gch)}_{1\mbox{\footnotesize r}}({\bar q}^2)$ 
at the threshold value ${\bar q}^2 = 0$. The analytic continuation
follows then from the usual prescription:
$\ln {\bar q}^2 \mapsto \ln ( -q^2- i {\epsilon} ) \mapsto
\ln q^2 - i {\pi}$, for $q^2 > 0 $. The real part of this term
was written in (\ref{kapGchr}) as a separate 
$\ln {\varepsilon}^2$-term. None of the 
integrals in (\ref{kapGchr}) is singular.

The expressions ${\kappa}^{(X)}_{j \mbox{\footnotesize r} }$
for the case of the covariant spherical (S) cutoff were written
in Ref.~\cite{cv} where we substituted for ${\varepsilon}^2$
the leading-$N_{\mbox{\footnotesize c}}$ gap equation
solution ${\varepsilon}_0^2$, instead of the ``bare''
mass solution ${\varepsilon}_{\mbox{\scriptsize gap}}^2$ 
of the NTL gap equation. 
This was formally an acceptable
approximation, because the renormalization contributions 
$\delta ( {\varepsilon}^2 )^{\mbox{\scriptsize (NTL)}}_{\mbox
{\scriptsize ren.}}$ are
$1/N_{\mbox{\footnotesize c}}$ effects, as are the NTL gap
contributions 
$\delta ( {\varepsilon}^2 )^{\mbox{\scriptsize (NTL)}}_{\mbox
{\scriptsize gap}} = {\varepsilon}_{\mbox{\scriptsize gap}}^2 -
{\varepsilon}_0^2$. However, we investigate the
borderline cases where the NTL gap contributions are large:
$|\delta ( {\varepsilon}^2 )^{\mbox{\scriptsize (NTL)}}_{\mbox
{\scriptsize gap}}| \stackrel{<}{\sim} {\varepsilon}_0^2$, in general
larger than $|\delta ( {\varepsilon}^2 )^{\mbox{\scriptsize (NTL)}}
_{\mbox{\scriptsize ren.}}|$. Therefore,
it would be more realistic also in the S case of~\cite{cv}
to insert in the integrands for 
${\kappa}^{(X)}_{j \mbox{\footnotesize r}}$'s
the ``bare'' mass solution of the NTL gap equation (\ref{NTLgapS}):
${\varepsilon}^2 = {\varepsilon}_{\mbox{\scriptsize gap}}^2 
( < {\varepsilon}_0^2 )$. However, once we do that,
the formulas for ${\kappa}^{(X)}_{j \mbox{\footnotesize r}}$'s
($X=Gn, Gch$) become singular. As mentioned at the end
of the previous Section, the remedy for this formal problem is known:
the replacement of the ``coated'' (bubble-chain-corrected)
S-regularized scalar propagators
$ ( 1 - a {\cal{J}}_X )^{-1} $ by 
$ a^{-1} 
[ 1 - {\varepsilon}^2 \ln \left( {\varepsilon}^{-2} + 1 \right)
- {\cal{J}}_X ]^{-1}$, resulting formally in an NNTL modification
(thus legitimate at the NTL level) and giving the completely
nonsingular integrals for ${\kappa}^{(X)}_{j \mbox{\footnotesize r}}$'s
in the S case:
\begin{equation}
\kappa _{1\mbox{\footnotesize r}}^{(H)}=-\frac{1}{4}
\int_0^{{\Lambda }_{\mbox{\scriptsize b}}^2/
{\Lambda}_{\mbox{\scriptsize f}}^2}
\frac{ d{\bar p}^2 }
{ \left[ 1- {\varepsilon}^2 \ln \left(
{\varepsilon}^{-2} + 1 \right)
- {\cal{J}}_{H} \left( {\bar p}^2 ; {\varepsilon}^2 \right)
\right] }
\left[ \left( \sqrt{ {\bar p}^2 ( {\bar p}^2+4{\varepsilon}^2 ) }
-{\bar p}^2 \right) 
\left( 2+\frac{{\bar p}^2}{2{\varepsilon}^2}\right) 
- {\bar p}^2 \right] \ ,  
\label{kapSHr}
\end{equation}
\begin{equation}
\kappa _{1\mbox{\footnotesize r}}^{(Gn)}=
+ \frac{1}{4} \int_0^{{\Lambda }_{\mbox{\scriptsize b}}^2/
{\Lambda}_{\mbox{\scriptsize f}}^2}
\frac{ d{\bar p}^2 }
{ \left[ 1 - {\varepsilon}^2 \ln \left(
{\varepsilon}^{-2} + 1 \right)
- {\cal J}_{Gn}\left( {\bar p}^2;{\varepsilon}^2\right) 
\right] }
\frac{1}{2{\varepsilon}^2}
{\bar p}^2
\left[ {\bar p}^2+2{\varepsilon}^2
-\sqrt{ {\bar p}^2 ( {\bar p}^2+4{\varepsilon}^2 ) }
\right] \ ,  
\label{kapSGnr}
\end{equation}
\begin{eqnarray}
\kappa _{1\mbox{\footnotesize r}}^{(Gch)} &=&
+\frac{1}{4}
{\Bigg \{}
\int_0^{-{\varepsilon}^2} 
\frac{ d {\bar p}^2 {\bar p}^2 
 \left[ 2+{\bar p}^2/{\varepsilon}^2 \right] }
{ \left[ 1 - {\varepsilon}^2 \ln \left(
{\varepsilon}^{-2} + 1 \right)
- {\cal J}_{Gch} \left( {\bar p}^2; {\varepsilon}^2
\right) \right] }
\nonumber\\
& & -{\varepsilon}^2 
 \int_{-{\varepsilon}^2}^{ {\Lambda}_{\mbox{\scriptsize b}}^2/
{\Lambda}_{\mbox{\scriptsize f}}^2}
d {\bar p}^2 \left[ 
\frac{1}{ \left( 
1 - {\varepsilon}^2 \ln \left(
{\varepsilon}^{-2} + 1 \right) -
 {\cal J}_{Gch}\left( {\bar p}^2; {\varepsilon}^2\right) 
\right) }
-\frac{2}{ {\bar p}^2 
\ln \left( 1+1/{\varepsilon}^2\right) }
\right]   
\nonumber\\
& & -\frac{2{\varepsilon}^2}
{ \ln \left( 1+1/{\varepsilon}^2\right) }
\left[ -\ln {\varepsilon}^2+
 \ln \left( {\Lambda }_{\mbox{\footnotesize b}}^2/
{\Lambda}_{\mbox{\footnotesize f}}^2\right) \right] 
{\Bigg \}} \ .  
\label{kapSGchr}
\end{eqnarray}
We insert in these expressions the NTL ``bare'' mass value
${\varepsilon}^2 = {\varepsilon}_{\mbox{\scriptsize gap}}^2$.

Finally, we include the leading part of QCD effects. This was
already done in Refs.~\cite{cpv},~\cite{cv} and we cite here only
the results. The leading ``gap'' part of QCD
is represented by the contributions coming from the 
diagrams of Fig.~2, where the internal dashed lines represent now 
the gluon propagators (in Landau gauge). Since QCD effects
turn out to be only of minor numerical importance in our framework,
we decided to regulate the corresponding QCD integrals
by means of only one specific approach -- by the proper time cutoff
$1/{\Lambda}_{\mbox{\footnotesize f}}^2$ for the quarks and
$1/{\Lambda }_{\mbox{\footnotesize b}}^2$ for the gluons.
The corresponding contribution to $\Xi ^{(1)}$ to be added in
(\ref{Xi1}) was derived in~\cite{cpv}
\begin{equation}
\Xi^{(1;\mbox{\scriptsize gl})}
\left( {\varepsilon}^2;
{\Lambda }_{\mbox{\footnotesize b}}^2/{\Lambda}
_{\mbox{\footnotesize f}}^2;a_{\mbox{\scriptsize gl}}\right) 
= 2 \int_0^{{\Lambda }_{\mbox{\scriptsize b}}^2/{\Lambda}
_{\mbox{\scriptsize f}}^2}
d{\bar p}^2{\bar p}^2
\ln \left[ 1-a_{\mbox{\scriptsize gl}}
 {\cal {J}}_{\mbox{\scriptsize gl}}\left( {\bar p}^2;
 {\varepsilon}^2\right)
\right] \ ,  
\label{Xigl}
\end{equation}
Here, $a_{\mbox{\scriptsize gl}}$ is the relevant
QCD coupling parameter:
$a_{\mbox{\scriptsize gl}}=3\alpha _s(m_t)/\pi \approx 0.105$. 
The (proper time regulated) two-point Green function 
${\cal {J}}_{\mbox{\scriptsize gl}}$ appearing in (\ref{Xigl}) is
\begin{eqnarray}
{\cal {J}}_{\mbox{\scriptsize gl}}
\left( {\bar p}^2,{\varepsilon}^2\right)  &=&
 \frac 16\ln {\varepsilon}^2+\frac 29 
-\frac 16 ( 2 w - 1 ) \left[ -2
+ \sqrt{4 w + 1} \ln \left( \frac{ \sqrt{4w+1} + 1}{ \sqrt{4w+1} - 1}
\right) \right] {\Bigg |}_{w= {\varepsilon}^2/{\bar p}^2 }
\nonumber \\
&&
 -\frac 16\left( \frac{{\bar p}^2}5+{\varepsilon}^2\right) 
 +\frac 14\left( 
 \frac{{\bar p}^4}{140}+\frac{{\bar p}^2{\varepsilon}^2}{15}
 +\frac{{\varepsilon}^4}6 \right) 
 +{\cal {O}}\left( {\bar p}^6,{\varepsilon}^6 \right) 
\ .  
\label{Jglpt}
\end{eqnarray}
It is worth mentioning that QCD expression (\ref{Xigl}), unlike (\ref{Xi1}),
turns out to be numerically almost equal to its two-loop approximation
(obtained by the replacement: 
$\ln [ 1 - a_{\mbox{\scriptsize gl}} 
 {\cal {J}}_{\mbox{\scriptsize gl}} ({\bar p}^2, {\varepsilon}^2 )]
\mapsto - a_{\mbox{\scriptsize gl}} 
{\cal {J}}_{\mbox{\scriptsize gl}}({\bar p}^2, {\varepsilon}^2 ) $),
the difference being only a fraction of a percent. This has to do
with the small values of $a_{\mbox{\scriptsize gl}}(E)$ at the
relevant energies $E \stackrel{>}{\sim} m_t$. That's why the
QCD contributions to 
$\delta ({\varepsilon}^2)_{\mbox{\scriptsize gap}}$
turn out to be quite small (almost negligible) in comparison to
the NTL contributions of the scalars in the present framework.

The leading QCD $m_t$-mass renormalization effect, which
is numerically more important than the QCD contribution
to $\delta ({\varepsilon}^2)_{\mbox{\scriptsize gap}}
={\varepsilon}_{\mbox{\scriptsize gap}}^2 - {\varepsilon}_0^2$,
comes from the 
version of the diagrams of Fig.~3, where the dashed line 
with a blob is now the gluonic propagator. With the proper time cutoff 
we obtain 
(cf.~\cite{cpv}) 
\begin{equation}
\delta ({\varepsilon}^2)_{\mbox{\scriptsize  ren.}}
^{\mbox{\scriptsize QCD}} \approx
\frac{2}{3} a_{\mbox{\scriptsize gl}}{\varepsilon}^2
\left[ 
\ln \left( {\varepsilon}^{-2}\right) 
 +\ln \left( {\Lambda }_{\mbox{\footnotesize b}}^2/
{\Lambda}_{\mbox{\footnotesize f}}^2\right) 
 +0.256 \ldots
 +\frac{5{\Lambda}^2_{\mbox{\footnotesize f}}}
{9{\Lambda}^2_{\mbox{\footnotesize b}}}{\varepsilon}^2
 +{\cal {O}}({\varepsilon}^4)
\right] \ ,
\label{kapglr}
\end{equation}
where we insert again for ${\varepsilon}^2$ the ``bare''
mass value ${\varepsilon}_{\mbox{\scriptsize gap}}^2$.
This expression is to be added to (\ref{kappar}) in order to obtain 
the QCD modified 
$\delta({\varepsilon}^2)_{\mbox{\scriptsize  ren.}}
= {\varepsilon}_{\mbox{\scriptsize ren.}}^2 -
{\varepsilon}_{\mbox{\scriptsize gap}}^2$.

In our original paper~\cite{cpv}, we regarded only the scalar
sector contributions to $V_{\mbox{\footnotesize eff}}$ and
$\delta V_{\mbox{\footnotesize eff}} / \delta {\varepsilon}^2$
as being organized in an $1/N_{\mbox{\footnotesize c}}$
expansion, and assumed that QCD contributions can be organized
in a perturbative series in powers of ${\alpha}_s(m_t)$.
However, as argued in~\cite{mty},~\cite{king} 
and~\cite{yamawaki}, the dominant part of the
QCD contributions is
formally a leading-$N_{\mbox{\footnotesize c}}$ contribution.
The reason lies in the fact that ${\alpha}_s = 
{\cal{O}}(N_{\mbox{\footnotesize c}}^{-1})$. 
In ${\Xi}^{(1; {\mbox{\scriptsize gl}})}$ in (\ref{Xigl}),
the factor $2$ is replaced in the case of a 
general $N_{\mbox{\footnotesize c}}$
by $(N_{\mbox{\footnotesize c}}^2-1)/4 = 
{\cal{O}}(N_{\mbox{\footnotesize c}}^2)$; furthermore,
$\ln ( 1 - a_{\mbox{\scriptsize gl}} {\cal{J}}_{\mbox{\scriptsize gl}} )
\approx - a_{\mbox{\scriptsize gl}} {\cal{J}}_{\mbox{\scriptsize gl}} 
\propto a_{\mbox{\scriptsize gl}} \propto {\alpha}_s =
{\cal{O}}(N_{\mbox{\footnotesize c}}^{-1})$. 
Therefore,
${\Xi}^{(1; {\mbox{\scriptsize gl}})} = 
{\cal{O}}(N_{\mbox{\footnotesize c}})$ when added
to the NTL part ${\Xi}^{(1)} = {\cal{O}}(N_{\mbox{\footnotesize c}}^0)$ 
of (\ref{Xi1}). Hence, we see really that
${\Xi}^{(1; {\mbox{\scriptsize gl}})}$ is formally not
NTL, but leading-$N_{\mbox{\footnotesize c}}$ contribution
to $\Xi_{\mbox{\footnotesize eff}}$ in the 
expansion (\ref{defka2}). Also the QCD renormalization
contribution (\ref{kapglr}) is formally 
leading-$N_{\mbox{\footnotesize c}}$. Namely, the
factor $2/3$ in (\ref{kapglr}) is in the general case replaced by
$(N_{\mbox{\footnotesize c}}^2-1)/(4 N_{\mbox{\footnotesize c}})
= {\cal{O}}(N_{\mbox{\footnotesize c}})$, and since
$a_{\mbox{\scriptsize gl}} = {\cal{O}}(N_{\mbox{\footnotesize c}}^{-1})$,
we have $\delta ({\varepsilon}^2)_{\mbox{\scriptsize ren.}}^
{\mbox{\scriptsize QCD}} = {\cal{O}}(N_{\mbox{\footnotesize c}}^0)$,
i.e., formally leading-$N_{\mbox{\footnotesize c}}$ effect.
However, it will turn out that for the cases considered in
the present paper (with cutoffs
${\Lambda} \sim 1$ TeV when the NTL contributions of the
scalars become comparable to the leading-$N_{\mbox{\footnotesize c}}$
quark loop contributions), QCD contributions are numerically
almost an order of magnitude
smaller than both the leading-$N_{\mbox{\footnotesize c}}$
quark loop contributions and
the NTL scalar contributions. That's why we included
in our formulas QCD contributions in the NTL parts.

\section{Numerical evaluations}

The inputs for the integrations are the values of the parameter 
$a=N_{\mbox{\footnotesize c}}G{\Lambda}_
{\mbox{\footnotesize f}}^2/8\pi ^2$ of 
(\ref{defka1}), which is essentially a 
dimensionless measure of the strength of the original 
four-fermion coupling $G$ in (\ref{TSM}), as well as the values of
the ratio ${\Lambda }_{\mbox{\footnotesize b}}/
{\Lambda}_{\mbox{\footnotesize f}}$ ($ \sim 1$)
of the bosonic and fermionic cutoff parameters.
The diagrams of Fig.~2 suggest
${\bar p}_{\mbox{\footnotesize max}}^2
\leq {\bar k}_{\mbox{\footnotesize max}}^2$, implying
the input values ${\Lambda }_{\mbox{\footnotesize b}}/
{\Lambda}_{\mbox{\footnotesize f}} \stackrel{<}{\sim} 1$, at least
in the more intuitive cases of the fermionic S or PTC cutoff. 
Therefore, we have made three input choices for these
ratios in the PTC case:
${\Lambda }_{\mbox{\footnotesize b}}/
{\Lambda}_{\mbox{\footnotesize f}}^{\mbox{\scriptsize (PTC)}} =
$0.5$, 1/\sqrt{2}$ ($\approx 0.707$), $1$. In fact, it will
turn out that  
${\Lambda }_{\mbox{\footnotesize b}}/
{\Lambda}_{\mbox{\footnotesize f}}^{\mbox{\scriptsize (PTC)}} 
 > 1$ does not lead to physically acceptable results.

It turns out that both the NTL gap equation (\ref{NTLgap}) [with QCD
contributions from (\ref{Xigl}) included] and the (NTL) mass
renormalization effects (\ref{kappar}), (\ref{kapHr})-(\ref{kapGchr}),
(\ref{kapglr}) decrease the ratio ${\varepsilon}^2$ when compared
to the leading-$N_{\mbox{\footnotesize c}}$ value ${\varepsilon}_0^2$
of (\ref{gaplead}). In other words, we have:
$\delta ( {\varepsilon}^2 )_{\mbox{\scriptsize gap}} =
{\varepsilon}_{\mbox{\scriptsize gap}}^2 - {\varepsilon}_0^2
< 0$ and
$\delta ( {\varepsilon}^2 )_{\mbox{\scriptsize ren.}} =
{\varepsilon}_{\mbox{\scriptsize ren.}}^2 - 
{\varepsilon}_{\mbox{\scriptsize gap}}^2 < 0$. In general, we have
$|\delta ( {\varepsilon}^2 )_{\mbox{\scriptsize gap}}| > 
 |\delta ( {\varepsilon}^2 )_{\mbox{\scriptsize ren.}}|$. 
When ${\varepsilon}_{\mbox{\scriptsize gap}}^2 \to 0$, 
$\delta ({\varepsilon}^2)_{\mbox{\scriptsize gap}}$
remains relatively stable while
$|\delta ({\varepsilon}^2)_{\mbox{\scriptsize  ren.}}| \
( < {\varepsilon}_{\mbox{\scriptsize gap}}^2 ) \to 0$. Hence,
$\delta ({\varepsilon}^2)_{\mbox{\scriptsize  gap}}$ is 
identified as the source of the observed
``$1/N_{\mbox{\footnotesize c}}$-nonperturbative'' behavior,
unlike $\delta ({\varepsilon}^2)_{\mbox{\scriptsize  ren.}}$,
when ${\varepsilon}_{\mbox{\scriptsize gap}}^2 \equiv 
m_t^2(\Lambda)/{\Lambda}_{\mbox{\footnotesize f}}^2 \to 0$.
Therefore, if we require $1/N_{\mbox{\footnotesize c}}$
expansion in our framework to have at least some qualitatively
predictive power, than the NTL gap change
$|\delta ( {\varepsilon}^2 )_{\mbox{\scriptsize gap}}|$
should not be too large in comparison to ${\varepsilon}_0^2$,
i.e., $m_t({\Lambda})/m_t^{(0)} = 
\sqrt{ {\varepsilon}_{\mbox{\scriptsize gap}}^2 / {\varepsilon}_0^2 }$
should not decrease beyond some critical small value. Consequently, the 
leading-$N_{\mbox{\footnotesize c}}$ ratio ${\varepsilon}_0^2$,
or equivalently the input parameter $a$ [cf.~(\ref{gaplead})],
should not decrease beyond certain corresponding critical values.
In Table 1, we chose in the PTC case, for given ratio 
${\Lambda}_{\mbox{\footnotesize b}}/{\Lambda}_{\mbox{\footnotesize f}}$
(second column), various ``critical'' small values for
$m_t^2({\Lambda})/m_t^{(0)2} = {\varepsilon}_{\mbox{\scriptsize gap}}^2/
{\varepsilon}_0^2$: $1/4$, $1/3$, $1/2$, $2/3$ (third column). 
There is a certain arbitrariness in deciding which of these values
represents most realistically the breakdown of the 
$1/N_{\mbox{\footnotesize c}}$
expansion approach in the PTC case. In the fourth column, the 
corresponding ratio of the renormalized mass vs.~$m_t^{(0)}$
is given. The last two columns contain the corresponding
values of the cutoff parameters 
${\Lambda}_{\mbox{\footnotesize b}}$ and
${\Lambda}_{\mbox{\footnotesize f}}^{\mbox{\scriptsize (PTC)}}$,
which are obtained from the calculated values of 
${\varepsilon}_{\mbox{\scriptsize ren.}}^2 = 
( m_t^{\mbox{\scriptsize ren.}}/{\Lambda}_{\mbox{\footnotesize f}} )^2$
and the approximate experimental value
$m_t^{\mbox{\scriptsize ren.}} = 180 \mbox{ GeV}$~\cite{CDFD0}.
In the first column, the
corresponding values of the coupling input parameter
$a(\mbox{\scriptsize PTC})$ are given. We note that the values 
in the last two columns are the upper bounds of the cutoff
parameters, once we take the stand that the NTL gap
equation effects should not be stronger than
in the specific case, i.e., that the ratio
$m_t({\Lambda})/m_t^{(0)}$ should not be smaller than
the chosen critical value in the third column.

As we see from Table 1, in the discussed PTC case the cutoff parameters
must be quite low, of order $1$ TeV or less, for $1/N_{\mbox
{\footnotesize c}}$ expansion to have some predictive power.
Furthermore, once we increase the cutoff parameter ratio 
${\Lambda}_{\mbox{\footnotesize b}}/{\Lambda}_{\mbox{\footnotesize f}}$
to the value $1/\sqrt{2} \approx 0.707$, or $1$, the value of the ratio 
${\varepsilon}_{\mbox{\scriptsize gap}}^2/{\varepsilon}_0^2$ 
in the PTC case cannot even be larger
than $0.545$, $0.314$, respectively, as displayed in Table 1.
This shows that the larger ratios of 
${\Lambda}_{\mbox{\footnotesize b}}/{\Lambda}_{\mbox{\footnotesize f}}$
lead to even larger NTL gap effects. Thus, for
${\Lambda}_{\mbox{\footnotesize b}}/{\Lambda}_{\mbox{\footnotesize f}}
> 1$, the ratio 
${\varepsilon}_{\mbox{\scriptsize gap}}^2/{\varepsilon}_0^2$
in the PTC case is always smaller than $0.314$, indicating that
${\Lambda}_{\mbox{\footnotesize b}} >
{\Lambda}_{\mbox{\footnotesize f}}$ is more or less
unacceptable in the
present framework, i.e., $1/N_{\mbox{\footnotesize c}}$ expansion
loses predictability in this case.

In addition, in columns 5-6 we included the values of
$m_t({\Lambda})/m_t^{(0)}$ and $m_t^{\mbox{\scriptsize ren.}}/
m_t^{(0)}$ in the corresponding cases for the PV, and
in columns 7-8 for the S regularization.
By the ``corresponding values'' we mean those values
which correspond to the same values of the four-fermion coupling
$G$ of Eq.~(\ref{TSM}) and the same values of the 
cutoff $\Lambda_{\mbox{\footnotesize b}}$, since this parameter
represents always the covariant spherical cutoff for the
bosonic momenta and is not influenced by the regularization
choice (PTC, PV, S) for the fermionic momenta. Technically, to
find these corresponding values, i.e., to find the corresponding
input parameters
$y(X)={\Lambda}_{\mbox{\footnotesize b}}/
{\Lambda}_{\mbox{\footnotesize f}}(X)$ and $a(X)$
($X$ denotes PV or S), we have to require in the
numerical program that the two numbers
$E_1 = a(X)* y^2(X) = {\Lambda}_{\mbox{\footnotesize b}}^2
G N_{\mbox{\footnotesize c}}/ (8 {\pi}^2) $ and 
$E_2 = a(X)*{\varepsilon}_{\mbox{\scriptsize ren.}}^2(X) =
(m_t^{\mbox{\scriptsize ren.}})^2 G N_{\mbox{\footnotesize c}}/ 
(8 {\pi}^2) $  be the same as in the PTC case.

It should be mentioned that certain entries in Table 1
should be regarded with additional reservations --
those for which the leading-$N_{\mbox{\footnotesize c}}$
values 
${\varepsilon}_0 = m_t^{(0)}/{\Lambda}_{\mbox{\footnotesize f}}$
are larger than $1$. These are the entries of the last line for
the PTC case, the last three lines for the PV case, 
and for the S case the lines corresponding to
$a({\mbox{\scriptsize PTC}}) = 5.091$ and $7.935$.
On the other hand, all the entries in Table 1 have
the values of 
${\varepsilon}_{\mbox{\scriptsize gap}}^{\mbox{\scriptsize (NTL)}} = 
m_t({\Lambda})/{\Lambda}_{\mbox{\footnotesize f}}$ and
${\varepsilon}_{\mbox{\scriptsize ren.}}^{\mbox{\scriptsize (NTL)}} = 
m_t^{\mbox{\scriptsize ren.}}/{\Lambda}_{\mbox{\footnotesize f}}$ 
smaller than $1$.

A few additional technical remarks: 
In the PV case, the series
for $({\tilde{\beta}}^{(1)}_1 \pm {\tilde{\beta}}^{(1)}_2)$ 
in inverse powers of ${\Lambda}_{\mbox{\footnotesize f}}$
[Eq.~(\ref{bePVn})] is, unfortunately, quite slowly convergent
for low values of ${\Lambda}_{\mbox{\footnotesize f}}$
($\leq 0.5 \mbox{ TeV}$), unlike the series for
$({\alpha}^{(2)} + {\tilde{\beta}}^{(2)})$ 
[Eq.~(\ref{bePVch})]. Therefore, we used in our calculations
in the PV case for
$({\tilde{\beta}}^{(1)}_1 \pm {\tilde{\beta}}^{(1)}_2)$  
the closed analytic expression (\ref{PVann}).
For $({\alpha}^{(2)} + {\tilde{\beta}}^{(2)})$ in
the PV case we used
the closed analytic expresssion (\ref{PVanch}), except
in Eq.~(\ref{kapGchr}). In this equation for 
$\kappa^{(Gch)}_{\mbox{\footnotesize 1r}}$, describing the
mass renormalization effects due to the charged Goldstone
degrees of freedom, a part of the integration is performed
over a region with negative ${\bar p}^2$, for which formula
(\ref{PVanch}) is not suitable. Therefore, we applied there
in the PV case
the series (\ref{bePVch}), including the terms in the sum
up to $n=8$.

In all regularization cases, we are 
led to the same qualitative conclusion: the cutoff 
${\Lambda}_{\mbox{\footnotesize b}}$ does not surpass 
${\cal{O}}(1\mbox{ TeV})$ as long as we demand that
the NTL gap effects not drastically ``wash out'' the
leading-$N_{\mbox{\footnotesize c}}$ effects. 
Comparing the ratios $(m_t(\Lambda)/m_t^{(0)})
= \sqrt{ {\varepsilon}_{\mbox{\scriptsize gap}}^2/{\varepsilon}_0^2 }$
for the corresponding cases of the PTC, PV and S regularization,
as well as the ratios $(m_t^{\mbox{\scriptsize ren.}}/m_t^{(0)})
= \sqrt{ {\varepsilon}_{\mbox{\scriptsize ren.}}^2/{\varepsilon}_0^2 }$,
we conclude the following:
\begin{itemize}
\item The S cases give somewhat smaller gap NTL changes  
$|\delta ( {\varepsilon}^2 )_{\mbox{\scriptsize gap}} /
{\varepsilon}_0^2|$ 
of the solution of the gap equation 
than the corresponding PTC cases; this implies that the
NTL-tolerable values of the cutoff parameter
${\Lambda}_{\mbox{\footnotesize b}}$ are 
somewhat higher in the S cases. On the other hand, the full
NTL changes $| [ \delta ( {\varepsilon}^2 )_{\mbox{\scriptsize gap}} 
+ \delta ( {\varepsilon}^2 )_{\mbox{\scriptsize ren.}}] /
{\varepsilon}_0^2 |$ are almost the same in both cases.

\item The PV cases give somewhat larger NTL changes 
$| \delta ( {\varepsilon}^2 )_{\mbox{\scriptsize gap}} /
{\varepsilon}_0^2| $ than the corresponding PTC cases;
this implies that the NTL-tolerable values of the
cutoff parameter 
${\Lambda}_{\mbox{\footnotesize b}}$ 
are somewhat lower in the PV cases.
Also the full NTL changes 
$1 - {\varepsilon}_{\mbox{\scriptsize ren.}}^2/{\varepsilon}_0^2
= | [ \delta ( {\varepsilon}^2 )_{\mbox{\scriptsize gap}}
+ \delta ( {\varepsilon}^2 )_{\mbox{\scriptsize ren.}}] /
{\varepsilon}_0^2 |$ are somewhat lower in the PV cases.
\end{itemize}

When inspecting more closely the separate contributions of the 
various degrees of freedom to the NTL gap shift 
$\delta ({\varepsilon}^2)_{\mbox{\scriptsize gap}} =
{\varepsilon}_{\mbox{\scriptsize gap}}^2 - {\varepsilon}_0^2$ 
of (\ref{NTLgap}) and to the mass renormalization NTL shift 
$\delta ({\varepsilon}^2)_{\mbox{\scriptsize  ren.}} =
{\varepsilon}_{\mbox{\scriptsize  ren.}}^2 -
{\varepsilon}_{\mbox{\scriptsize gap}}^2 $ of 
(\ref{kappar}), (\ref{kapHr})-(\ref{kapGchr}), (\ref{kapglr}), 
for the cases displayed in Table 1, we see the following~\footnote{
To estimate roughly the separate contributions to
$\delta ({\varepsilon}^2)_{\mbox{\scriptsize  gap}}$, we use
the approximate NTL gap equation (\ref{deps2gap}), where these
contributions are purely additive.}:
the Higgs and each one of the three Goldstone degrees of freedom 
contribute comparable negative values to 
$\delta ({\varepsilon}^2)_{\mbox{\scriptsize gap}}$, and
gluons a small positive value which is by an order of magnitude
smaller than the absolute values of the separate scalar
contributions; 
the Higgs and the charged Goldstone degrees of freedom
contribute each a negative value and the 
neutral Goldstone and gluons  weaker positive values
to $\delta ({\varepsilon}^2)_{\mbox{\scriptsize  ren.}}$, 
resulting thus in a negative
$\delta ({\varepsilon}^2)_{\mbox{\scriptsize  ren.}}$.
Therefore, both
$\delta ({\varepsilon}^2)_{\mbox{\scriptsize gap}}$ 
and $\delta ({\varepsilon}^2)_{\mbox{\scriptsize ren.}}$ 
are negative,
and $|\delta ({\varepsilon}^2)_{\mbox{\scriptsize  gap}}|$ is
usually larger than
$|\delta ({\varepsilon}^2)_{\mbox{\scriptsize ren.}}|$
by more than a factor of $2$.

In Ref.~\cite{cv} we calculated the NTL effects for the S 
case of regularization of fermionic momenta.
Comparing them with the results of Table 1 for the S case,
we find that the cutoffs ${\Lambda}_{\mbox{\footnotesize b}}$
in Table 1 are somewhat larger in the corresponding cases.
These differences arise because, unlike in~\cite{cv}, here we
inserted in the nonsingular (``regularized'')
expressions of the renormalization contributions
(\ref{kapSHr})-(\ref{kapSGchr}) and (\ref{kapglr}) the NTL 
corrected ``bare'' mass parameter 
${\varepsilon}_{\mbox{\scriptsize gap}}^2$, as mentioned earlier,
and not ${\varepsilon}_0^2$ (${\varepsilon}_0^2 >
{\varepsilon}_{\mbox{\scriptsize gap}}^2$), resulting thus
in numerically smaller numbers for 
$\delta ( {\varepsilon}^2 )_{\mbox{\scriptsize ren.}}
= {\varepsilon}_{\mbox{\scriptsize ren.}}^2 - 
{\varepsilon}_{\mbox{\scriptsize gap}}^2$.
In~\cite{cv}, on the other hand, we used singular integral
expressions for ${\kappa}_{1\mbox{\footnotesize r}}^{(X)}$
($X=H$, $Gn$, $Gch$), and we had no other choice there but to insert
for ${\varepsilon}^2$ the value ${\varepsilon}_0^2$
to avoid singularities in the integrals over bosonic momenta.
If we inserted ${\varepsilon}^2 = {\varepsilon}_0^2$ in the
nonsingular (``regularized'') 
expressions (\ref{kapSHr})-(\ref{kapSGchr}), these would
give us results for ${\kappa}_{1\mbox{\footnotesize r}}$
identical with those in~\cite{cv}, as it should be.
In addition, in~\cite{cv} we used the approximate
NTL gap equation (\ref{deps2gap}) to obtain
$\delta ({\varepsilon}^2)_{\mbox{\scriptsize  gap}}$, and
not the ``exact'' NTL gap equation (\ref{NTLgapS}) for
the S cutoff. This resulted in~\cite{cv} in somewhat larger
values for $|\delta ({\varepsilon}^2)_{\mbox{\scriptsize gap}}|$, 
and contributed thus additionally to smaller values of
${\Lambda}_{\mbox{\footnotesize b}}$ (and
${\Lambda}_{\mbox{\footnotesize f}}$).
Incidentally, in~\cite{cv} we regarded 
$\left({\delta}({\varepsilon}^2)_{\mbox{\scriptsize gap}} +
{\delta}({\varepsilon}^2)_{\mbox{\scriptsize ren.}}\right)/
{\varepsilon}_0^2 \ [ \mapsto (m_t^{\mbox{\scriptsize ren.}}/
m_t^{(0)} )^2 ]$ as a measure of the NTL changes. On the other
hand, in the present paper we chose to regard only the
NTL gap effects as the genuine NTL effects, i.e., we chose
as a measure of the NTL changes the ratio
${\delta}({\varepsilon}^2)_{\mbox{\scriptsize gap}}/ 
{\varepsilon}_0^2 \ [ \mapsto (m_t({\Lambda})/m_t^{(0)})^2 ]$.

\section{Conclusions and comparison to other works}

In this work, we calculated the next-to-leading (NTL) terms
in $1/N_{\mbox{\footnotesize c}}$ expansion
of the effective potential of ${\bar t} t$ condensate
and of the corresponding gap equation, in the effective
non-gauged $SU(2)_L \times U(1)_Y$ Nambu--Jona-Lasinio
type model (called also the top-mode standard model -- TSM)
of dynamical symmetry breaking (DSB). Furthermore, we calculated
also the (NTL) mass renormalization effects after the DSB.
We included all the 
degrees of freedom that are relevant at the NTL level of
this framework: the Higgs and the three Goldstone condensates,
and the quarks of the third generation (top and bottom).
In addition, we included also the dominant part of 
QCD contributions. The latter turned out to be numerically
less important in the present framework.
We considered the effective potential
as a function of a hard mass term ${\sigma}_0$ of the top
quark, i.e., of the expectation value ${\sigma}_0$ of the
composite scalar isodoublet field $\hat{\sigma} =
\sqrt{g} \hat{\Phi}$. 
We concentrated in particular
on the question of regularizing the integrals over the fermionic
Euclidean momenta in a way that is mutually consistent at the 
leading-$N_{\mbox{\footnotesize c}}$ and at the NTL level,
and is free of the momentum branching ambiguities.
These ends are achieved by employing the proper time regularization 
techniques, and we specifically employed the proper time cutoff (PTC)
and the Pauli-Villars (PV) two-subtractions regulator for the
fermionic momenta within the proper time framework.
Furthermore, we discussed in detail how to
ensure the validity of Goldstone theorem at the
NTL level -- in the proper time regularization framework
and in the case of the simple covariant spherical (S)
cutoff for the fermionic momenta. For integrals over the bosonic 
momenta, no branching ambiguity problem appeared, but the
proper time approach doesn't regularize them. We
always employed covariant spherical cutoff for them.
The dependence of our results on the various regularization
schemes (PTC, PV, S) for the fermionic momenta was investigated.
The basic conclusions of the previous paper~\cite{cv},
in which a (simplified) S regularization approach was 
applied, remain unchanged: as long as the cutoff energy
${\Lambda}$ ($\sim {\Lambda}_{\mbox{\footnotesize f}}
\sim {\Lambda}_{\mbox{\footnotesize b}}$), at which the
${\bar t}t$ condensation is assumed to take place, is larger
than ${\cal{O}}(1 \mbox{ TeV})$, then the negative
NTL gap corrections
$\delta ( {\varepsilon}^2 )_{\mbox{\scriptsize gap}}$
to the DSB (the gap equation solution)
have absolute values quite close to the values of the positive
leading-$N_{\mbox{\footnotesize c}}$ quark loop
contributions ${\varepsilon}_0^2$,
thus essentially ``washing out'' the latter ones and
making the model difficult or impossible to interpret.
On the other hand, for ${\Lambda} < {\cal{O}}(1 \mbox{ TeV})$,
explicit calculations here show that the PTC, PV and S regularizations,
when physically acceptable,
give similar numerical results, the PV regularization
having somewhat stronger and the S regularization somewhat weaker
NTL gap corrections than the PTC
in the corresponding cases.

As argued in the previous Section,
we identify $|\delta ({\varepsilon}^2)_{\mbox{\scriptsize gap}}|$
as the source of the $1/N_{\mbox{\footnotesize c}}$ expansion
breakdown, in contrast to the smaller
$|\delta ({\varepsilon}^2)_{\mbox{\scriptsize ren.}}|$.
The conclusion that the NTL gap
contributions can easily become strong [for ${\Lambda}
> {\cal{O}}(1 \mbox{ TeV})$ too strong for the applicability
of $1/N_{\mbox{\footnotesize c}}$ expansion] is not quite
implausible. The NTL contributions come primarily from
the coupling of the composite scalar sector itself to
its constituent top quarks (some kind of ``feedback
effect''). This coupling must be relatively strong because
otherwise the condensation cannot occur in the first place.

The conclusions of this paper seem to contrast with those
of Bardeen, Hill and Lindner (BHL)~\cite{bhl} and
Miransky, Tanabashi and Yamawaki (MTY)~\cite{mty};
however, they don't necessarily exclude them.
In the following, we briefly describe the approaches and results
of BHL and MTY, compare them with our approach and results,
and point out analogies, differences,
and those points that remain unclear and deserve further
investigation.

The authors of~\cite{bhl} employed the one-loop renormalization
group equation (RGE) of the minimal SM for the top quark
Yukawa coupling $g_t$ and demanded that it diverge
or become large at the energy of condensation $\Lambda$.
This is motivated by the compositeness condition which
says that the renormalization constants of the composite
scalar fields should vanish in a theory with the
cutoff as high as $E={\Lambda}$, i.e., that the composite particles
disappear (disintegrate) at that energy. This RGE approach
implicitly assumes that ${\Lambda}$ is large, i.e.,
that $\ln ({\Lambda}/E_{\mbox{\scriptsize ew}}) \gg 1$,
and that the details of the condensation mechanism get
decoupled from the minimal SM behavior at energies
which are, on logarithmic scale, quite
close to (but below) the energy ${\Lambda}$.
Their approach results in very large ${\Lambda}$'s;
the larger the ${\Lambda}$, the smaller the 
$m_t^{\mbox{\scriptsize ren.}}$. For $\Lambda
\sim \Lambda_{\mbox{\scriptsize Planck}}$
($\sim 10^{19}$ GeV), they obtain $m_t^{\mbox{\scriptsize ren.}}
\approx 220$ GeV, still substantially higher than
the measured $m_t^{\mbox{\scriptsize phys.}}
\approx 170-180$ GeV~\cite{CDFD0}. This approach concentrates 
on the $\delta (m_t)_{\mbox{\scriptsize ren.}}$ effects --
the one-loop RGE for $g_t$ contains the
leading-$N_{\mbox{\footnotesize c}}$ (including the QCD)
and at least
one part of the NTL scalar contributions to
$\delta (g_t)_{\mbox{\scriptsize ren.}}$ [$\mapsto
\delta ({\varepsilon}^2)_{\mbox{\scriptsize ren.}}$]~\footnote{
If only the leading-$N_c$ quark loop effects are included
in the RGE of the minimal SM, the running of the logarithms of
the Yukawa coupling
and the vacuum expectation value are exactly opposite to
each other, so that $m_t$ is non-running. This is in agreement
with the well known fact that the quark loop
leading-$N_{\mbox{\scriptsize c}}$ solution to the gap
equation in TSM does not get modified by renormalization
at that level of approximation.}.
The authors of~\cite{bhl} argue that, due to the
quasi infrared (IR) fixed-point behavior of the RGE,
their prediction of $m_t^{\mbox{\scriptsize ren.}} =
g_t^{\mbox{\scriptsize ren.}} v/ \sqrt{2}$
is quite stable against any details of the actual
condensation mechanism. More specifically,
their $m_t^{\mbox{\scriptsize ren.}}$ is quite
stable when $g_t^2({\Lambda})/(4 \pi)$ is changed
between the values $1$ and $\infty$, and/or
$\ln ({\Lambda}/E_{\mbox{\scriptsize ew}})$ is
changed by values of ${\cal{O}}(1)$. This is
reasonably true as long as ${\Lambda} > 10^{10}$ GeV,
limiting thus the applicability of their method to the
scenarios with thus large cutoffs. Including the two-loop effects 
in the then coupled RGEs for $g_t$ and for the composite scalar
self-coupling $\lambda$ doesn't change much the results
of this method~\cite{lavoura} -- the predicted 
$m_t^{\mbox{\scriptsize ren.}}$ is then increased by a few
GeV.

The approach by MTY~\cite{mty}, and subsequently by
King and Mannan~\cite{king}, is closer to the approach of the
present paper. Unlike BHL, they investigate the actual
condensation mechanism of TSM, by considering the
Dyson-Schwinger (DS) integral equation for the
top quark mass function ${\Sigma}_t ({\bar q}^2)$
and the Pagels--Stokar relation (PS)~\cite{PS}. The
DS equation was applied at the 
leading-$N_{\mbox{\footnotesize c}}$ order (including
QCD), and is basically the variational version of the
usual gap equation; the gap equation at the 
leading-$N_{\mbox{\footnotesize c}}$ order can be recovered
from this DS equation by replacing 
${\Sigma}_t ({\bar q}^2)$ by $m_t({\Lambda})$ and
the one-loop running
${\alpha}_s({\bar p}^2)$ by a constant ${\alpha}_s(E_0)$
($E_0 \geq m_t$). The mass function
${\Sigma}_t ({\bar q}^2)$ appears in the top quark
propagator and
is essentially the running mass
of the top quark, in the non-perturbative sense.

In our approach, on the other hand,
we took into account the effects of the
running of mass $m_t({\bar q}^2)$ between ${\bar q}^2 =
{\Lambda}_{\mbox{\footnotesize f}}^2$ and the electroweak 
scale $E^2_{\mbox{\footnotesize ew}}$ by calculating in an
``$1/N_{\mbox{\footnotesize c}}$-perturbative'' way
$\delta ({\varepsilon}^2)_{\mbox{\scriptsize ren.}}$
of QCD and the NTL scalar effects,
but only {\it after} solving the NTL gap equation.
However, unlike MTY, we didn't include in our
investigation the PS relation or an NTL-improved 
analogous relation. PS relation contains 
leading-$N_{\mbox{\footnotesize c}}$ effects of the
Bethe-Salpeter (BS) bound state equation for the
composite Goldstones~\footnote{
The problem of incorporating systematically NTL
effects in the BS equation has to our knowledge 
not been investigated yet.},
and connects the low energy
vacuum expectation value (VEV) $v \approx
246$ GeV (or equivalently: $M_W$) with ${\Sigma}_t({\bar q}^2)$
and ${\Lambda}$.
On the other hand,
the DS (or: gap) equation connects
the four-quark strength parameter $a = G N_{\mbox{\footnotesize c}}
{\Lambda}_{\mbox{\footnotesize f}}^2/(8 \pi^2) \sim 1$ with 
${\Sigma}^2_t({\bar q}^2)/{\Lambda}_{\mbox{\footnotesize f}}^2$
(or: $m_t^2({\Lambda})/{\Lambda}_{\mbox{\footnotesize f}}^2
= {\varepsilon}^2_{\mbox{\scriptsize gap}}$). Therefore,
the combination of DS and BS predicts, 
for a given value of the four-quark strength parameter $a$
(and of the ratio ${\Lambda}_{\mbox{\footnotesize b}}/
{\Lambda}_{\mbox{\footnotesize f}} \sim 1$), in principle both the value
of $m_t^{\mbox{\scriptsize ren.}}$ and of the onset scale
${\Lambda}$ ($\sim {\Lambda}_{\mbox{\footnotesize f}} \sim 
{\Lambda}_{\mbox{\footnotesize b}}$)
of the underlying
physics. Looking the other way around, since 
$m_t^{\mbox{\scriptsize phys.}}$ 
is experimentally known ($m_t^{\mbox{\scriptsize phys.}} 
 \approx 170-180$ GeV), such calculations predict
${\Lambda}$ and $a$ ($G$). 

MTY approach, as performed by~\cite{mty} and~\cite{king}
at the leading-$N_{\mbox{\footnotesize c}}$ level, was shown
to be numerically~\cite{mahanta} and analytically~\cite{yamawaki}
equivalent to the leading-$N_{\mbox{\footnotesize c}}$ version
of BHL approach. Furthermore, the results
of BHL and the (leading-$N_{\mbox{\footnotesize c}}$)
results of MTY are relatively close to each other: 
$(m_t^{\mbox{\scriptsize ren.}})_{\mbox{\scriptsize MTY}}
\stackrel{>}{\approx} 240$ GeV for ${\Lambda} \leq 
{\Lambda}_{\mbox{\scriptsize Planck}}$. This would
suggest that MTY approach could be continued to include
NTL corrections for very large ${\Lambda}$'s, 
if we trust BHL approach for such ${\Lambda}$'s. On the other
hand, the results of the present paper suggest that
$1/N_{\mbox{\footnotesize c}}$ expansion fails for
${\Lambda} > {\cal{O}} (1 \mbox{ TeV})$. Therefore, the
first question to be asked is: Would the modification of
our approach, by changing the NTL gap equation (\ref{NTLgap})
[with QCD effects from (\ref{Xigl}) included] to its
variational version of MTY-type, change the basic conclusions
of the paper? 
We have indications that it would not.
Namely, the ``running'' of the mass function
${\Sigma}_t({\bar q}^2)$ can be approximately described by
varying ${\varepsilon}^2$ between 
${\varepsilon}^2_{\mbox{\scriptsize gap}}$ ($=m_t^2({\Lambda})/
{\Lambda}_{\mbox{\footnotesize f}}^2$) and
${\varepsilon}^2_{\mbox{\scriptsize ren.}}$
($ \approx m_t^2(E_{\mbox{\scriptsize ew}})/   
{\Lambda}_{\mbox{\footnotesize f}}^2$). For the entries
in Table 1, however, these values differ roughly only
by $5$, $10$ and $20$ percent, for
${\Lambda}_{\mbox{\footnotesize f}}/
{\Lambda}_{\mbox{\footnotesize b}} = 1/2$, $1/\sqrt{2}$,
$1$, respectively. Also QCD contributions are
quite small in all these cases. 

However, our calculations do not exclude the possibility
of much higher values of the cutoff ${\Lambda}$, i.e.,
values of several orders of magnitude larger than
$1$ TeV. The reason may lie in our separation of the calculation
of the mass gap and of the mass renormalization effects. It is
possible that this approach may not work well for very 
large ${\Lambda}$'s [$\ln ({\Lambda}/1 \mbox{ TeV}) \gg 1$]
where the interplay between the two effects may become important.
In such a case, the variational (DS) version 
of the gap equation at the NTL level, which takes into 
account both the mass gap and the mass running
effects simultaneously and thus also their interplay,
would be the appropriate tool to apply. Then we could 
possibly see whether 
the scalar NTL corrections to ${\varepsilon}^2({\bar q}^2)
= {\Sigma}_t^2 ({\bar q}^2)/{\Lambda}_{\mbox{\footnotesize f}}^2$
for ${\Lambda}_{\mbox{\footnotesize f}} \gg {\cal{O}}(1 \mbox{ TeV}) $
can become tolerably small for $1/N_{\mbox{\footnotesize c}}$
expansion to make sense.
One possible signal that our paper cannot exclude the
entire region ${\Lambda} > {\cal{O}}(1 \mbox{ TeV})$
from the $1/N_{\mbox{\footnotesize c}}$ expansion approach
lies in the fact that the leading QCD contributions to
${\varepsilon}^2_{\mbox{\scriptsize ren.}}$, while for
${\Lambda} \leq {\cal{O}}( 1 \mbox{ TeV} )$ being substantially
smaller than the quark loop (bubble-chain) contributions,
start for much higher ${\Lambda}$'s to compete with the
bubble-chain contributions and substantially change the
results in the $\mbox{DS}+\mbox{PS}$ approach~\cite{yamawaki}. 
Therefore, we believe that performing the
calculations at the NTL level
with the variational (DS) approach would be
important at this stage, in order to investigate the region
of very high momenta. This could be done by employing
the formalism of Cornwall, Jackiw and Tomboulis (CJT)~\cite{cjt}
in which it is possible to calculate the effective potential
as a functional of the top quark propagator, and hence of the
mass function ${\Sigma}_t({\bar q}^2)$. In addition,
the inclusion of the
BS equation at the NTL level
would represent an additional
important step in investigating the whole realm of
the NTL effects of the strong composite scalar sector on the
condensation mechanism.

One may also ask how the results would change when including
the pure (i.e., transverse) components of the electroweak
gauge bosons in the calculations. The RGE approach of BHL
indicates that these contributions, at least as to the
mass renormalization, are of minor importance, due to
the relatively small $SU(2)_L \times U(1)_Y$ gauge couplings.
Furthermore, the condensation occurs primarily due
to the strong enough four-quark attractive coupling G of Eq.~(\ref{TSM}),
so that the NTL contributions of the strong composite scalar sector
to itself, i.e., the effects of the composite scalar
couplings to its own constituent top quarks (``feedback effects''),
are expected to be stronger and more important than those
of the weak $SU(2)_L \times U(1)_Y$ couplings.

As a point independent of the discussion above,
we stress that the simple TSM framework (\ref{TSM}),
in which the 6-dimensional four-quark contact term 
triggers the top quark condensation, might not be sufficient for
a fully realistic picture when the energy ${\Lambda}$
at which the condensation takes place is as low as
${\cal{O}}(1 \mbox{ TeV})$. This is because higher dimensional
femionic contact terms~\cite{hasenfratz}-\cite{suzuki}, which come
from the details of the underlying physics at $E >
{\Lambda}$, might contribute 
to the physical quantities relative corrections as high as 
$1/\ln({\Lambda}^2/m_t^2) \approx 0.3$.
However, these corrections can be, 
for specific models of the underlying physics,
substantially smaller~\cite{hill}.

Other authors have investigated NTL effects 
in NJL-type frameworks without gauge bosons
~\cite{nikolov},~\cite{dmitras}-~\cite{lurie}.
The authors of~\cite{nikolov} calculated  the NTL effects
with an effective action formalism, and those of~\cite{dmitras}
with diagrammatic methods. Both groups regarded their
discussed $SU(2)$ symmetric NJL type model as a framework
of the low energy QCD and took particular care that 
Goldstone theorem remains valid. 

The authors of~\cite{hands}
calculated NTL contributions to critical exponents of the
fields at the fixed point, i.e., at the location of the
nontrivial zero of $\beta$ function, for various dimensions
$d \leq 4$. The implications of~\cite{hands} for
physical predictions of four-dimensional NJL models with
finite cutoff are not clear from these works and 
would deserve investigation.
On the other hand, Akama~\cite{akama} investigated the NTL
contributions by considering the compositeness condition
which says that the renormalization constants of composite
scalar fields and their self-interaction parameters should
be zero. Also, Luri\'e and Tupper~\cite{lurie} had earlier
investigated the compositeness condition, taking into account
at least some of the NTL effects. Both Akama and Luri\'e
and Tupper conclude that the compositeness condition implies
that the NTL contributions to physical quantities
for $N_{\mbox{\footnotesize c}}=3$ are larger than the 
leading-$N_{\mbox{\footnotesize c}}$ contributions, indicating
that $1/N_{\mbox{\footnotesize c}}$ expansion diverges.
We stress that these three authors treated TSM as a
renormalizable Yukawa type model without gauge bosons plus
the compositeness condition, an approach similar (not identical)
to the approach of BHL~\cite{bhl}. Hence, implicitly they 
assumed that the cutoffs ${\Lambda}$ are large, i.e., that
$\ln {\Lambda}$ terms entirely dominate over the 
${\Lambda}$-independent terms. Therefore, their results
apparently don't contradict the results of the present
paper and~\cite{cv} -- i.e., that TSM without electroweak gauge
bosons can make sense at the NTL level only if the cutoffs
are quite low: ${\Lambda} \sim {\Lambda}_{\mbox{\footnotesize f}}
\sim {\Lambda}_{\mbox{\footnotesize b}} = {\cal{O}}(1 \mbox{ TeV})$.

\section{Acknowledgment}
This work was supported in part by the Deutsche Forschungsgemeinschaft
and in part by European Union Science Project No.~SC1-CT91-0729.

\begin{appendix}
\section[]{Diagonalization of 
${\hat{\cal{D}}}^{\dagger}{\hat{\cal{D}}} \left( \lbrace \sigma_j \rbrace
\right)$ }
\setcounter{equation}{0}
The non-negative definite hermitean operator
${\hat{\cal{D}}}^{\dagger}{\hat{\cal{D}}} \left( \lbrace 
{\tilde{\sigma}}_j \rbrace
\right)$, in the Euclidean metric,
is obtained directly from (\ref{Ddef})-(\ref{notat1}):
\begin{equation}
{\hat{\cal{D}}}^{\dagger}{\hat{\cal{D}}} \left( \lbrace \sigma_j \rbrace
\right) = {\hat{\bar{P}}} \cdot {\hat{\bar{P}}} + 
+ {\hat{\cal{M}}}^{\dagger}{\hat{\cal{M}}}
+ i {\bar{\gamma}}_{\mu} \frac{\partial}{\partial{\bar{x}}_{\mu}}
{\hat{\cal{M}}} \ ,
\label{DD}
\end{equation}
where we denoted the Euclidean quantities:
${\bar x}_0 = i x^0$, ${\bar x}_j = x^j$, ${\bar{\gamma}}_0=
i{\gamma}^0$, ${\bar{\gamma}}_j = {\gamma}^j$, and ${\hat{\bar{P}}}_{\mu}
= i {\partial}/{\partial{\bar{x}}}_{\mu}$.
The $8 \times 8$ matrix ${\hat{\cal{M}}} \left( \lbrace
{\tilde{\sigma}}_j \rbrace \right)$ is the one written
in Eq.~(\ref{Ddef}). Since ${\tilde{\sigma}}_j({\bar x})
= {\sigma}_j + s_j({\bar x})$ [cf.~(\ref{fluct})], we obtain for the
part without the fluctuations $s_j({\bar x})$
\[
{\hat{\cal{D}}}^{\dagger} {\hat{\cal{D}}} \left( \lbrace
\sigma_j \rbrace \right) \equiv {\hat{\triangle}}_0 \left( \lbrace
\sigma_j \rbrace \right) =
{\hat{\bar{P}}} \cdot {\hat{\bar{P}}} +
{\hat{\cal{M}}}_0^{\dagger} {\hat{\cal{M}}}_0 \ ,
\]
where
\begin{equation}
{\hat{\cal{M}}}_0^{\dagger} {\hat{\cal{M}}}_0 =
\frac{1}{2} \left[ 
\begin{array}{ll}
\left[ 2{\sigma}_0^2+2{\sigma}_1^2+ 
\left( {\sigma}_2^2+{\sigma}_3^2 \right)
\left( 1+{\gamma}_5 \right) \right] &
\left[ - \left( {\sigma}_0 - i {\sigma}_1 \right) 
\left( {\sigma}_2 + i {\sigma}_3 \right)  
\left( 1- {\gamma}_5 \right)
\right] \\
 \left[ - \left( {\sigma}_0 + i {\sigma}_1 \right) 
\left( {\sigma}_2 - i {\sigma}_3 \right) 
\left( 1- {\gamma}_5  \right) \right] &
\left[ \left( {\sigma}_2^2 + {\sigma}_3^2 \right)
\left( 1- {\gamma}_5  \right) \right]
\end{array}
\right] \ .
\label{MM0}
\end{equation}
In order to calculate the various parts of the effective potential,
in particular the leading-$N_{\mbox{\footnotesize c}}$ part 
$N_{\mbox{\footnotesize c}} V^{(0)}$ of
(\ref{veff0}), it is very convenient to work in the
rotated basis in which the $8 \times 8$ matrix (\ref{MM0}) becomes
diagonal. There exist many unitary $8 \times 8$ matrices which
accomplish such rotations. We will use the following
one which has the convenient property that it goes to the
identity matrix when the expectation values of the charged Goldstones 
go to zero ($\sigma_2$ and $\sigma_3 \to 0$):
\begin{equation}
U^{\dagger} {\hat{\cal{M}}}_0^{\dagger}{\hat{\cal{M}}}_0 U
= \sum_{j=0}^3 \sigma_j^2
\left[
\begin{array}{cc}
1 & 0 \\
0 & 0  
\end{array}
\right] \ ,
\label{MM0diag}
\end{equation}
where the above $8 \times 8$ diagonal matrix is a block matrix made up of
four blocks ($1$, $0$) of dimension $4 \times 4$, and $U$ is the 
following unitary matrix:
\begin{equation}
U =
\left[ 
\begin{array}{ll}
\left( k_0^{(+)} + {\gamma}_5 k_0^{(-)} \right) &
  g_0^{(+)} \left( 1- {\gamma}_5 \right) \\
 -g_0^{(-)} \left( 1- {\gamma}_5 \right) &
\left( k_0^{(+)} + {\gamma}_5 k_0^{(-)} \right) 
\end{array}
\right] \ .
\label{U}
\end{equation}
Here we used the following notations:
\begin{eqnarray} 
k_0^{(\pm)} & = & \frac{1}{2} \left[ 1 \pm 
\sqrt{ \frac{ {\sigma}_0^2+ {\sigma}_1^2 }{
\sum_{j=0}^3 {\sigma}_j^2 } } \right] \ ,
\nonumber\\
g_0^{(\pm)} & = &  \frac{ \left( {\sigma}_2 \pm i {\sigma}_3 \right) }{
2 \sqrt{ \sum_{j=0}^3 {\sigma}_j^2 } } e^{ \mp i {\delta}_0 } \ ,
\quad \mbox{where: } {\delta}_0 = 
\arg \left( {\sigma}_0 + i {\sigma}_1 \right)
\ .
\label{notatU}
\end{eqnarray}
From (\ref{MM0diag}) we see that the trace $\mbox{Tr} 
{\hat{\cal{D}}}^{\dagger}{\hat{\cal{D}}} \left( \lbrace \sigma_j \rbrace
\right)$, which is relevant for calculation of various parts of the
effective potential, is a function of the $SU(2)_L \times U(1)_Y$ 
invariant combination ${\sigma}^2 = \sum_0^3 {\sigma}_j^2 = 
g {\Phi}_0^{\dagger} {\Phi}_0$ of the
scalar field expectation values. In Sec.~II [cf.~Eq.~(\ref{Utransf})],
the use of (\ref{MM0diag}) was crucial to derive the general formula
(\ref{veff0res}) for the leading-$N_{\mbox{\footnotesize c}}$ part
$N_{\mbox{\footnotesize c}} V^{(0)}$ of the effective potential.

\section[]{Explicit calculation of $V^{(0)}$ and $V^{(1)}$}
\setcounter{equation}{0}

The general expression for the leading-$N_{\mbox{\footnotesize c}}$
part $N_{\mbox{\footnotesize c}} V^{(0)}$ in terms of ${\sigma}^2
= g {\Phi}_0^{\dagger} {\Phi}_0$ is given in (\ref{veff0res})
and can be rewritten in terms of ${\sigma}^2$ and the
four-fermion coupling $G$
\begin{equation}
N_{\mbox{\footnotesize c}} V^{(0)} ( {\sigma}^2; G ) =
\frac{ {\sigma}^2 }{G} + 
\frac{ N_{\mbox{\footnotesize c}} }{ 8 {\pi}^2 }
\int_0^{\infty} \frac{ d {\tau} }{ {\tau}^3 }
{\rho}_{\mbox{\scriptsize f}} ( {\tau} )
\left( e^{- {\tau} {\sigma}^2 } - 1 \right) \ .
\label{v0}
\end{equation}
Consequently, when using
the dimensionless notation (\ref{defka1})-(\ref{Xi0}),
we obtain for the leading-$N_{\mbox{\footnotesize c}}$ part
of the effective potential in the PTC case (\ref{PTC}) the
following expression:
\begin{eqnarray}
\lefteqn{
{\Xi}^{(0)} ( {\varepsilon}^2; a )^{\mbox{\scriptsize (PTC)}} 
 = \frac{ 8 {\pi}^2 }{ {\Lambda}_{\mbox{\footnotesize f}}^4
N_{\mbox{\footnotesize c}} }
 N_{\mbox{\footnotesize c}} V^{(0)} 
( {\sigma}^2 = {\varepsilon}^2
{\Lambda}_{\mbox{\footnotesize f}}^2; 
G )^{\mbox{\scriptsize (PTC)}} =
\frac{ {\varepsilon}^2 }{a}+ \int_1^{\infty} \frac{dz}{z^3} 
\left( e^{- z {\varepsilon}^2 } - 1 \right)
} \nonumber\\
& = & \frac{ {\varepsilon}^2 }{a} + \frac{ {\varepsilon}^4 }{2} 
\left[ e^{ -{\varepsilon}^2 } {\varepsilon}^{-2} 
\left( {\varepsilon}^{-2} - 1 \right) - 
\mbox{li} ( e^{ -{\varepsilon}^2 } ) - {\varepsilon}^{-4} \right]
 =  \frac{1}{2} {\Bigg \{ } {\varepsilon}^2 2 \left( \frac{1}{a} - 1 
\right) 
\nonumber\\
& & - {\varepsilon}^4 \ln ( {\varepsilon}^2 )
+ {\varepsilon}^4 \left( \frac{3}{2} - {\cal{C}} \right) +
{\varepsilon}^6 \left( \frac{2}{3! 1} \right) 
+ \cdots
+ \left( -1 \right)^{n+1} {\varepsilon}^{2n} 
\left( \frac{2}{n! (n-2)} \right) + \cdots {\Bigg \} } \ .
\label{Xi0PTC}
\end{eqnarray}
In the above formula, $\mbox{li}$ is the conventional Logarithm-integral
function, for which we used the conventional expansion in powers
of ${\varepsilon}^2$ (i.e., in inverse powers of 
${\Lambda}_{\mbox{\footnotesize f}}^2$); ${\cal{C}}$ appearing in this
expansion is Euler's constant (${\cal{C}}=0.577215\ldots$).
The PV cutoff case (\ref{PV}) gives analogously
\begin{eqnarray}
\lefteqn{
{\Xi}^{(0)} ( {\varepsilon}^2; a )^{\mbox{\scriptsize (PV)}} 
 = \frac{ 8 {\pi}^2 }{ {\Lambda}_{\mbox{\footnotesize f}}^4
N_{\mbox{\footnotesize c}} }
 N_{\mbox{\footnotesize c}} V^{(0)} 
( {\sigma}^2 = {\varepsilon}^2 
{\Lambda}_{\mbox{\footnotesize f}}^2; 
G )^{\mbox{\scriptsize (PV)}} 
 =  \frac{ {\varepsilon}^2 }{a}+ \int_0^{\infty} \frac{dz}{z^3}
\left( 1 - e^{-z} \right)^2
\left( e^{ - z {\varepsilon}^2 } - 1 \right)
} \nonumber\\
& = & \frac{ {\varepsilon}^2 }{a} + 
 \lim_{ w \to {\infty} }
\int_{1/w}^{\infty} \frac{dz}{z^3} \left( 1 - 2 e^{-z} + e^{ - 2 z}
\right)
\left( e^{ -z {\varepsilon}^2 } - 1 \right) = \cdots
\nonumber\\
& = & \left[ 
\frac{ {\varepsilon}^2 }{a} 
- \frac{ {\varepsilon}^4 }{2} \ln ( {\varepsilon}^2 )
+ ( 1 + {\varepsilon}^2 )^2 
\ln ( 1 + {\varepsilon}^2 ) 
 - \frac{1}{2} ( 2 + {\varepsilon}^2 )^2
\ln ( 2 + {\varepsilon}^2 ) + 2 \ln 2 
\right] \ .
\label{Xi0PV}
\end{eqnarray}
From here we get the first derivative 
in the two regularization cases:
\begin{eqnarray}
\lefteqn{
\frac{ {\partial} {\Xi}^{(0)} ( {\varepsilon}^2 ; a ) }
{ {\partial} {\varepsilon}^2 }^{\mbox{\scriptsize (PTC)}} 
= \frac{ 8 {\pi}^2 }{ {\Lambda}_{\mbox{\footnotesize f}}^2 }
\frac{ {\partial} }{ {\partial} {\sigma}^2 }
V^{(0)} ( {\sigma}^2  
{\Lambda}_{\mbox{\footnotesize f}}^2 {\varepsilon}^2; G )^{
\mbox{\scriptsize (PTC) }}  
}
\nonumber\\   
& & = {\Bigg \{ }
\frac{1}{a} - 1 - {\varepsilon}^2 \ln ( {\varepsilon}^2 )
+ {\varepsilon}^2 ( 1 -  {\cal{C}} ) + 
\frac{ {\varepsilon}^4 }{2}
- \frac{ {\varepsilon}^6 }{12} + 
\cdots + \left( -1 \right)^n
\frac{ {\varepsilon}^{2n} }{ n! \left( n - 1 \right) } + \cdots 
{\Bigg \} } \ ,
\label{dXi0PTC}
\end{eqnarray}
\begin{equation}
\frac{ {\partial} {\Xi}^{(0)} ( {\varepsilon}^2 ; a ) }
{ {\partial} {\varepsilon}^2 }^{\mbox{\scriptsize (PV)}} =
{\Bigg \{ }
\frac{1}{a} - {\varepsilon}^2 \ln ( {\varepsilon}^2 )
+ 2 ( 1 + {\varepsilon}^2 ) \ln ( 1 + {\varepsilon}^2 ) 
 - ( 2 + {\varepsilon}^2 ) \ln ( 2 + {\varepsilon}^2 )
{\Bigg \} } \ .
\label{dXi0PV}
\end{equation}
These expressions are zero at the corresponding 
leading-$N_{\mbox{\footnotesize c}}$ gap equation solutions
${\varepsilon}^2 = {\varepsilon}_0^2$.

Now we turn to the calculation of the NTL part of the effective
potential. The operators ${\hat{\triangle}}_1$ and
${\hat{\triangle}}_2$ of (\ref{expanDD}), which are linear and 
quadratic in the scalar fields
fluctuations $\lbrace s_j \left( {\bar x} \right) \rbrace$, respectively,
are obtained from (\ref{DD}) by simply using in the
${\hat{\cal{M}}} = {\hat{\cal{M}}}_0 + \delta{\hat{\cal{M}}}$ 
matrix of (\ref{Ddef}) the full
fluctuating values ${\tilde{\sigma}}_j \left( {\bar x} \right) = \sigma_j 
+ s_j \left( {\bar x} \right)$, and extracting from there the terms
linear and quadratic in $\lbrace s_j \left( {\bar x} \right) \rbrace$
\begin{eqnarray}
\lefteqn{
{\hat{\triangle}}_1 \left( \lbrace \sigma_j ; s_j \left(
{\bar x} \right) \rbrace \right) =
{\hat{\cal{M}}}_0^{\dagger} \delta{\hat{\cal{M}}} +
\delta{\hat{\cal{M}}}^{\dagger} {\hat{\cal{M}}}_0 +
i {\bar{\gamma}}_{\mu} 
\left( \frac{ {\partial} }{ {\partial}{\bar x}_{\mu} }
\delta{\hat{\cal{M}}} \right)
} \nonumber\\
& = & 
\left[ 
\begin{array}{ll}
\left[ 2 {\sigma}_0 s_0 + 2{\sigma}_1 s_1 + 
\left( {\sigma}_2 s_2 + {\sigma}_3 s_3 \right)
\left( 1+ {\gamma}_5 \right) \right] &
\left[ \left( - {\sigma}_n^{(-)} s_c^{(+)}  
              - {\sigma}_c^{(+)} s_n^{(-)} \right)
\left( 1- {\gamma}_5 \right) \right]
\\
\left[  \left( - {\sigma}_n^{(+)} s_c^{(-)}
               - {\sigma}_c^{(-)} s_n^{(+)} \right)
\left( 1- {\gamma}_5  \right) \right] &
\left[ \left( {\sigma}_2 s_2 + {\sigma}_3 s_3 \right)
\left( 1-  {\gamma}_5 \right) \right]
\end{array}
\right] 
\nonumber\\
&& + i {\bar{\gamma}}_{\mu} \left[
\begin{array}{ll}
\left[ \frac{ {\partial} }{ {\partial}{\bar x}_{\mu} } \left(
s_0 - i {\gamma}_5 s_1 \right) \right] &
\left[ - \frac{1}{ {\sqrt{2}} }
 \frac{ {\partial} }{ {\partial}{\bar x}_{\mu} } 
s_c^{(+)}
\left( 1 - {\gamma}_5 \right) \right] \\
\left[ - \frac{1}{ {\sqrt{2}} }
 \frac{ {\partial} }{ {\partial}{\bar x}_{\mu} } 
s_c^{(-)}
\left( 1 + {\gamma}_5  \right) \right] & 0
\end{array}
\right] \ ,
\label{triang1}
\end{eqnarray}
\begin{eqnarray}
\lefteqn{
{\hat{\triangle}}_2 \left( \lbrace {\sigma}_j ; s_j \left(
{\bar x} \right) \rbrace \right) =
\delta {\hat{\cal{M}}}^{\dagger} \delta{\hat{\cal{M}}} 
} \nonumber\\
& = & 
\frac{1}{2} \left[ 
\begin{array}{ll}
\left[ 2 s_0^2+2 s_1^2+ \left( s_2^2+ s_3^2 \right)
\left(  1+ {\gamma}_5 \right) \right] &
\left[ - 2 s_n^{(-)} s_c^{(+)}
 \left(  1- {\gamma}_5  \right)
\right] \\
 \left[ - 2 s_n^{(+)} s_c^{(-)}
 \left( 1- {\gamma}_5  \right) 
\right] &
\left[ \left( s_2^2 + s_3^2 \right)
\left( 1- {\gamma}_5  \right) \right]
\end{array}
\right] \ ,
\label{triang2}
\end{eqnarray}
where we used shorthand notations
\begin{eqnarray}
{\sigma}_n^{(\pm)} & = & \frac{1}{ \sqrt{2} }
\left( {\sigma}_0 \pm i {\sigma}_1 \right) \ , \qquad
s_n^{(\pm)} = \frac{1}{ \sqrt{2} }
\left( s_0 \pm i s_1 \right) \ , 
\nonumber\\
{\sigma}_c^{(\pm)} & = & \frac{1}{ \sqrt{2} }
\left( {\sigma}_2 \pm i {\sigma}_3 \right) \ , \qquad
s_c^{(\pm)} = \frac{1}{ \sqrt{2} }
\left( s_2 \pm i s_3 \right) \ .
\label{notattri}
\end{eqnarray}
We use the expressions (\ref{triang1})-(\ref{triang2})
to calculate the proper time integrals of the bosonic
effective action
in the curly brackets of the exponent in the path integral of 
(\ref{veff1}). The first question appearing at this point is:
in which basis is it most convenient to calculate the traces
over the 8-dimensional spinor/isospin degrees of freedom?
It seems that the most convenient basis is the one given by
(\ref{MM0diag})-(\ref{U}), in which the matrix 
${\hat{\triangle}}_0$, and hence the matrix 
$\exp ( - \tau_j {\hat{\triangle}}_0 )$, 
are diagonal and explicitly known
through (\ref{MM0diag}). The matrices ${\hat{\triangle}}_1$
and ${\hat{\triangle}}_2$, on the other hand, are in this basis
not diagonal, but can be calculated explicitly by help of
(\ref{U}) and (\ref{triang1})-(\ref{triang2}). The tracing
over the configuration space is most easily carried out
by using for the $U$-rotated operators
$U^{\dagger} {\hat{\triangle}}_0  U $ the momentum basis,
and for the $U$-rotated operators $U^{\dagger} 
{\hat{\triangle}}_j U$ ($j=1,2$) the coordinate basis
\[
{\langle} {\bar q}^{\prime} {\Big | } U^{\dagger} 
e^{ - {\tau}_j {\hat{\triangle}}_0 }
U {\Big | } {\bar q} {\rangle} =
\delta \left( {\bar q} - {\bar q}^{\prime} \right)
\exp \left( - {\tau}_j {\bar q}^2 \right) \left[
\begin{array}{cc}
\exp \left( - {\tau}_j \sum_{k=0}^3 {\sigma}_k^2 \right) & 0 \\
0 & 1 
\end{array}
\right] \ ,
\]
\[
{\langle} {\bar x}^{\prime} {\Big | } U^{\dagger} {\hat{\triangle}}_j U
{\Big | } {\bar x} {\rangle} =
\delta \left( {\bar x} - {\bar x}^{\prime} \right)
U^{\dagger} {\hat{\triangle}}_j \left( {\bar x} \right) U \ .
\]
In the above matrix, 
$1$, $0$ and $\mbox{exp}( \cdots )$ represent
the corresponding $4 \times 4$ blocks. First we perform
the trivial integrations over those
variables which appear in the $\delta$ functions.
Subsequently, integrations by parts
in the expression $\mbox{Tr} [ e^{ - {\tau}_1 {\hat{\triangle}}_0 }
{\hat{\triangle}}_1 
e^{ - {\tau}_2 {\hat{\triangle}}_0 } {\hat{\triangle}}_1 ]$
result in the replacement of the partial derivatives in
(\ref{triang1}) by the corresponding (differences of) momenta.
Integrations in the expression
$\mbox{Tr} [ e^{ - {\tau} {\hat{\triangle}}_0 } {\hat{\triangle}}_2 ]$
are simpler. 
The tracing over the color degrees of freedom is trivial and gives factor
$N_{\mbox{\footnotesize c}}$, due to the $N_{\mbox{\footnotesize c}}
\times N_{\mbox{\footnotesize c}}$ identity matrix structure of
these operators in the color subspace.
After longer, but straightforward algebraic manipulations along the
lines outlined here, we obtain explicit integrals for the traces and
end up with the following expression for the effective action
$\Gamma$ in the curly brackets of the exponent of (\ref{veff1})
\begin{equation}
{\Gamma}( {\sigma}^2; \lbrace v_j \rbrace; \lbrace I_j \rbrace ) =
- \frac{1}{2} \int \int d^4 {\bar x} d^4 {\bar y}
\sum_{j=0}^3 v_j \left( {\bar y} \right) 
{\hat{\cal{A}}}_j \left( {\bar y}, {\bar x}; \sigma^2 \right)
v_j \left( {\bar x} \right)
- i \sum_{j=0}^3 I_j \int d^4 {\bar x} v_j \left( {\bar x} \right) \ ,
\label{act2}
\end{equation}
where $\sigma^2 =
{\sum}_0^3 {\sigma}_j^2 = g {\Phi}_0^{\dagger} {\Phi}_0$,
and the kernels ${\hat{\cal{A}}}_j$ have the structure
(\ref{veff1argx}), with the terms there being the following
explicit integrals over the proper time variables:
\begin{eqnarray}
\alpha^{(1)} ( {\sigma}^2 ; G ) & = & \frac{2}{G}
- \frac{ N_{\mbox{\footnotesize c}} }{ 4 \pi^2 }
\int_0^{\infty} \frac{ d {\tau} }{ {\tau^2} } 
\rho_{\mbox{\scriptsize f}}
( {\tau} ) e^{- {\tau} {\sigma}^2 } \ ,
\nonumber\\
\alpha^{(2)} ( {\sigma}^2 ) & = &
- \frac{ N_{\mbox{\footnotesize c}} }{ 8 \pi^2 }
\int_0^{\infty} \frac{ d {\tau} }{ {\tau}^2 } 
\rho_{\mbox{\scriptsize f}}
\left( {\tau} \right) \left( 1 - e^{- {\tau} {\sigma}^2 } \right) \ ,
\label{alphas}
\end{eqnarray}
\begin{eqnarray}
\left( {\beta}_1^{(1)} \pm {\beta}_2^{(1)} \right) 
( {\bar x} ; {\sigma}^2 )
& = &  \frac{ N_{\mbox{\footnotesize c}} }{ 8 \left( 2 \pi \right)^4 }
\int_0^{\infty} \int_0^{\infty} 
\frac{ d {\tau}_1 d {\tau}_2 }{ {\tau}_1^2 {\tau}_2^2 } 
\rho_{\mbox{\scriptsize f}}
\left( {\tau}_1 + {\tau}_2 \right) 
\exp \left[ - \left( {\tau}_1 + {\tau}_2 \right) {\sigma}^2 \right]
\times \nonumber\\
& & \exp \left[ - \frac{ \left( {\tau}_1 + {\tau}_2 \right) }
{ 4 {\tau}_1 {\tau}_2 } {\bar x}^2 \right]
\left[ \frac{ \left( {\tau}_1 + {\tau}_2 \right) }{ 2 {\tau}_1 {\tau}_2 }
\left( 4 - \frac{ \left( {\tau}_1 + {\tau}_2 \right) }
{ 2 {\tau}_1 {\tau}_2 }
{\bar x}^2 \right) + 2 {\sigma}^2 \left( 1 \pm 1 \right) \right] \ ,
\nonumber\\
{\beta}^{(2)} ( {\bar x}; {\sigma}^2 ) & = &
  \frac{ N_{\mbox{\footnotesize c}} }{ 16 \left( 2 \pi \right)^4 }
\int_0^{\infty} \int_0^{\infty} 
\frac{ d {\tau}_1 d {\tau}_2 }{ {\tau}_1^2 {\tau}_2^2 } 
\rho_{\mbox{\scriptsize f}}
\left( {\tau}_1 + {\tau}_2 \right) 
\exp \left[ - {\tau}_1 {\sigma}^2 \right]
\times \nonumber\\
& & \exp \left[ - \frac{ \left( {\tau}_1 + {\tau}_2 \right) }
{ 4 {\tau}_1 {\tau}_2 } {\bar x}^2 \right]
\left[ \frac{ \left( {\tau}_1 + {\tau}_2 \right) }
{ 1 {\tau}_1 {\tau}_2 }
\left( 4 - \frac{ \left( {\tau}_1 + {\tau}_2 \right) }
{ 2 {\tau}_1 {\tau}_2 }
{\bar x}^2 \right) + 2 {\sigma}^2 \right] \ .
\label{betasx}
\end{eqnarray}
The scalar field fluctuations $\lbrace v_j ( {\bar x} ) \rbrace$ are specific
orthonormal combinations of the original scalar field fluctuations
$\lbrace s_j( {\bar x} ) \rbrace$
\begin{equation}
v_j \left( {\bar x} \right) = {\cal{O}}_{jk} s_k \left( {\bar x} \right) \ ,
\label{vs}
\end{equation}
where ${\cal{O}}$ is the following orthonormal $4 \times 4$ matrix:
\begin{equation}
{\cal{O}} = 
\frac{1}{ \sqrt{ \sum_{k=0}^3 {\sigma}_k^2 } }
\left[
\begin{array}{llll}
{\sigma}_0 & {\sigma}_1 & {\sigma}_2 & {\sigma}_3 \\
-{\sigma}_1 & {\sigma}_0 & -{\sigma}_3 & {\sigma}_2 \\
{\varsigma}^{-1} {\sigma}_0 & {\varsigma}^{-1} {\sigma}_1 & 
- {\varsigma} {\sigma}_2 & - {\varsigma} {\sigma}_3 \\
- {\varsigma}^{-1} {\sigma}_1 & {\varsigma}^{-1} {\sigma}_0 &
{\varsigma} {\sigma}_3 & -{\varsigma} {\sigma}_2
\end{array}
\right] \ ,
\label{O}
\end{equation}
and we denoted here:
\[ 
{\varsigma} = 
\sqrt{ \frac{ {\sigma}_0^2 + {\sigma}_1^2 }{ {\sigma}_2^2+{\sigma}_3^2 } }
\ .
\]
In the special case of our interest, only the neutral scalar (Higgs)
component acquires nonzero value, i.e., ${\sigma}_1 =
{\sigma}_2 = {\sigma}_3 = 0$. In this limiting case (we have the
freedom to additionally
require: ${\sigma}_3/{\sigma}_2 \to 0$), the fluctuations
$\lbrace v_j ( {\bar x} ) \rbrace$ reduce to the old 
fluctuations of the non-rotated fields: $v_0 =s_0$,
$v_1=s_1$, $v_2=-s_2$, $v_3=-s_3$, and the scalar expectation
value $\sigma$ reduces to ${\sigma}_0$.

Comparing (\ref{v0}) and (\ref{alphas}), 
we see immediately that $\alpha^{(1)} ( {\sigma}^2 ; G )$
is exactly twice the derivative with respect to ${\sigma}^2$
of the leading-$N_{\mbox{\footnotesize c}}$
part of the effective potential $N_{\mbox{\footnotesize c}} V^{(0)}
\left( {\sigma}^2 ; G \right) = {\Lambda}_{\mbox{\footnotesize f}}^4
N_{\mbox{\footnotesize c}} {\Xi}^{(0)}/(8 {\pi}^2)$, i.e., relation
(\ref{alpha1}). In the PTC case (\ref{PTC}), the relation
(\ref{alphas}) gives
\begin{eqnarray}
{\alpha}^{(1)} ( {\sigma}^2; G )^{\mbox{\scriptsize (PTC)}}
& = &
\frac{2}{G} - \frac{ N_{\mbox{\footnotesize c}} }{ 4 {\pi}^2 }
{\Bigg \{ }
{\Lambda}_{\mbox{\footnotesize f}}^2
- {\sigma}^2 \ln \left( \frac{ {\Lambda}_{\mbox{\footnotesize f}}^2 }
{ {\sigma}^2 } \right) - (1 - {\cal{C}}) {\sigma}^2
- \frac{1}{2} \frac{ {\sigma}^4 }{
{\Lambda}_{\mbox{\footnotesize f}}^2 }
\nonumber\\
& & + \frac{1}{12} \frac{ {\sigma}^6 }{
{\Lambda}_{\mbox{\footnotesize f}}^4 } + \cdots
+ \left( -1 \right)^{n+1} \frac{1}{ n! \left( n - 1 \right) }
\frac{ {\sigma}^{2n} }{
{\Lambda}_{\mbox{\footnotesize f}}^{2n-2} } + \cdots
{\Bigg \} } \ ,
\label{al1PTC}
\end{eqnarray}
which is compatible with (\ref{dXi0PTC}).
The Fourier transformed quantities ${\tilde{\beta}}^{(i)}_j
\left( {\bar p}^2; {\sigma}^2 \right)$, appearing in the
NTL part (\ref{veff1mom}),
can be calculated from (\ref{betasx}) by first
carrying out explicitly the integration over $d^4 {\bar x}$
\begin{eqnarray} 
\left( {\tilde{\beta}}^{(1)}_1 \pm {\tilde{\beta}}^{(1)}_2 \right)
( {\bar p}^2; {\sigma}^2 ) & = &
\frac{ N_{\mbox{\footnotesize c}} }{ 8 \pi^2 }
\left[ {\bar p}^2 + 2 \left( 1 \pm 1 \right) {\sigma}^2 \right]
\int_0^{\infty} \int_0^{\infty} 
\frac{ d {\tau}_1 d {\tau}_2 }{ \left( {\tau}_1 + {\tau}_2 \right)^2 }
\rho_{\mbox{\scriptsize f}} \left( {\tau}_1 + {\tau}_2 \right)
\exp \left[ - \left( {\tau}_1 + {\tau}_2 \right) {\sigma}^2 \right]
\nonumber\\
& & \times \exp \left[ - \frac{  {\tau}_1 {\tau}_2 {\bar p}^2 }
{ \left( {\tau}_1 + {\tau}_2 \right) } \right] \ ,
\label{FTB1}
\end{eqnarray}
\begin{equation}
\alpha^{(2)} ( {\sigma}^2 ) +
{\tilde{\beta}}^{(2)} ( {\bar p}^2; {\sigma}^2 ) = 
\frac{ N_{\mbox{\footnotesize c}} }{ 4 \pi^2 }
{\bar p}^2 
\int_0^{\infty} \int_0^{\infty} 
\frac{ d {\tau}_1 d {\tau}_2 }{ \left( {\tau}_1 + {\tau}_2 \right)^3}
{\tau}_1 \rho_{\mbox{\scriptsize f}} \left( {\tau}_1 + {\tau}_2 \right)
 \exp \left( - {\tau}_1 {\sigma}^2 \right)
\exp \left[ - \frac{ {\tau}_1 {\tau}_2 {\bar p}^2 }{
\left( {\tau}_1 + {\tau}_2 \right) } 
 \right] \ .
\label{FTB2}
\end{equation}
In the last formula, after the Fourier transformation we performed 
the substitution $\tau= \tau_1 + \tau_2$
and carried out integration by parts over 
$d {\tau}_1$, and then reintroduced ${\tau}_2$.

We now introduce the new variables $z$ and $\tau$:
${\tau}_1 = {\tau} z$, ${\tau}_2 = {\tau} (1 - z)$, 
where $\tau$ and $z$ run
through the intervals $[0,{\infty}]$ and $[0,1]$, respectively. The
above quantities (\ref{FTB1}) and (\ref{FTB2}) can then be written
in the form
\begin{equation}
\left( {\tilde{\beta}}^{(1)}_1 \pm {\tilde{\beta}}^{(1)}_2 \right)
( {\bar p}^2; {\sigma}^2 ) = 
\frac{ N_{\mbox{\footnotesize c}} }{ 8 \pi^2 }
\left[ {\bar p}^2 + 2 \left( 1 \pm 1 \right) {\sigma}^2 \right]
\int_0^1 dz \int_0^{\infty} \frac{ d {\tau} }{ {\tau} }
\rho_{\mbox{\scriptsize f}} \left( {\tau} \right)
\exp {\Big \{ } 
- {\tau} \left[ {\bar p}^2 z \left( 1 - z \right) + {\sigma}^2 \right] 
{\Big \} }  \ ,
\label{FT2B1}
\end{equation}
\begin{equation}
{\alpha}^{(2)} ( {\sigma}^2 ) +
{\tilde{\beta}}^{(2)} ( {\bar p}^2; {\sigma}^2 ) = 
\frac{ N_{\mbox{\footnotesize c}} }{ 4 \pi^2 }
{\bar p}^2 
\int_0^1 dz z \int_0^{\infty} \frac{ d {\tau} }{ {\tau} }
\rho_{\mbox{\scriptsize f}} \left( {\tau} \right)
\exp {\Big \{ } - {\tau} \left[ {\bar p}^2 z \left( 1 - z \right)
+ {\sigma}^2 z \right] {\Big \} }  \ .
\label{FT2B2}
\end{equation}
In the PTC case, the integration over $d {\tau}$ gives us the 
Logarithm-integral
\begin{eqnarray}
\int_{1/{\Lambda}_{\mbox{\scriptsize f}}^2}^{\infty}
\frac{ d {\tau} }{ {\tau} } 
\exp \left[ - {\tau} {\cal{F}} \left( z ; {\bar p}^2,
{\sigma}^2 \right) \right] & =&
- \mbox{li} \left( e^{ - {\cal{F}}/{\Lambda}_{\mbox{\scriptsize f}}^2 }
\right) 
\nonumber\\
&=& - \left[ {\cal{C}} + \ln x + \sum_{k=1}^{\infty}
\left( -1 \right)^k \frac{x^k}{k! k} \right] {\Bigg |}_{
x={\cal{F}}\left(z; {\bar p}^2, {\sigma}^2 \right)/
{\Lambda}_{\mbox{\scriptsize f}}^2 } \ ,
\label{li}
\end{eqnarray}
where ${\cal{C}}$ is Euler's constant (${\cal{C}}= 0.577215\ldots$).
The integration over $dz$ can then be carried out term by term,
and we end up with the following series in inverse powers of the
cutoff $\Lambda_{\mbox{\footnotesize f}}$ for the PTC case:
\begin{eqnarray}
\lefteqn{
\frac{ 8 \pi^2 }{ N_{\mbox{\footnotesize c}} }
\frac{
\left( {\tilde \beta}_1^{(1)} \pm {\tilde \beta}_2^{(1)} \right)
\left( {\bar p}^2; {\sigma}^2 \right)^{\mbox{\scriptsize (PTC)}} }
{ \left[ {\bar p}^2 + 2 {\sigma^2} \left( 1 \pm 1 \right) \right] }
= {\Bigg \{ }
\ln \left( \frac{ {\Lambda}_{\mbox{\footnotesize f}}^2 }
{ {\sigma}^2 } \right)
+ \left[ 
 - {\cal{C}}- \frac{2}{3} z \mbox{F} \left( z \right) 
{\Bigg |}_{ z={\bar p}^2/\left( {\bar p}^2 + 4 {\sigma}^2 \right) }
\right]
 }  \nonumber\\
 & &
 + \left[
  \frac{ {\sigma}^2 }{ {\Lambda}_{\mbox{\footnotesize f}}^2 }
 + \frac{1}{6} \frac{ {\bar p}^2 }{ 
{\Lambda}_{\mbox{\footnotesize f}}^2 }
   \right]
 - \frac{1}{4} \left[
 \left( \frac{ {\sigma}^2 }
{ {\Lambda}_{\mbox{\footnotesize f}}^2 } \right)^2
 + \frac{1}{3}
 \frac{ {\sigma}^2 {\bar p}^2 }{ {\Lambda}_{\mbox{\footnotesize f}}^4 }
 + \frac{1}{30}
 \left( \frac{ {\bar p}^2 }{
 {\Lambda}_{\mbox{\footnotesize f}}^2 } \right)^2
  \right] 
+ \frac{1}{18} {\Bigg [ } \left( 
\frac{ {\sigma}^2 }{ {\Lambda}_{\mbox{\footnotesize f}}^2 } 
\right)^3
 + \frac{1}{2}
 \frac{  {\sigma}^4 {\bar p}^2 }{ {\Lambda}_{\mbox{\footnotesize f}}^6 }
\nonumber\\ 
& &+ \frac{1}{10}
 \frac{  {\sigma}^2 \left( {\bar p}^2 \right)^2 }{ 
{\Lambda}_{\mbox{\footnotesize f}}^6 }
 + \frac{1}{140}
 \left( 
\frac{ {\bar p}^2 }{ {\Lambda}_{\mbox{\footnotesize f}}^2 } 
\right)^3
{\Bigg ] }
+ \left( -2 \right) \sum_{k=4}^{\infty}
\frac{ \left( -1 \right)^k }{k! k}
\left( 
\frac{ {\bar p}^2 }{ {\Lambda}_{\mbox{\footnotesize f}}^2 } 
\right)^k
\int_0^{1/2} du \left( \frac{ {\sigma}^2 }{ {\bar p}^2 } +
\frac{1}{4} - u^2 \right)^k 
{\Bigg \} } \ ,
\label{be1PTC}
\end{eqnarray}
\begin{eqnarray}
\lefteqn{
\frac{ 8 \pi^2 }{ N_{\mbox{\footnotesize c}} {\bar p}^2 }
\left[
{\alpha}^{(2)} \left( {\sigma}^2 \right) +
 {\tilde{\beta}}^{(2)}
\left( {\bar p}^2; {\sigma}^2 \right)
\right]^{\mbox{\footnotesize (PTC)}}
=
{\Bigg \{ }
\ln \left( \frac{ {\Lambda}_{\mbox{\footnotesize f}}^2 }
{ {\sigma}^2 } \right)
+ \left[ - {\cal{C}} + 2 +
 \frac{ {\sigma}^2 }{ {\bar p}^2 }
- \left( 1 + \frac{ {\sigma}^2 }{ {\bar p}^2 } \right)^2
\ln \left( 1 + \frac{ {\bar p}^2 }{ {\sigma}^2 } \right) 
\right]
} \nonumber\\
& &
 + \left[ 
\frac{2}{3} \frac{ {\sigma}^2 }{ {\Lambda}_{\mbox{\footnotesize f}}^2 }
+ \frac{1}{6} \frac{ {\bar p}^2 }{ {\Lambda}_{\mbox{\footnotesize f}}^2 }
\right]
- \frac{1}{4} \left[
 \frac{1}{2} 
\left( \frac{ {\sigma}^2 }
{ {\Lambda}_{\mbox{\footnotesize f}}^2 } \right)^2
 + \frac{1}{5}
 \frac{ {\sigma}^2 {\bar p}^2 }{ {\Lambda}_{\mbox{\footnotesize f}}^4 }
 + \frac{1}{30}
 \left( \frac{ {\bar p}^2 }
{ {\Lambda}_{\mbox{\footnotesize f}}^2 } \right)^2
 \right]
+ \frac{1}{45} {\Bigg [ }
\left( \frac{ {\sigma}^2 }{ {\Lambda}_{\mbox{\footnotesize f}}^2 } 
\right)^3
+ \frac{1}{2} \frac{ {\sigma}^4 {\bar p}^2 }{ 
{\Lambda}_{\mbox{\footnotesize f}}^6  } 
\nonumber\\
& &
+ \frac{1}{7} \frac{ {\sigma}^2 \left( {\bar p}^2 \right)^2 }{
{\Lambda}_{\mbox{\footnotesize f}}^6  } +
\frac{1}{56} \left( \frac{ {\bar p}^2 }{ 
{\Lambda}_{\mbox{\footnotesize f}}^2 } \right)^3
{\Bigg ] }
- 2 \sum_{k=4}^{\infty} \frac{1}{k! k}
\left( \frac{ {\bar p}^2 }{ {\Lambda}_{\mbox{\footnotesize f}}^2 }
\right)^k \int_0^1 du u^{k+1} 
\left( u - 1 - \frac{ {\sigma}^2 }{ {\bar p}^2 } \right)^k
{\Bigg \} } \ .
\label{be2PTC}
\end{eqnarray}
The explicit expression for the function $\mbox{F}(z)$ appearing
in the ${\Lambda}_{\mbox{\footnotesize f}}$-independent part
of (\ref{be1PTC}) is 
\begin{equation}
\mbox{F}(z) = \frac{3}{2} z^{-3/2} \ln
\left( \frac{ 1 + \sqrt{z} }{ 1- \sqrt{z} } \right) - \frac{3}{z}
= 1 + \frac{3}{5} z + \frac{3}{7} z^2 + \frac{3}{9} z^3 +
\cdots \ .
\label{F}
\end{equation}
For the PV cutoff case (\ref{PV}), we obtain the corresponding
results, as given in (\ref{bePVn}) and (\ref{bePVch}),
by very similar procedure: first we carry out integration over
$d {\tau}$ in (\ref{FT2B1})-(\ref{FT2B2})
according to (\ref{li}), where we set $1/{\Lambda}^2$
for the lower bound of integration; then we take the limit
${\Lambda}^2 \to \infty$; the resulting integrands 
(sums of logarithms) are finite, and integration over $dz$
can be performed analytically; expansion of the results in
inverse powers of ${\Lambda}_{\mbox{\footnotesize f}}^2$
gives then the expressions given in
(\ref{bePVn}) and (\ref{bePVch}).
It is also possible to follow this procedure without
expanding in inverse powers of ${\Lambda}_{\mbox{\footnotesize f}}^2$.
We then obtain, for ${\bar p}^2 > 0$, the following solutions
for $({\tilde{\beta}}^{(1)}_1 \pm {\tilde{\beta}}^{(1)}_2)$
and $({\alpha}^{(2)} + {\tilde{\beta}}^{(2)})$ in the PV case:
\begin{equation}
\frac{ 8 \pi^2 }{ N_{\mbox{\footnotesize c}} }
\frac{
\left( {\tilde \beta}_1^{(1)} \pm {\tilde \beta}_2^{(1)} \right)
\left( {\bar p}^2; {\sigma}^2 \right)^{\mbox{\scriptsize (PV)}} }
{ \left[ {\bar p}^2 + 2 {\sigma^2} \left( 1 \pm 1 \right) \right] }
 = 
- \sum_{j=0}^2 a_j \left[ (1 - 2 {\delta}_j )
\ln \left( \frac{ {\sigma}^2 + {\lambda}_j^2  }{ {\bar p}^2 } \right)
+ 4 {\delta}_j \ln \left( {\delta}_j + \frac{1}{2} \right) \right] 
\ ,
\label{PVann}
\end{equation}
\begin{eqnarray}
\lefteqn{
\frac{ 8 \pi^2 }{ N_{\mbox{\footnotesize c}} {\bar p}^2 }
\left[
{\alpha}^{(2)} \left( {\sigma}^2 \right) +
 {\tilde{\beta}}^{(2)}
\left( {\bar p}^2; {\sigma}^2 \right)
\right]^{\mbox{\footnotesize (PV)}}
=
{\Bigg \{ }  
- \left( 1 + \frac{ {\sigma}^2 }{ {\bar p}^2 } \right)^2 
\ln \left( 1 + \frac{ {\bar p}^2 }{ {\sigma}^2 } \right)
}
\nonumber\\
& & 
- \sum_{j=0}^2 a_j \ln \left( \frac{ {\sigma}^2 + {\lambda}_j^2 }
{ {\bar p}^2 } \right) +
\sum_{j=1}^2 a_j \left[ \frac{1}{2} 
\left( 1 + \frac{ {\sigma}^2 }{ {\bar p}^2 } \right)^2
+ \frac{ {\lambda}_j^2 }{ {\bar p}^2 } +
\left( 1 + \frac{ {\sigma}^2 }{ {\bar p}^2 } \right)
{\eta}_j \right] \ln \left( 1 + \frac{ {\sigma}^2 }{ {\lambda}_j^2 }
\right)
\nonumber\\
&&
- 2 \left( 1 + \frac{ {\sigma}^2 }{ {\bar p}^2 } \right)
\sum_{j=1}^2 a_j {\eta}_j \left[
\ln \left( {\eta}_j - \frac{ {\sigma}^2 }{ 2 {\bar p}^2 }
+ \frac{1}{2} \right) -
\ln \left( {\eta}_j - \frac{ {\sigma}^2 }{ 2 {\bar p}^2 }
- \frac{1}{2} \right) \right]
{\Bigg \} } \ ,
\label{PVanch}
\end{eqnarray}
where we use the notation
\[
a_0  = a_1 = 1 \ , \ a_2=-2 \ ; \qquad
{\lambda}_0^2 = 0 \ , \
{\lambda}_1^2 = 2 {\Lambda}_{\mbox{\footnotesize f}}^2 \ , \
{\lambda}_2^2 = {\Lambda}_{\mbox{\footnotesize f}}^2 \ ;
\]
\begin{equation}
{\delta}_j = \sqrt{ \frac{ {\sigma}^2 + {\lambda}_j^2 }{ {\bar p}^2 }
+ \frac{1}{4} } \ ; \qquad
{\eta}_j  = \sqrt{ \frac{1}{4} \left( 1 + \frac{ {\sigma}^2 }{{\bar p}^2}
\right)^2 + \frac{ {\lambda}_j^2 }{ {\bar p}^2 } } \ .
\label{PVannot}
\end{equation}

\end{appendix}

\newpage

\vspace{4cm}

\oddsidemargin-2.8cm 
\evensidemargin-2.8cm

\begin{table}[h]
\vspace{0.3cm}
\par
\begin{center}
Table 1 \\\vspace{0.5cm} 
\begin{tabular}{|c|l|l||c||c|c||c|c||c|c|}
\hline
$a$ &
${\Lambda}_{\mbox{\footnotesize b}}/
 {\Lambda}_{\mbox{\footnotesize f}}$ & 
$\frac{ m_t({\Lambda}) }{ m_t^{(0)} } $ & 
$\frac{ m_t^{\mbox{\scriptsize ren}} }{ m_t^{(0)} } $ &
$\frac{ m_t({\Lambda}) }{ m_t^{(0)} }$ & 
$\frac{ m_t^{\mbox{\scriptsize ren}} }{ m_t^{(0)} } $ &
$\frac{ m_t({\Lambda}) }{ m_t^{(0)} } $ & 
$\frac{ m_t^{\mbox{\scriptsize ren}} }{ m_t^{(0)} } $ &
${\Lambda}_{\mbox{\footnotesize b}}$ & 
${\Lambda}_{\mbox{\footnotesize f}}^{ \mbox{\scriptsize (PTC)} }$ \\
 $\mbox{\scriptsize (PTC)}$ &    $\mbox{\scriptsize (PTC)}$ &  
  $\mbox{\scriptsize (PTC)}$ &    $\mbox{\scriptsize (PTC)}$ &  
  $\mbox{\scriptsize (PV)}$ &    $\mbox{\scriptsize (PV)}$ &  
  $\mbox{\scriptsize (S)}$ &    $\mbox{\scriptsize (S)}$ &  
  $\mbox{\scriptsize [TeV]}$ & $\mbox{\scriptsize [TeV]}$ \\ 
\hline\hline
1.273 & 1/2 & 0.500 ($=\sqrt{1/4}$) & 0.475 & 0.481 & 0.445 & 0.517 & 0.488 &
0.724 & 1.448 \\
1.314 & 1/2 & 0.577 ($=\sqrt{1/3}$) & 0.549 & 0.555 & 0.513 & 0.594 & 0.559 & 
0.581 & 1.161 \\
1.458 & 1/2 & 0.707 ($=\sqrt{1/2}$) & 0.669 & 0.678 & 0.623 & 0.720 & 0.673 & 
0.387 & 0.773 \\
1.982 & 1/2 & 0.816 ($=\sqrt{2/3}$) & 0.760 & 0.775 & 0.699 & 0.824 & 0.757 & 
0.231 & 0.462 \\
\hline
1.665 & 1/$\sqrt{2}$  & 0.500 ($=\sqrt{1/4}$) & 0.453 & 0.473 & 0.407 &
0.531 & 0.470 & 0.663 & 0.938 \\
1.815 & 1/$\sqrt{2}$          & 0.577 ($=\sqrt{1/3}$) & 0.523 & 0.542 & 0.465 &
0.606 & 0.533 & 0.520 & 0.735 \\
2.736 & 1/$\sqrt{2}$          & 0.707 ($=\sqrt{1/2}$) & 0.630 & 0.641 & 0.532 &
0.724 & 0.618 & 0.306 & 0.433 \\
5.091 & 1/$\sqrt{2}$        & 0.738 ($=\sqrt{0.545}$) & 0.643 & 0.627 & 0.489 &
0.739 & 0.603 & 0.219 & 0.310 \\
\hline
3.613 & 1             & 0.500 ($=\sqrt{1/4}$) & 0.410 & 0.437 & 0.310 &
0.563 & 0.395 & 0.568 & 0.568 \\
7.935 & 1           & 0.560 ($=\sqrt{0.314}$) & 0.443 &  0.421 & 0.264 &
0.582 & 0.348 & 0.385 & 0.385 \\
\hline
\end{tabular}
\end{center}
\end{table}

\vspace{1cm}

\oddsidemargin-0.5cm 
\evensidemargin-0.5cm
{\footnotesize
\noindent {\bf Table 1}: 
The bosonic cutoffs ${\Lambda}_{\mbox{\scriptsize b}}$ and
the quark (fermion) cutoff parameters 
${\Lambda}_{\mbox{\scriptsize f}}$, 
which result when we impose 
the requirement that the ratio 
${\varepsilon}^2/{\varepsilon}_0^2= (m_t({\Lambda})/m_t^{(0)})^2$
of the solution of the NTL gap equation
with the solution of the leading-$N_{\mbox{\scriptsize c}}$ quark loop
gap equation not be smaller than: $1/4$, $1/3$, $1/2$, $2/3$,
for the proper time cutoff (PTC) case. The corresponding
ratios of the NTL-renormalized mass with the 
leading-$N_{\mbox{\scriptsize c}}$ mass are also given.
In addition, these mass ratios are given also for the
corresponding cases [i.e., with the same 
${\Lambda}_{\mbox{\scriptsize b}}$ and the same four-fermion
coupling $G$ of (\ref{TSM})] when the Pauli-Villars (PV)
regulator and the simple covariant spherical (S) cutoff are
applied to the fermionic momenta. We took 
$m_t^{\mbox{\scriptsize ren.}} = 180 \mbox{ GeV}$.
}

\vspace{1cm}

\newpage

\noindent

\hspace*{0.cm}\epsfig{file=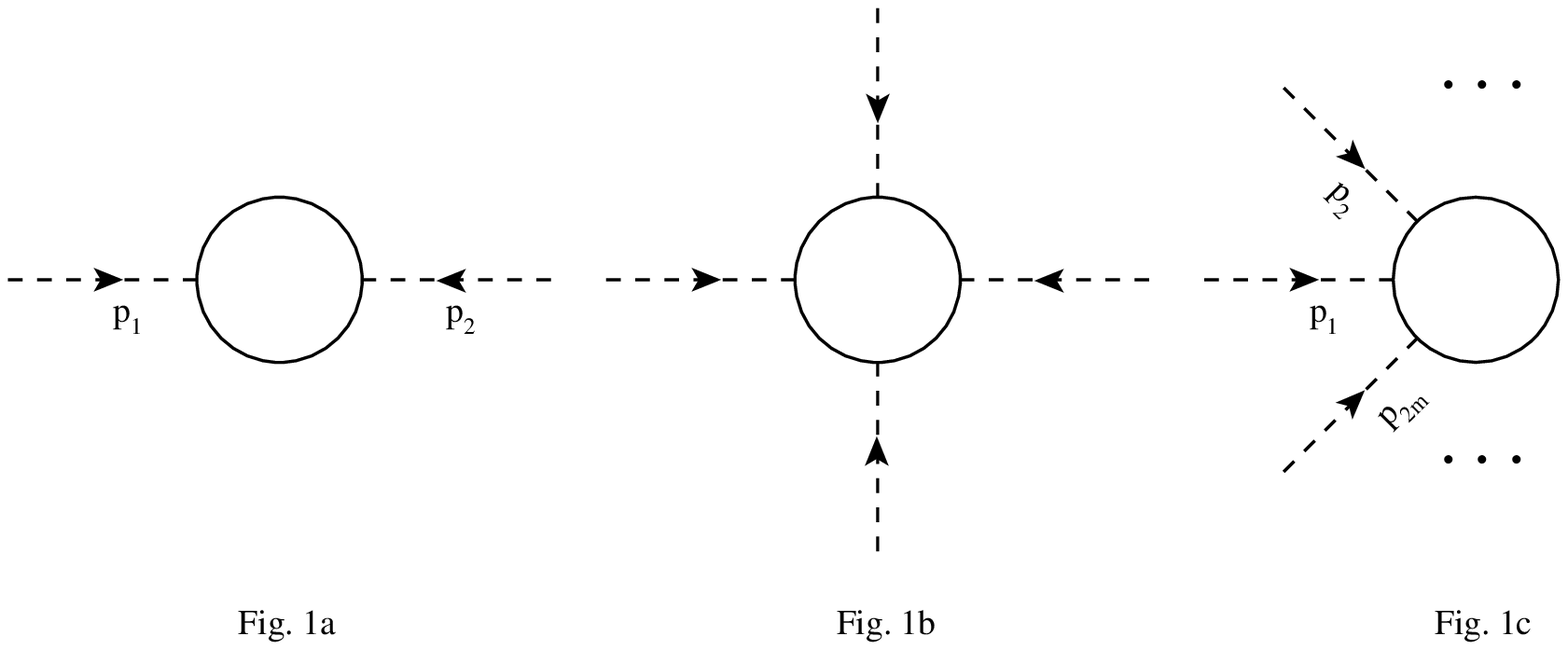,width=15.5cm}

\vspace{-6.5cm}

{\footnotesize
\noindent {\bf Figs.~1(a)-(c)}: The 1-loop 1-PI diagrams 
contributing to 1-PI Green functions 
$\tilde \Gamma_{H} ^{(2m; 1)}(p_1, \ldots, p_{2m})$,
which in turn yield the leading-$N_{\mbox{\scriptsize c}}$ part 
$N_{\mbox{\scriptsize c}} V^{(0)}$ in 
$1/N_{\mbox{\scriptsize c}}$ expansion of 
$V_{\mbox{\scriptsize eff}}$. Full lines represent massless top 
quarks, and dotted external lines the scalar nondynamical Higgs 
of the Lagrangian (\ref{efflagr}).
}

\newpage

\noindent

\hspace*{0.cm}\epsfig{file=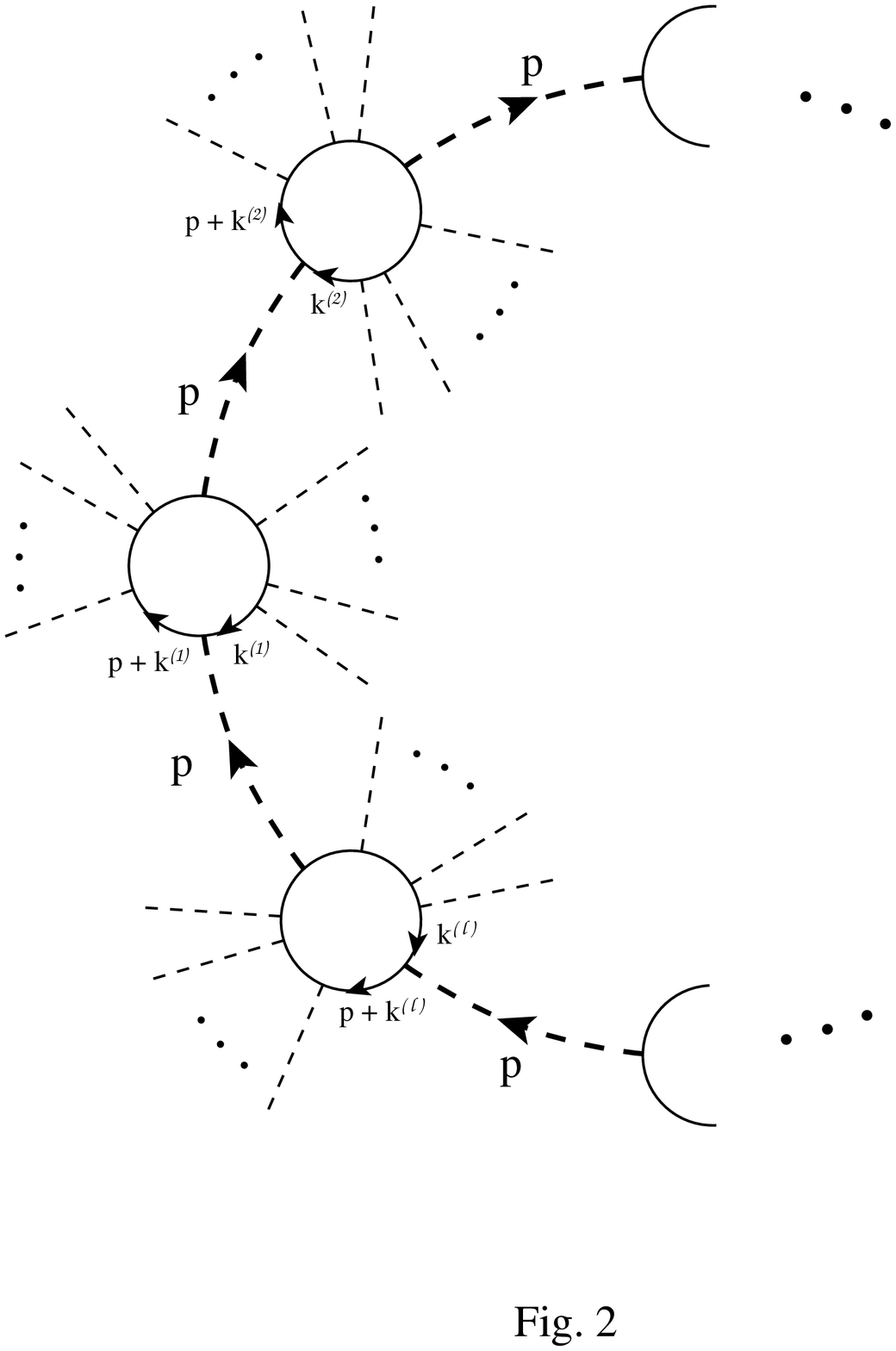,width=13.5cm}

\vspace{-0.5cm}

{\footnotesize
\noindent {\bf Fig.~2}: The $(\ell + 1)$-loop 1-PI diagrams which 
contribute to the 1-PI Green functions which in turn yield the 
NTL part $V^{(1)}$ (beyond one loop) in 
$1/N_{\mbox{\scriptsize c}}$ expansion of 
$V_{\mbox{\scriptsize eff}}$. The diagrams contain $\ell$ loops 
of (massless) quarks. These loops are connected into another circle 
by $\ell$ propagators of the (nondynamical) scalars (all either 
Higgs, or neutral Goldstone, or charged Goldstones). In the case of 
charged Goldstone propagators, the quark loops are made up of the 
top and the bottom quark.
}

\newpage

\noindent

\hspace*{0.cm}\epsfig{file=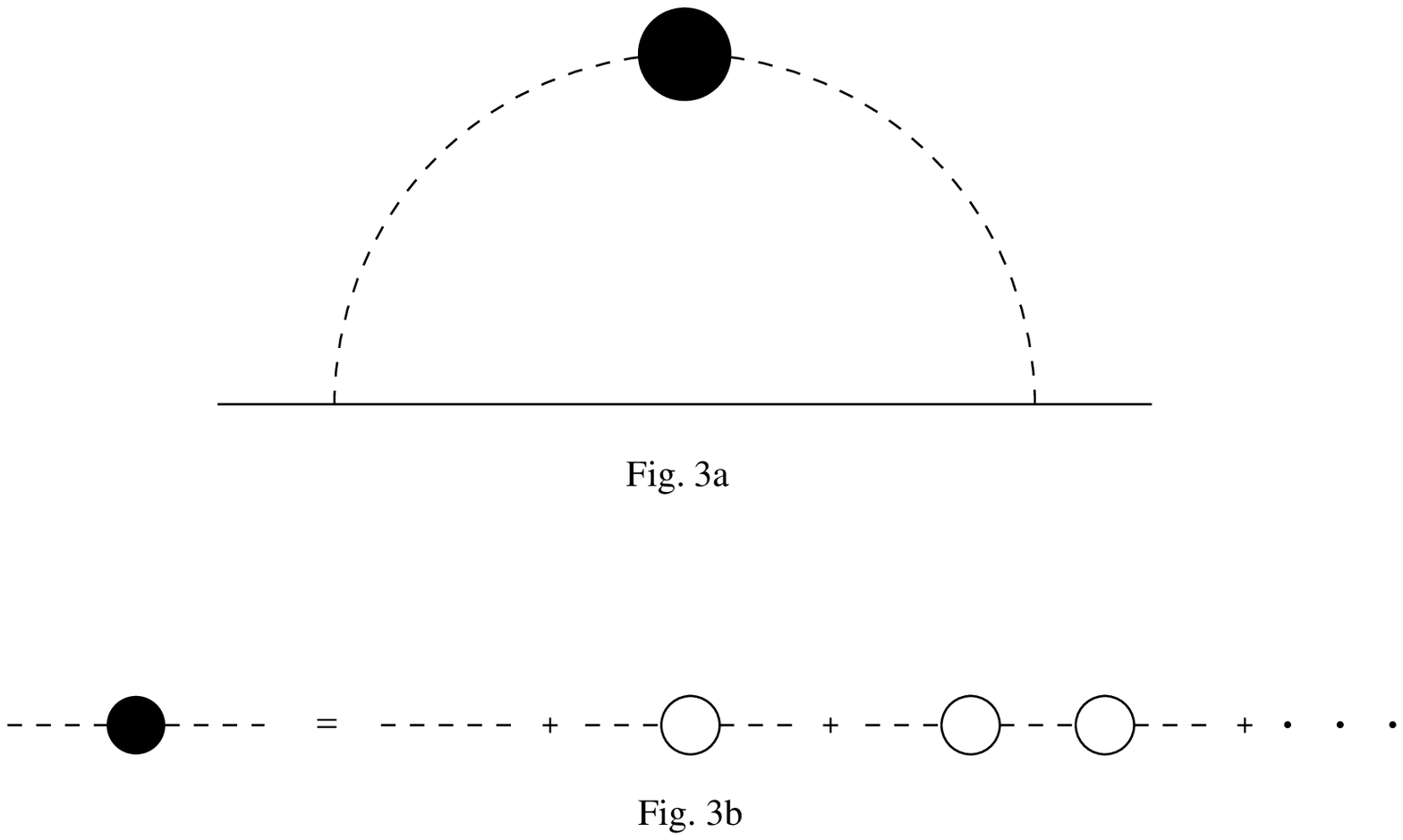,width=15.5cm}

\vspace{-7.5cm}

{\footnotesize
\noindent {\bf Fig.~3}: the 1-PI diagrams with two external 
top quark legs which give the leading 
(${\cal {O}}(1/N_{\mbox{\scriptsize c}})$) contribution to the 
renormalization of the mass $m_{t} $. Unlike the diagrams of 
Figs.~1-2, the top quark propagators here contain the nonzero
bare mass $m_{t}({\Lambda}) $ which was the solution to the 
NTL gap equation. The dashed lines are the propagators of the
nondynamical scalars 
(either the Higgs, or the neutral Goldstones, or the charged 
Goldstones), while the dashed lines with a black blob are
the propagators of the corresponding scalars that became dynamical
through the NTL quantum effects.
For the case of charged Goldstone propagators, the 
fermion loops contain one massive top quark and one massless 
bottom quark.
}

\end{document}